\documentclass[twocolumn]{aastex63}

\usepackage{xfrac}
\usepackage{float}
\usepackage{graphicx}

\graphicspath{{./}{figures/}}

\received{}
\revised{}
\accepted{May 21, 2021}
\submitjournal{ApJ}

\shorttitle{Variations in the FUor-type star V1057~Cyg}
\shortauthors{Szab\'o et al.}

\graphicspath{{./}{figures/}}

\begin{document}

\title{A study of the photometric and spectroscopic variations of the prototypical FU~Orionis-type star V1057~Cyg}

\correspondingauthor{Zs\'ofia M. Szab\'o}
\email{szabo.zsofia@csfk.org}

\author[0000-0001-9830-3509]{Zs. M. Szab\'o}
\affiliation{Konkoly Observatory, Research Centre for Astronomy and Earth Sciences, E\"otv\"os Lor\'and Research Network (ELKH), Konkoly-Thege Mikl\'os \'ut 15-17, 1121 Budapest, Hungary}
\affiliation{E\"otv\"os Lor\'and University, Department of Astronomy, P\'azm\'any P\'eter s\'et\'any 1/A, 1117 Budapest, Hungary}

\author[0000-0001-7157-6275]{\'A. K\'osp\'al}
\affiliation{Konkoly Observatory, Research Centre for Astronomy and Earth Sciences, E\"otv\"os Lor\'and Research Network (ELKH), Konkoly-Thege Mikl\'os \'ut 15-17, 1121 Budapest, Hungary}
\affiliation{Max Planck Institute for Astronomy, K\"onigstuhl 17, D-69117 Heidelberg, Germany}
\affiliation{ELTE Eötvös Loránd University, Institute of Physics, Pázmány Péter sétány 1/A, H-1117 Budapest, Hungary}

\author[0000-0001-6015-646X]{P. \'Abrah\'am}
\affiliation{Konkoly Observatory, Research Centre for Astronomy and Earth Sciences, E\"otv\"os Lor\'and Research Network (ELKH), Konkoly-Thege Mikl\'os \'ut 15-17, 1121 Budapest, Hungary}
\affiliation{ELTE Eötvös Loránd University, Institute of Physics, Pázmány Péter sétány 1/A, H-1117 Budapest, Hungary}

\author[0000-0003-4099-1171]{S. Park}
\affiliation{Konkoly Observatory, Research Centre for Astronomy and Earth Sciences, E\"otv\"os Lor\'and Research Network (ELKH), Konkoly-Thege Mikl\'os \'ut 15-17, 1121 Budapest, Hungary}

\author[0000-0001-5018-3560]{M. Siwak}
\affiliation{Konkoly Observatory, Research Centre for Astronomy and Earth Sciences, E\"otv\"os Lor\'and Research Network (ELKH), Konkoly-Thege Mikl\'os \'ut 15-17, 1121 Budapest, Hungary}

\author[0000-0003-1665-5709]{J. D. Green}
\affiliation{Space Telescope Science Institute, 3700 San Martin Dr., Baltimore, MD 21218, USA}

\author{A. Mo\'or}
\affiliation{Konkoly Observatory, Research Centre for Astronomy and Earth Sciences, E\"otv\"os Lor\'and Research Network (ELKH), Konkoly-Thege Mikl\'os \'ut 15-17, 1121 Budapest, Hungary}
\affiliation{ELTE Eötvös Loránd University, Institute of Physics, Pázmány Péter sétány 1/A, H-1117 Budapest, Hungary}

\author[0000-0001-5449-2467]{A. P\'al}
\affiliation{Konkoly Observatory, Research Centre for Astronomy and Earth Sciences, E\"otv\"os Lor\'and Research Network (ELKH), Konkoly-Thege Mikl\'os \'ut 15-17, 1121 Budapest, Hungary}
\affiliation{E\"otv\"os Lor\'and University, Department of Astronomy, P\'azm\'any P\'eter s\'et\'any 1/A, 1117 Budapest, Hungary}
\affiliation{ELTE Eötvös Loránd University, Institute of Physics, Pázmány Péter sétány 1/A, H-1117 Budapest, Hungary}
\affiliation{MIT Kavli Institute for Astrophysics and Space Research, 70 Vassar Street, Cambridge, MA 02109, USA}

\author[0000-0002-0433-9656]{J. A. Acosta-Pulido}
\affiliation{Instituto de Astrofísica de Canarias, Avenida Vía Láctea, Tenerife, Spain}
\affiliation{Departamento de Astrofísica, Universidad de La Laguna, Tenerife, Spain}

\author[0000-0003-3119-2087]{J.-E. Lee}
\affiliation{School of Space Research, Kyung Hee University, 1732, Deogyeong-daero, Giheung-gu, Yongin-si, Gyeonggi-do 17104, Republic of Korea}

\author{B. Cseh}
\affiliation{Konkoly Observatory, Research Centre for Astronomy and Earth Sciences, E\"otv\"os Lor\'and Research Network (ELKH), Konkoly-Thege Mikl\'os \'ut 15-17, 1121 Budapest, Hungary}
\author{G. Csörnyei}
\affiliation{Konkoly Observatory, Research Centre for Astronomy and Earth Sciences, E\"otv\"os Lor\'and Research Network (ELKH), Konkoly-Thege Mikl\'os \'ut 15-17, 1121 Budapest, Hungary}
\author{O. Hanyecz}
\affiliation{Konkoly Observatory, Research Centre for Astronomy and Earth Sciences, E\"otv\"os Lor\'and Research Network (ELKH), Konkoly-Thege Mikl\'os \'ut 15-17, 1121 Budapest, Hungary}
\author[0000-0002-8770-6764]{R. Könyves-Tóth}
\affiliation{Konkoly Observatory, Research Centre for Astronomy and Earth Sciences, E\"otv\"os Lor\'and Research Network (ELKH), Konkoly-Thege Mikl\'os \'ut 15-17, 1121 Budapest, Hungary}

\author[0000-0002-8813-4884]{M. Krezinger}
\affiliation{Konkoly Observatory, Research Centre for Astronomy and Earth Sciences, E\"otv\"os Lor\'and Research Network (ELKH), Konkoly-Thege Mikl\'os \'ut 15-17, 1121 Budapest, Hungary}
\affiliation{E\"otv\"os Lor\'and University, Department of Astronomy, P\'azm\'any P\'eter s\'et\'any 1/A, 1117 Budapest, Hungary}

\author{L. Kriskovics}
\affiliation{Konkoly Observatory, Research Centre for Astronomy and Earth Sciences, E\"otv\"os Lor\'and Research Network (ELKH), Konkoly-Thege Mikl\'os \'ut 15-17, 1121 Budapest, Hungary}
\affiliation{ELTE Eötvös Loránd University, Institute of Physics, Pázmány Péter sétány 1/A, H-1117 Budapest, Hungary}

\author{A. Ordasi}
\affiliation{Konkoly Observatory, Research Centre for Astronomy and Earth Sciences, E\"otv\"os Lor\'and Research Network (ELKH), Konkoly-Thege Mikl\'os \'ut 15-17, 1121 Budapest, Hungary}
\author[0000-0003-0926-3950]{K. Sárneczky}
\affiliation{Konkoly Observatory, Research Centre for Astronomy and Earth Sciences, E\"otv\"os Lor\'and Research Network (ELKH), Konkoly-Thege Mikl\'os \'ut 15-17, 1121 Budapest, Hungary}
\author[0000-0002-3658-2175]{B. Seli}
\affiliation{Konkoly Observatory, Research Centre for Astronomy and Earth Sciences, E\"otv\"os Lor\'and Research Network (ELKH), Konkoly-Thege Mikl\'os \'ut 15-17, 1121 Budapest, Hungary}
\affiliation{E\"otv\"os Lor\'and University, Department of Astronomy, P\'azm\'any P\'eter s\'et\'any 1/A, 1117 Budapest, Hungary}

\author[0000-0002-1698-605X]{R. Szakáts}
\affiliation{Konkoly Observatory, Research Centre for Astronomy and Earth Sciences, E\"otv\"os Lor\'and Research Network (ELKH), Konkoly-Thege Mikl\'os \'ut 15-17, 1121 Budapest, Hungary}
\author{A. Szing}
\affiliation{Konkoly Observatory, Research Centre for Astronomy and Earth Sciences, E\"otv\"os Lor\'and Research Network (ELKH), Konkoly-Thege Mikl\'os \'ut 15-17, 1121 Budapest, Hungary}
\author[0000-0002-6471-8607]{K. Vida}
\affiliation{Konkoly Observatory, Research Centre for Astronomy and Earth Sciences, E\"otv\"os Lor\'and Research Network (ELKH), Konkoly-Thege Mikl\'os \'ut 15-17, 1121 Budapest, Hungary}

\begin{abstract}
Among the low-mass pre-main sequence stars, a small group called FU Orionis-type objects (FUors) are notable for undergoing powerful accretion outbursts. 
V1057~Cyg, a classical example of an FUor, went into outburst around 1969 -- 1970, after which it faded rapidly, making it the fastest fading FUor known. Around 1995, a more rapid increase in fading occurred. Since that time, strong photometric modulations have been present.
We present nearly 10 years of source monitoring at Piszkéstető Observatory, complemented with optical/near-infrared photometry and spectroscopy from the Nordic Optical Telescope, Bohyunsan Optical Astronomy Observatory, Transiting Exoplanet Survey Satellite, and the Stratospheric Observatory for Infrared Astronomy. Our light curves show continuation of significant quasi-periodic variability in brightness over the past decade.
Our spectroscopic observations show strong wind features, shell features, and forbidden emission lines. 
All of these spectral lines vary with time. We also report the first detection of [S\,{\footnotesize II}], [N\,{\footnotesize II}], and [O\,{\footnotesize III}] lines in the star.
\end{abstract}


\section{Introduction}
\label{sec:intro}

Photometric and spectroscopic monitoring of pre-main sequence (PMS) stars over a broad spectral range is crucial to understand the mechanisms leading to the formation of stars and ultimately planets. A small, but spectacular class of low-mass young stars are known as FU~Orionis-type stars (FUors), referring to the nova-like eruption of the archetype FU~Ori in 1936 \citep{wachmann1954}. \citet{herbig1966} argued that the outburst represented a newly uncovered phenomenon in the early protostellar evolution, rather than a classical nova (associated with an evolved star). A decade later, after a few similar outbursts were observed, \citet{herbig1977} defined the FUor class. These young eruptive stars are characterized by enormous increases in the brightness of their inner circumstellar disk, due to enhanced accretion from the disk onto the star caused by disk instabilities. These eruptions last for several decades and likely even centuries \citep{paczynski1976,lin1985, kenyon-and-hartmann1988,kenyon&hartmann1991, bell1995, turner1997, audard2014, kadam2020}. 

The members of this group, currently about 30 objects \citep{audard2014} show very similar optical spectra: F or G supergiants with wide absorption lines, H$\alpha$ P Cygni profiles, shell components, and strong Li\,{\footnotesize I} 670.7\,nm absorption. 
During a FUor eruption, the disk outshines the luminosity of the central star. Assuming that the bolometric luminosity calculated from the observed spectral energy distribution (SED) is dominated by the accretion luminosity, the accretion rate during the FUor stage can directly be obtained. Observations showed that the accretion rate rises from the average rate of a typical T Tauri star ($10^{-9}$ -- $10^{-7}$ \(M_\odot\) yr$^{-1}$) up to $10^{-5}$ -- $10^{-4}$ \(M_\odot\) yr$^{-1}$ in only a few months  \citep{kenyon&hartmann1996}.

V1057~Cyg became the second identified FUor in 1969, when it brightened by 6\,mag in the $V$-band \citep{welin1971a, welin1971b}. The source is located in the North America Nebula (NGC 7000), which, together with the Pelican Nebula (IC 5070) form a large HII region \citep{wendker1983, rebull2011}. Previous distance estimates for these regions vary between $520$ and $700$ pc \citep{laugalys2006, skinner2009, fischer2012}. In a recent work, \citet{kuhn2020} determined new distances for the members of the North America Nebula using Gaia DR2 astrometry \citep{gaiadr2}. They found that the main parts of the North America and Pelican Nebula are located at $\sim$795 pc, however V1057~Cyg, as a part of a smaller group of stars is located somewhat farther away. In this paper we adopt the Gaia DR2 distance value of $897$ pc from \citet{bailerjones1, bailerjones2} which was specifically determined for V1057~Cyg. 

\citet{herbig1977} studied V1057~Cyg in detail both photometrically and spectroscopically. They concluded that before its eruption, the object had shown the properties of a classical T Tauri-type star (CTTS). They also characterised a $1\farcm0 \times 1\farcm5$ ring-like nebula which appeared around the object after the outburst. Further observations showed that the ring faded with the central star in the following years, but its structure remained unchanged. This indicated that the ring was a reflection nebula: a structure already present before the eruption of V1057~Cyg, illuminated by the central source, and not material that had been blown out during the eruption.

Three decades later \citet{herbig2003} presented another detailed spectroscopic study focusing on this star. These high-resolution spectra, taken in 1996--2002, confirmed some of the previously observed features, such as the `doubling’ of low-excitation absorption lines which became more apparent between the 1980's and 1994. In this subsequent study, \citet{herbig2009} pointed out that V1057~Cyg has a long-lasting, high-velocity wind, which manifests itself through strong blueshifted absorption components at various optical lines. 

The last photometric analysis of V1057~Cyg was performed by \citet{kopatskaya2013}, who demonstrated that immediately after reaching the light maximum in 1970, the light curve started an exponential brightness decline until $\sim$1985, when the so called `first plateau' phase started and lasted for about 10 years. After that, the source faded by $\sim$0.5--1 mag in the optical within a year, and started to show quasi-periodic variations. The authors found that the variations could be characterized with two different periods: a longer period $1631\pm60$\,d, dominating the $BVR$ data, and a shorter one $523\pm40$\,d, dominating the $IJHK$ data. They initially concluded that these fluctuations reflected the binary nature of V1057~Cyg, which has also been proposed as a possible mechanism leading to enhanced accretion and the FUor phenomenon (e.g. \citealt{bonnell1992,bell1995}). Interestingly, using non-redundant aperture-masking interferometry, \citet{green2016} detected a faint companion star of V1057~Cyg, located at a projected separation of $58$\,mas with a brightness difference of $\Delta K=3.3$\,mag. Its distance from V1057~Cyg suggests that it could have triggered the original outburst with a close fly-by encounter \citep{vorobyov2021}.

\citet{connelley2018} published a near-infrared (NIR) spectroscopic survey including V1057~Cyg with observations from 2015, the latest NIR spectroscopic data of the source. They concluded that the CO absorption band was much weaker than in 1986. In contrast, the first high-resolution NIR spectroscopic observations of \citet{kenyon-and-hartmann1987} showed that the CO features have not changed much compared to \citet{mould1978}. \citet{biscaya1997} showed that the CO features became weaker in 1996 than in 1986 \citep{kenyon-and-hartmann1987} and interpreted that this weakening might be related to the brightness decline in 1995.

The infrared excess emission apparent in the SED of V1057~Cyg is due to a flared disk and envelope geometry \citep{kenyon1991}. The presence of an envelope was also confirmed by \citet{green2006} based on  $5-35\,\mu$m {\it Spitzer}/IRS observations with an estimated radius of 7000\,au. \citet{zhu2008} modeled the dust from {\it Spitzer}/IRS observations and found that an envelope typical of protostars is required for V1057~Cyg to match the observations. \citet{green2013} found the observed {\it Herschel} spectra were generally brighter than model predictions, which indicated an underestimate of the large scale reservoir of cold dust surrounding FUors. These works also suggested the idea of a large bipolar cavity in the envelope. \citet{feher2017} surveyed northern hemisphere FUors with the Plateau de Bure Interferometer (PdBI) and the IRAM 30 m telescope. Based on $^{13}$CO observations, they found a rotating envelope around V1057~Cyg which is roughly spherical with a radius of $5''$ ($3000$\,au) and a total circumstellar mass of $0.21\,M_{\odot}$.

Despite significant fading, the last visual spectrum of V1057~Cyg obtained in 2012 by \citet{lee2015} did not resemble that of a CTTS,
thus, further monitoring is key in tracing the gradual return of V1057~Cyg to quiescence.
We have occasionally observed our target in optical and infrared bands since 2005, but intensified our monitoring after 2011, due to increased telescope time.

We describe the new observations and our reduction methods in Section \ref{sec:obs}. Results obtained from the data analysis are presented in Sec.~\ref{sec:res} and discussed in Sec.~\ref{sec:discussion}. We summarize our findings in Sec~\ref{sec:conclusions}.

\section{Observations and data reduction} 
\label{sec:obs}
\subsection{Ground-based optical photometry}

We performed the majority of our photometric observations in $B$, $V$, $R_{\rm C}$, $I_{\rm C}$, $g'$, $r'$, and $i'$ filters at the Piszkéstető Mountain Station of Konkoly Observatory (Hungary) between 2005 and 2021. Three telescopes with three slightly different optical systems were used. In 2005--2007 we observed the star with the 1\,m Ritchey-Chrétien-coudé (RCC) telescope, equipped with a 1300$\times$1340 pixel Roper Scientific VersArray: 1300B CCD camera (pixel scale: 0$\farcs$306). The 60/90/180\,cm Schmidt telescope, equipped with a $4096\times4096$ pixel Apogee Alta U16 CCD camera (pixel scale: 1$\farcs$027), was used in 2011--2019. In each of the $BVR_{\rm C}I_{\rm C}$ filters, typically three images per night were taken. Since 2020 we started to use the Astro Systeme Austria 
AZ800 alt-azimuth direct drive 80-cm Ritchey-Chretien (RC80) telescope operating in fully autonomous mode. The optical setup with the effective focal length of $F=5600\,{\rm mm}$ yielded a pixel scale of 0$\farcs$55 and a field-of-view of $18{\farcm}8{\times}18{\farcm}8$ for a 2048$\times$2048 pixel FLI PL230 CCD camera. We obtained three images per night in $BVg'r'i'$ filters.

The frames were calibrated for bias, dark, and flatfield in the standard fashion. Photometry of V1057~Cyg and 12 comparison stars in its 8$'$ vicinity was extracted using an aperture radius of 4$\farcs$1 and sky annulus between 10$\farcs$3 -- 15$\farcs$4 for RCC and Schmidt frames, and 5$\farcs$5 and sky annulus between 11$''$ and 22$''$ for RC80 telescope frames. In order to eliminate system-related effects, photometric calibration was performed by fitting a color term using the magnitudes and colors of the comparison stars from the APASS DR9 catalog \citep{henden2016}, after converting them from the Sloan to the Bessel system using transformations from \citet{jordi2006}. We note that many Schmidt observations actually targeted another, fainter young eruptive star, HBC~722 \citep{kospal2011, kospal2016}, and V1057~Cyg just happened to be in the field of view. As a consequence, V1057~Cyg saturated the detector in some of the $R_{\rm C}$ and $I_{\rm C}$ images, which were discarded from further analyses.

Except of our national facilities, we occasionally used other telescopes. On 2006 July 20 and 2012 October 13 we obtained  $B$, $V$, $R_{\rm J}$ and $I_{\rm J}$ images of V1057~Cyg with the IAC80 telescope of the Instituto de Astrof\'\i{}sica de Canarias located at Teide Observatory (Canary Islands, Spain). It was equipped with the Tromsoe CCD Photometer (TCP) with a 9$\farcm$2$\times$9$\farcm$0 field of view and a 0$\farcs$537 pixel scale. After the standard reduction steps for bias, dark, and flatfield correction, aperture photometry was done by using the same aperture and sky annulus size as for the Schmidt and RCC data. Photometric calibration was done using the same comparison stars, except for the two that fell outside the smaller field of view of the telescope. During 2019 August--September, in parallel with {\it TESS}, we additionally observed V1057~Cyg at the Northern Skies Observatory (NSO). We used the 0.4\,m telescope equipped with $BVI$ filters, operated remotely through {\sc Skynet}. The calibration procedures and comparison stars were the same as above, but only the $VI$ NSO filter data was of analysis quality.

We also observed V1057~Cyg with the $2.56$\,m Nordic Optical Telescope (NOT) at the Roque de los Muchachos Observatory, La Palma in the Canary Islands (Plan ID 61--414, PI: Zs.~M.~Szabó). For optical imaging we used the Alhambra Faint Object Spectrograph and Camera (ALFOSC) on 2020 August 17. ALFOSC is a $2048 \times 2064$ pixel CCD231-42-g-F61 CCD camera with a field of view of $6{\farcm}4{\times}6{\farcm}4$ and pixel scale of $0\farcs21$. The Bessel $BVR$ filter set was supplemented by an $i$ interference filter, which is similar to the SLOAN $i'$, but with a slightly longer effective wavelength of $\lambda_{\rm eff}=0.789\mu$m. We obtained three images in each filter, with exposure times between $1.5-30$\,s. After the standard CCD reduction steps, we obtained aperture photometry using an aperture radius of $3\farcs2$ and a sky annulus between $6\farcs4$ and $8\farcs6$. Because of the small field of view, the magnitudes of V1057~Cyg were obtained based on only one comparison star.

Our photometric results are shown in Fig.~\ref{fig:lc} and~\ref{fig:lc-piszkes}, and listed in Tab.~\ref{tab:phot} in the Appendix The typical uncertainty of our measurements is 0.03\,mag in $B$ and 0.01\,mag in all other filters. 

\subsection{Space-based optical and infrared photometry}

During 2019 August 15 -- October 7, V1057~Cyg was observed with 30-minute cadence with Camera 1 of the Transiting Exoplanet Survey Satellite ({\it TESS}, \citealt{ricker2015}). The total coverage time of {\it Sectors 15 and 16} of the satellite is 50.5625 days, but the run was interrupted three times, each for about 3.1--3.4 days to download the data to the MAST archive\footnote{\url{https://mast.stsci.edu}}. The calibrated full-frame images were processed in two main steps using the FITSH package \citep{pal2012}. Firstly the plate solution was derived based on the Gaia DR2 catalogue -- details of this complex procedure are described by \citet{2020ApJS..247...26P}. As the part of this step, we derived the flux zero-point with respect to the $G_{\rm RP}$ magnitudes of the matched Gaia sources, utilizing the similarities between the {\sl TESS} and Gaia $G_{\rm RP}$ filters throughputs. By examining various {\sl TESS} fields observed in the first two sectors we found that the RMS of our zero-level calibration is $\sim$0.015\,mag. The photometry of the source was performed via differential image analysis using FITSH/\texttt{ficonv} and \texttt{fiphot} \citep{pal2012}. It requires a reference frame, which we constructed as a median of 11 individual 64$\times$64 subframes obtained close to the middle of the observing sequence. As reference fluxes, required to correct for various instrumental and intrinsic differences between the target and the reference frames, we used the Gaia DR2 magnitudes. Data points affected by momentum wheel desaturation or significant stray light were flagged and removed, what caused three additional 1.2--1.3~d breaks in the time coverage. The resulting typical formal uncertainties of the data are about 0.65\,mmag. The {\sl TESS} light curve of V1057\,Cyg is presented in Fig.~\ref{fig:tess}. 

We complemented our work with data from the Wide-field Infrared Survey Explorer (\textit{WISE}, \citep{wright2010}). We used data obtained in the $3.4\,\mu$m (W1) and $4.6\,\mu$m (W2) bands from 2010 up until the most recent data release in 2021 \citep{cutri2012,cutri2014}.
Since V1057~Cyg was saturated, we corrected the data points using the saturation bias correction curves for the appropriate survey phase available in the \textit{WISE} Explanatory Supplement\footnote{\url{https://wise2.ipac.caltech.edu/docs/release/neowise/expsup/sec2_1civa.html}}.
The corrected WISE data are shown in Fig.~\ref{fig:lc} and listed in Tab.~\ref{tbl_phot_wise} in the Appendix.

\subsection{Near-infrared photometry}

We obtained near-infrared images in the $J$, $H$ and $K_{\rm s}$ bands at six epochs between 2006 July 15 and 2012 October 13 using the 1.52\,m Telescopio Carlos Sanchez (TCS) at the Teide Observatory. This telescope is equipped with CAIN\,III, a $256\times256$ Nicmos 3 detector, which provided a pixel scale of $1^{\prime\prime}$ in the wide optics configuration. Observations were performed in a 5-point dither pattern in order to enable proper sky subtraction. The total integration time was typically 1\,min per dither position in each filter, split into $1.5-5$\,s exposures. The images were reduced using \texttt{caindr}, an {\sc IRAF}-based data reduction package written by J. A. Acosta-Pulido and R. Barrena\footnote{\url{http://vivaldi.ll.iac.es/OOCC/iac-managed-telescopes/telescopio-carlos-sanchez/cain-iii/}}, as well as our own {\sc IDL} routines. Data reduction steps included sky subtraction, flat-fielding, registration, and coadding exposures by dither position and filter. To calibrate our photometry, we used the Two Micron All Sky Survey (2MASS) catalog \citep{cutri2003}. The instrumental magnitudes of the target and all good-quality 2MASS stars in the field were extracted 
using an aperture radius of 2$''$ in all filters. We determined a constant offset between the instrumental and the 2MASS magnitudes by averaging typically $20-30$ stars by means of biweight\textunderscore mean -- an outlier-resistant averaging method. 

We also used the NOTCam instrument on the NOT on 2020 August 29. The instrument includes a 1024$\times$1024 pixel HgCdTe Rockwell Science Center `HAWAII' array and for wide field (WF) imaging it has a $4^{\prime}\times 4^{\prime}$ field-of-view (pixel scale: 0$\farcs$234). We obtained 9 images in each of the $JHK_s$ bands with 3.6 s exposures. Because of the brightness of our target in the infrared, we used a 5 mm diameter pupil mask intended for very bright objects to diminish the telescope aperture, which gave about 10\% transmission. The images were reduced using the same method as described above at the TCS data reduction. The instrumental magnitudes of the target and the comparison star in the field were extracted 
using aperture radius of $3\farcs3$ and a sky annulus between $6\farcs6$ and $9\farcs4$. The photometric calibration was performed in the same fashion as the TCS images. Typical photometric uncertainties are of $0.01-0.03$\,mag, and we present the results of the optical and infrared photometry in Appendix A, Tab.~\ref{tab:phot}.

\subsection{Optical spectroscopy}

We obtained a new optical spectrum of V1057~Cyg with the high-resolution FIbre-fed Echelle Spectrograph (FIES) instrument on the NOT on 2020 August 17. We used a fibre with a larger entrance aperture of $2\farcs5$  which provided a spectral resolution $R$=25\,000, covering the $370 - 900$\,nm wavelength range. We obtained two spectra, each with 1800 s exposure time. During our analysis we used the spectra reduced by the FIEStool software.

V1057~Cyg was also observed with the Bohyunsan Optical Echelle Spectrograph \citep[BOES;][]{kim2002} installed on the 1.8\,m telescope at the Bohyunsan Optical Astronomy Observatory (BOAO). It provides $R$=30\,000 in the wavelength range $\sim$~$400-900$\,nm. The first spectrum was obtained on 2012 September 11 and the last on 2018 December 18.
We reduced these spectra in a standard way within {\sc IRAF}: after standard calibrations on bias and flatfield, the ThAr lamp spectrum was used for wavelength calibration, and continuum fitting was performed by \texttt{continuum} task. Finally, heliocentric velocity correction was applied by the \texttt{rvcorrect} task and the published radial velocity of V1057 Cyg \citep[$-$16 km s$^{-1}$;][]{herbig2003}.

As no telluric standard stars were observed neither for FIES nor BOES, we performed the telluric correction using the molecfit software \citep{smette2015,kausch2015} by fitting the telluric absorption bands of O$_2$ and H$_2$O. This generally provided good correction except for the deepest lines where the detected signal was close to zero. 

We present the spectroscopic observing log in Tab.~\ref{tbl_obs_log}.

\subsection{Near-infrared spectroscopy}

On 2020 August 29, we used the NOTCam on the NOT to obtain new near-infrared spectrum of V1057~Cyg and Iot~Cyg (A5 V) as our telluric standard star in the $JHK_s$ bands. We used the low-resolution camera mode ($R$=2500) with ABBA dither positions, and exposure times ranged from 25 to 60 seconds (Tab.~\ref{tbl_obs_log}). For each image, flat-fielding, bad pixel removal, sky subtraction, aperture tracing, and wavelength calibration steps were performed within {\sc IRAF}. For the wavelength calibration, the Xenon lamp spectrum was used. The Hydrogen absorption lines in Iot~Cyg were removed by Gaussian fitting. Then the spectrum of V1057~Cyg was divided by the normalized spectrum of Iot~Cyg for telluric correction. Finally, flux calibration was performed by applying the accretion disk model obtained using the NOT $JHK_s$ photometry (Sec.~\ref{sec:accdisc}).

\subsection{Mid-infrared observations}

On 2018 September 6, we observed V1057~Cyg with the Stratospheric Observatory for Infrared Astronomy (SOFIA; \citealt{young12}) using the Faint Object infraRed CAmera for the SOFIA Telescope (FORCAST; \citealt{herter13}). We obtained mid-infrared imaging in a series of short exposures in band F111 ($10.6-11.6$ $\mu$m) totaling $\sim$30s; a single exposure in F056 (5.6 $\mu$m) for 37s and F077 ($7.5-8$ $\mu$m) for 42s, as well as R$\sim$~$100-200$ spectra with G063 ($5-8$ $\mu$m) and G227 ($17-27$ $\mu$m) (Plan ID 06\textunderscore062, PI: J.~D.~Green). The spectra were processed using the SOFIA pipeline and retrieved as Level 3 data products from the SOFIA Science Archive as ingested into the IRSA database\footnote{Further information can be found at \url{https://nbviewer.jupyter.org/github/SOFIAObservatory/Recipes/blob/master/FORCAST-Grism_Inspection.ipynb}}. The program was only partially observed in SOFIA Cycle 6, and thus the data do not cover the full $5-25$ $\mu$m spectral range. The observations were performed in `C2N' (2-position chop with nod) mode, using the 4$\farcs$7 slit, and the NMC (nod-match-chop) pattern in which the chops and nod are in the same direction and have the same amplitude. In each case, an off-source calibrator was selected, using the observation closest in zenith angle and altitude to the science target, as previously done with FU~Orionis in SOFIA Cycle 4 \citep{green2016}. We did not use dithering. 

\begin{deluxetable}{ccccc}
\tabletypesize{\scriptsize} 
\tablecaption{Log of Spectroscopic Observations \label{tbl_obs_log}}
\tablewidth{0pt}
\tablehead{
\colhead{Telescope} & \colhead{Instrument} & \colhead{Spectral} & \colhead{Observation} & \colhead{Exp. time} \\[-3mm]
&  & \colhead{Resolution} & \colhead{Date [UT]} & \colhead{[sec]}}
\startdata
BOAO & BOES & 30,000 & 2012 Sep 11 & 3600 \\
$\cdots$ & $\cdots$ & $\cdots$ & 2015 Dec 27 & 3600 \\
$\cdots$ & $\cdots$ & $\cdots$ & 2017 May 29 & 3600 \\
$\cdots$ & $\cdots$ & $\cdots$ & 2018 Oct 07 & 3600 \\
$\cdots$ & $\cdots$ & $\cdots$ & 2018 Dec 18 & 3600 \\
 \cline{1-5}
NOT & FIES & 25,000 & 2020 Aug 18 & 1800 $\times$ 2 \\ 
 \cline{1-5}
NOT & NOTCam (J) & 2,500 & 2020 Aug 29 &  240$^{a}$ \\ 
$\cdots$ & NOTCam (H) & $\cdots$ & 2020 Aug 29 & 140$^{a}$ \\ 
$\cdots$ & NOTCam (K) & $\cdots$ & 2020 Aug 29 & 120$^{a}$ \\ 
\enddata
\label{tab:spec_dates}
\tablenotetext{a}{Total integration time of each target (exposure time $\times$ the number of exposures (ABBA) = total integration time).}
\end{deluxetable}
\section{Results and Analysis} \label{sec:res}

\subsection{Light curves}
\label{sec:lc}

\begin{figure*}
\includegraphics[width=\textwidth]{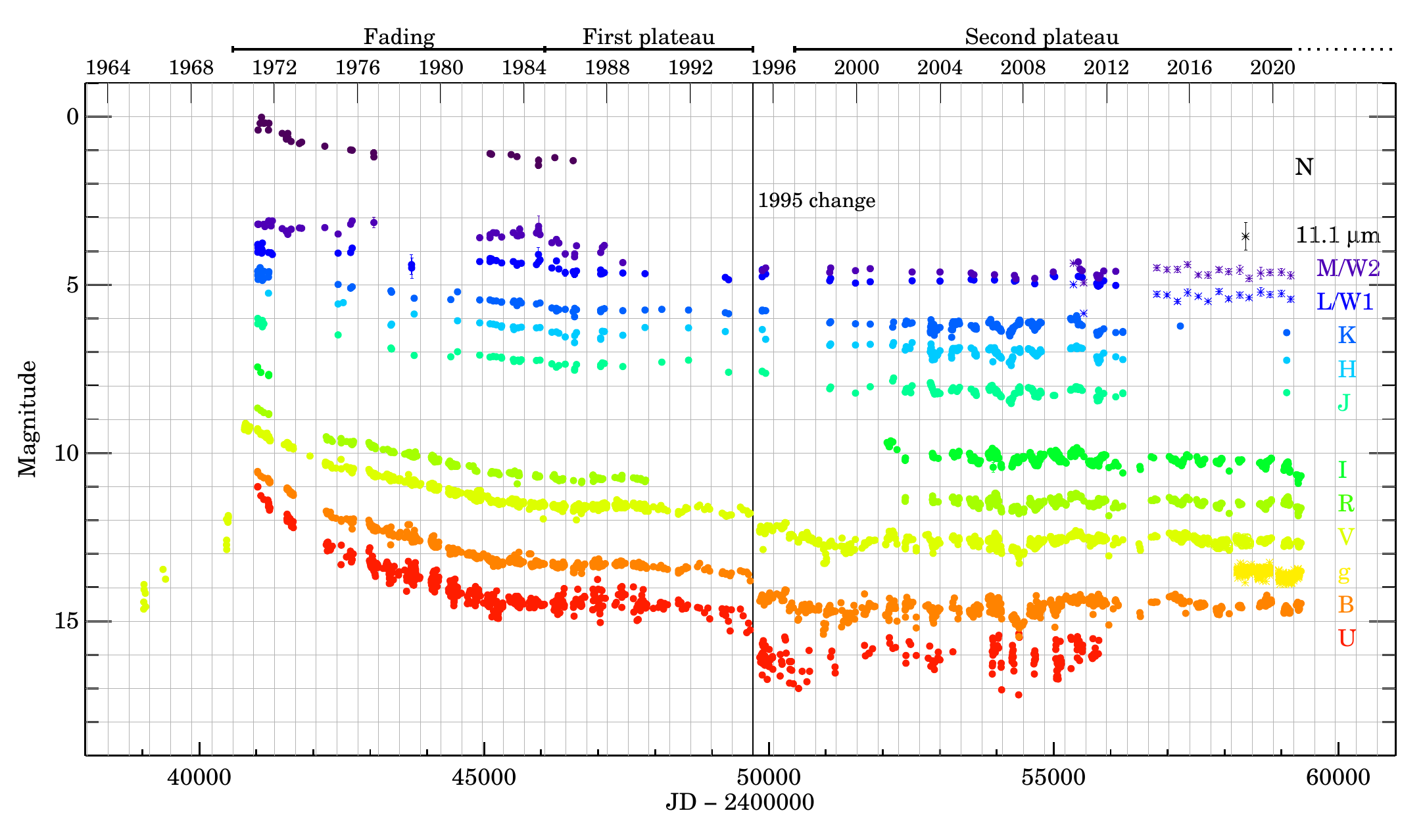}
\caption{Optical and infrared light curves of V1057~Cyg. We complemented our light curves with optical and infrared data prior to 2012 from \citet{mendoza1971, rieke1972, welin1975, welin1976, landolt1975, landolt1977, simon1975, simon1982, simon1988, kenyon1991, kopatskaya2013, green2016}.}
\label{fig:lc}
\end{figure*}

\begin{figure*}
\includegraphics[width=\textwidth]{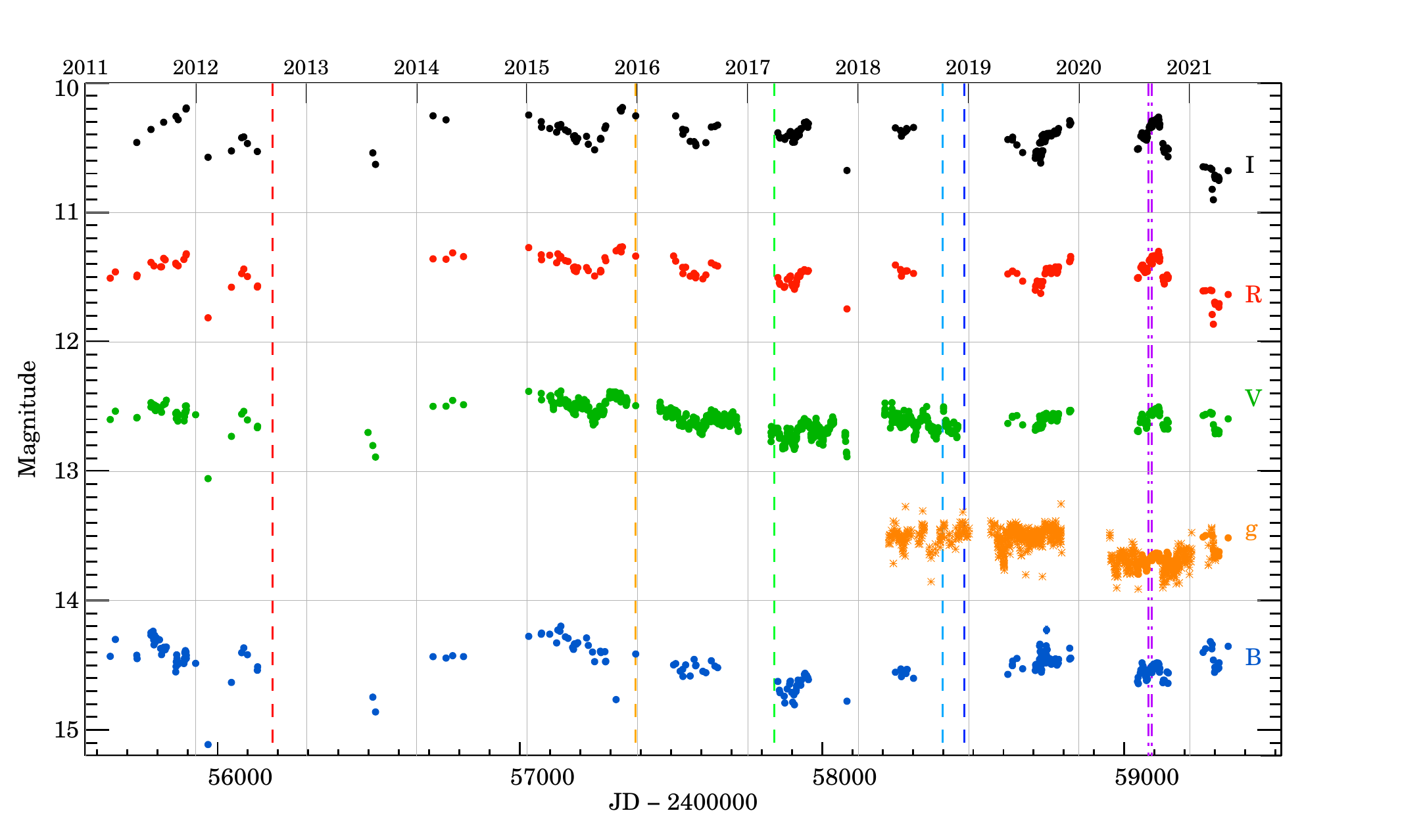}
\caption{Optical light curves of V1057~Cyg. The $BVR_CI_C$ data were obtained at Piszkéstető Observatory while some parts of the $V$ and $g$-band data are from the ASAS-SN archive. Vertical dotted lines mark our BOAO observations from 2012, 2015, 2017, 2018, while dash-dotted lines show our NOT observations in 2020. The colors are the same as the spectroscopic figures in Sec.~\ref{sec:optical_spectroscopy}.
}
\label{fig:lc-piszkes}
\end{figure*}

To study the long-term variability of V1057~Cyg, we complemented our work with data published in the literature \citep{mendoza1971,rieke1972,welin1975,welin1976,landolt1975,landolt1977,simon1975,simon1982,simon1988,kenyon1991,kopatskaya2013}. Our V1057~Cyg monitoring began in 2005 and overlapped with that of \citet{kopatskaya2013}. This enabled us to determine systematic shifts between filters utilised in these two data sets. We found systematic differences between the two sets of photometry, which may be due to different apertures, filters, detectors throughputs, and different comparison stars used. For plotting purposes, we shifted our $B$ band light curves by $+$0.12\,mag, $V$ band by $+$0.08\,mag, $R_{\rm C}$ band by $+$0.05\,mag and $I_{\rm C}$ band light curves by $-$0.14\,mag to be consistent with the earlier papers. In Appendix A Tab.~\ref{tab:phot} we present our original photometry, i.e. without these offsets. The resulting long-term light curves covering the 1965--2021 time period are shown in Fig.~\ref{fig:lc}, while in Fig.~\ref{fig:lc-piszkes} we show in detail our Piszkéstető optical monitoring (starting from 2011), complemented with $V$ and $g$-band observations from the All-Sky Automated Survey for Supernovae \citep[{\it ASAS-SN},][]{asas-sn-1,asas-sn-2}. In order to align the {\it ASAS-SN} $V$-band observations with our data, we applied a $-$0.026\,mag shift to the former ones. For consistency with the 1971--2019 data set, we also transformed our Sloan $r'i'$ data obtained in 2020 and 2021 into the Johnson--Cousins $R_{\rm C}I_{\rm C}$ system using the transformation equations given by \citet{jordi2006}. A brief summary of the data used for the construction of the long-term photometric light curve is presented in Tab.~\ref{tbl_phot_obs_log}. The table includes the dates of the observations, filters used, status of the source and the relevant papers.

\begin{deluxetable}{cccc}
\tabletypesize{\footnotesize} 
\tablecaption{Summary of the photometric data used for Fig.~\ref{fig:lc}\label{tbl_phot_obs_log}}
\tablewidth{0pt}
\tablehead{
\colhead{Date} & \colhead{Filters} & \colhead{Status of the source} & \colhead{Ref.}}
\startdata
1971    &   $JHKL$ &    Main fading phase  & 1 \\
1971    &   $UBVRI$ &     Main fading phase  & 2 \\
1971    &   $JHKLMN$ & Main fading phase  & 2 \\
1975    &   $UBV$ &     Main fading phase  & 3 \\
1976    &   $UBV$ &     Main fading phase  & 4 \\
1971--1974    &   $UBV$ &     Main fading phase  & 5 \\
1971--1974    &   $MN$ &    Main fading phase  & 6 \\
1975--1977    &   $UBV$ &     Main fading phase  & 7 \\
1981    &   $JHKLMN$ &     Main fading phase  & 8 \\
1971--1987    &   $JHKLMN$ &     Fading \& First plateau & 9 \\
1989--1991    &   $KMN$ &    First plateau & 10 \\
1985--2011    &   $UBVR$ &    First \& Second plateau & 11\\
1985--2011    &   $JHKLM$ &    First \& Second plateau & 11\\
2005--2007 & $BVR_CI_C$ & Second plateau & This work\\
2011--2019 & $BVR_CI_C$ & Second plateau & This work\\
2019--2020 & $BVg'r'i'$ & Second plateau & This work\\
2006, 2012 & $BVR_JI_J$ & Second plateau & This work\\ 
2020 & $BVRi^{a}$ & Second plateau & This work \\ 
2019 & TESS I & Second plateau & This work \\
2006, 2012 & $JHK_s$ & Second plateau & This work\\
2020 & $JHK_s$ & Second plateau & This work\\ 
\enddata
\tablenotetext{a}{$i$ interference filter, which is similar to the SLOAN $i'$, but with a slightly longer effective wavelength of $\lambda_{\rm eff}=0.789\mu$m \\
1: \citet{mendoza1971};2: \citet{rieke1972};3: \citet{welin1975};4: \citet{welin1976};5: \citet{landolt1975};6: \citet{simon1975};7: \citet{landolt1977};8: \citet{simon1982};9: \citet{simon1988};10: \citet{KH91};11: \citet{kopatskaya2013}}
\end{deluxetable}


Both the archival and our new light curves firmly indicate that the post-outburst brightness evolution of V1057~Cyg is exceptional as compared to other FUOrs. \citet{kolotilov1990} noticed that after the phase of exponential decay, in 1984--1988 the brightness of V1057~Cyg has stabilized at nearly constant level in all used filters. This was the so called `first plateau' phase, which lasted until 1995. As mentioned in Section~\ref{sec:intro}, $UBV$ measurements taken in 1995-1996 revealed a sudden fading by about 1\,mag in these bands and this process (indicated by the vertical line in Fig.~\ref{fig:lc}) stopped in 1997 \citep{kolotilov&kenyon1997, ibragimov1997, kopatskaya2002}. Since 1997, the average brightness of V1057~Cyg has remained practically constant in all bands and this phase is known as the `second plateau' \citep{kopatskaya2013}. This plateau is also still present in the infrared region, as inferred from comparison of our $JHK_s$ observations with the latest data points found in the literature \citep{kopatskaya2013}. 

\begin{figure}
\includegraphics[width=\columnwidth]{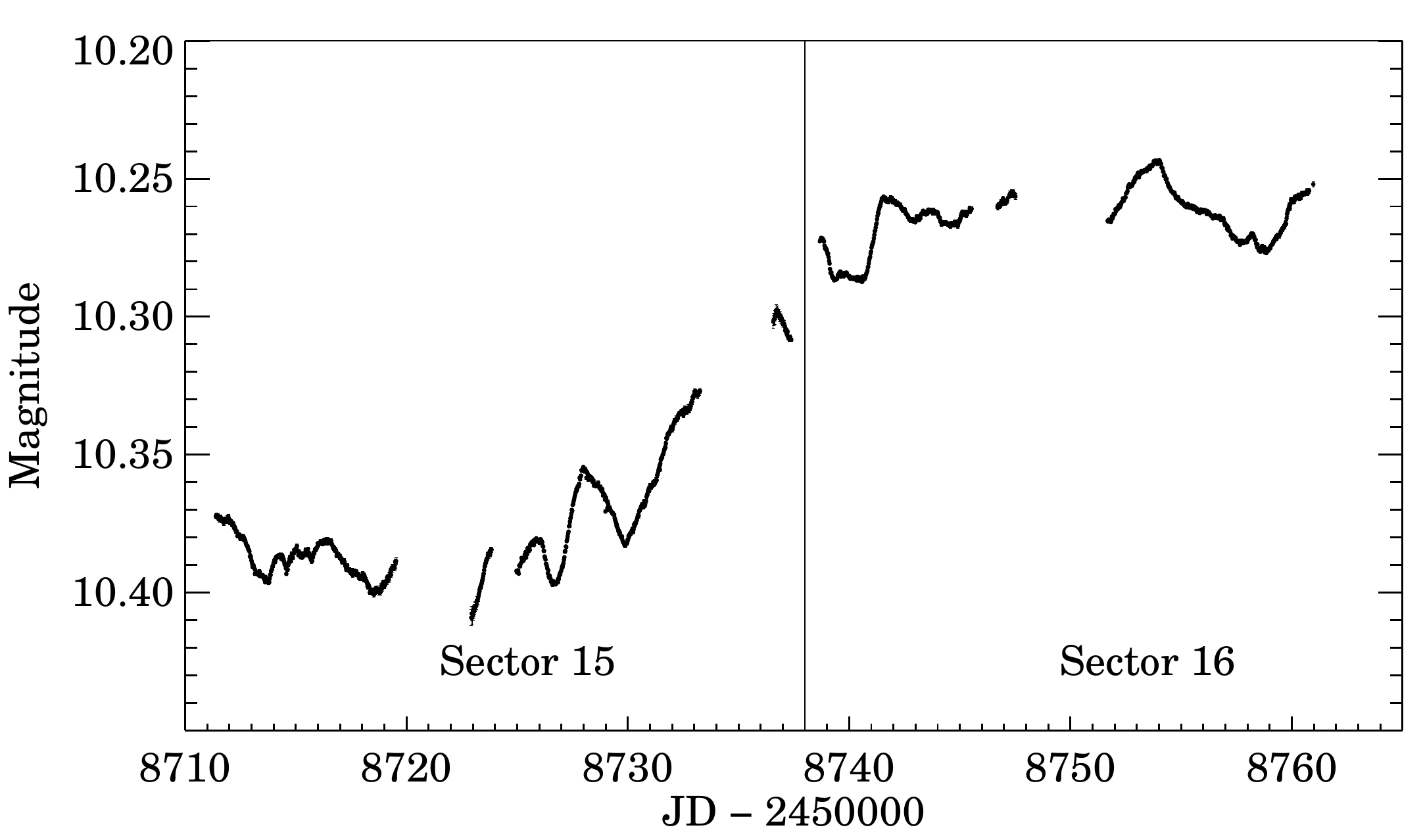}
\caption{The TESS light curve of V1057~Cyg.}
\label{fig:tess}
\end{figure}

The TESS light curve is presented in Fig.~\ref{fig:tess} and we shifted Sector 15 and Sector 16 to match our light curve in the $I_{\rm C}$ band. We performed interpolation to shift Sector~15 by $+$0.07\,mag and Sector~16 by $-$0.05\,mag. The cause of the six major breaks in the data acquisition were described in Section~\ref{sec:obs}. This precise light curve clearly shows the brightness changes occurring on a daily time-scale, whose detailed investigation remains beyond the capabilities of the ground-based telescopes.

\begin{figure*}
\includegraphics[width=0.33\textwidth]{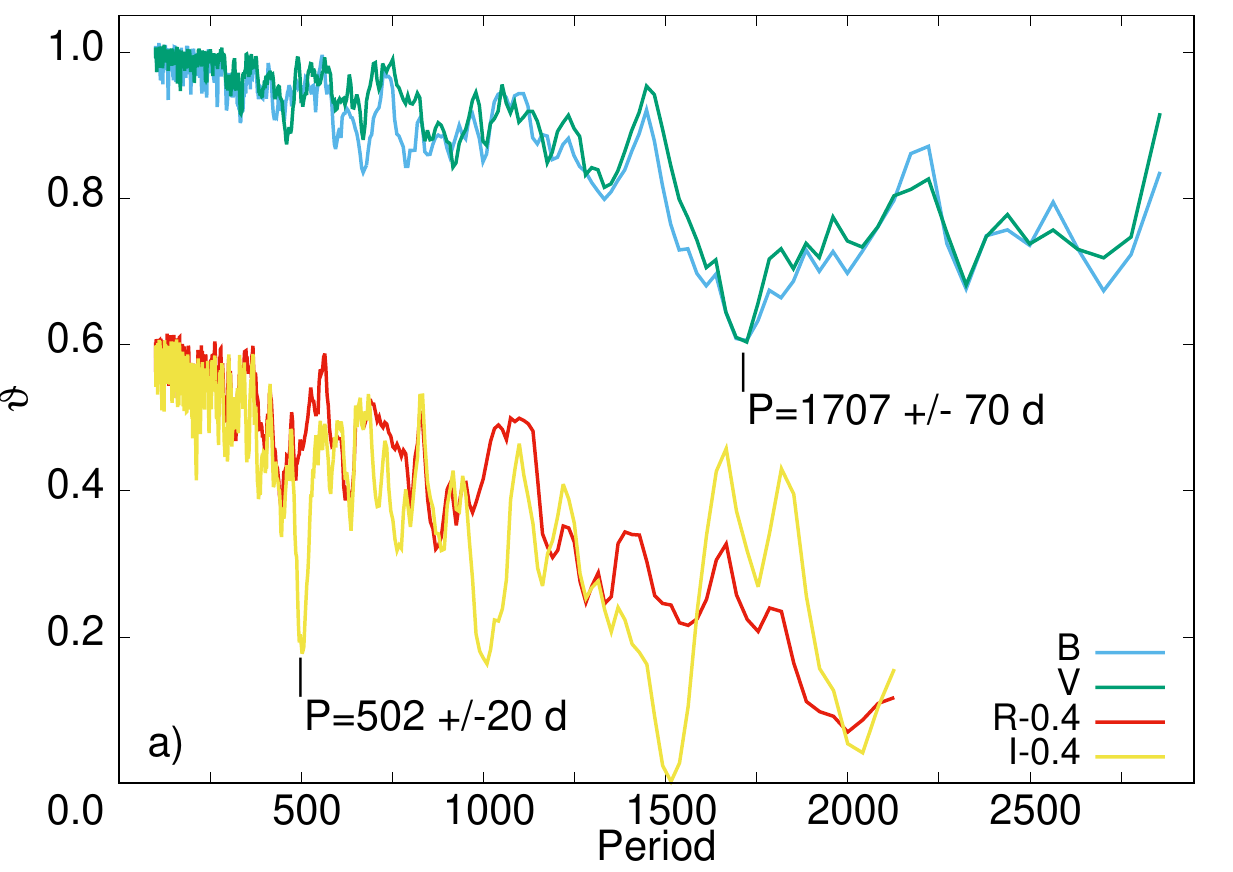}
\includegraphics[width=0.33\textwidth]{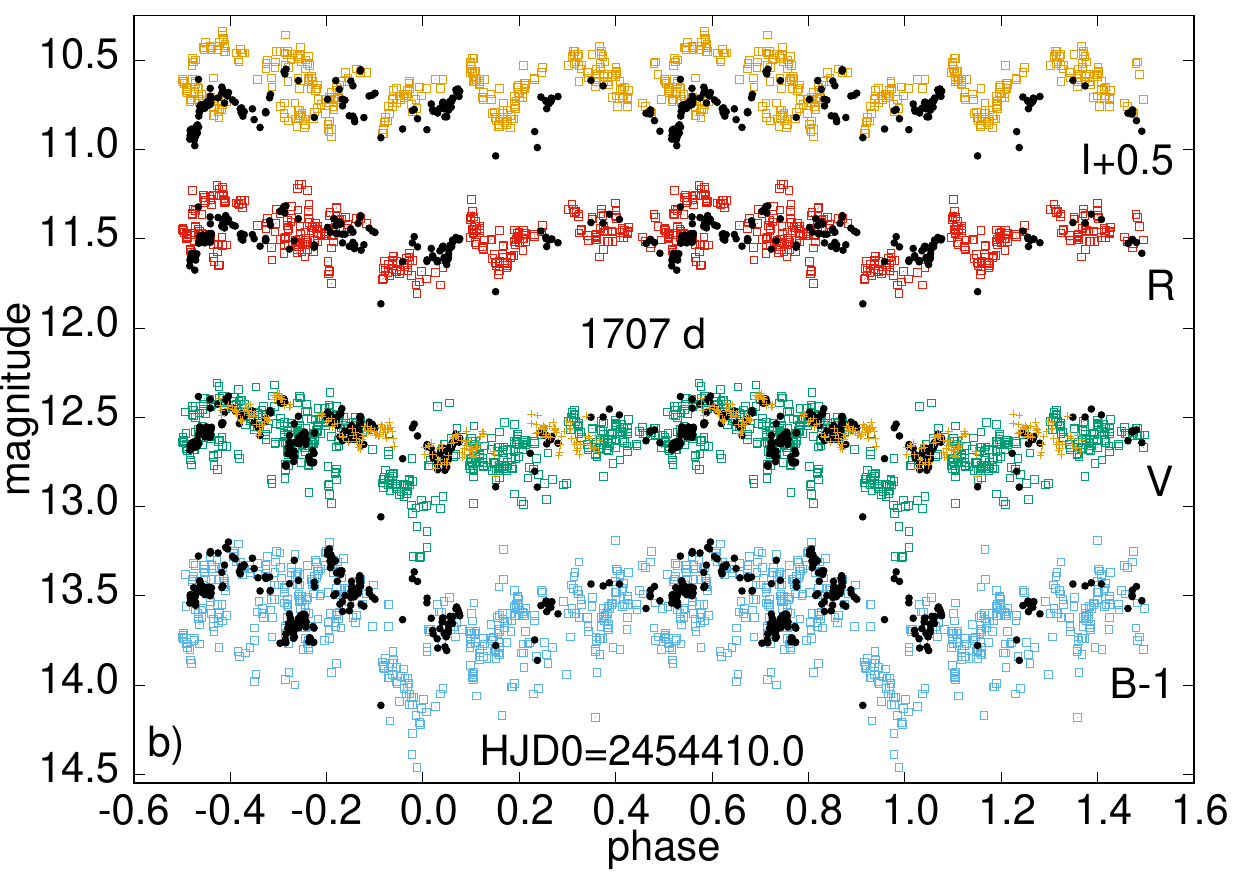}
\includegraphics[width=0.33\textwidth]{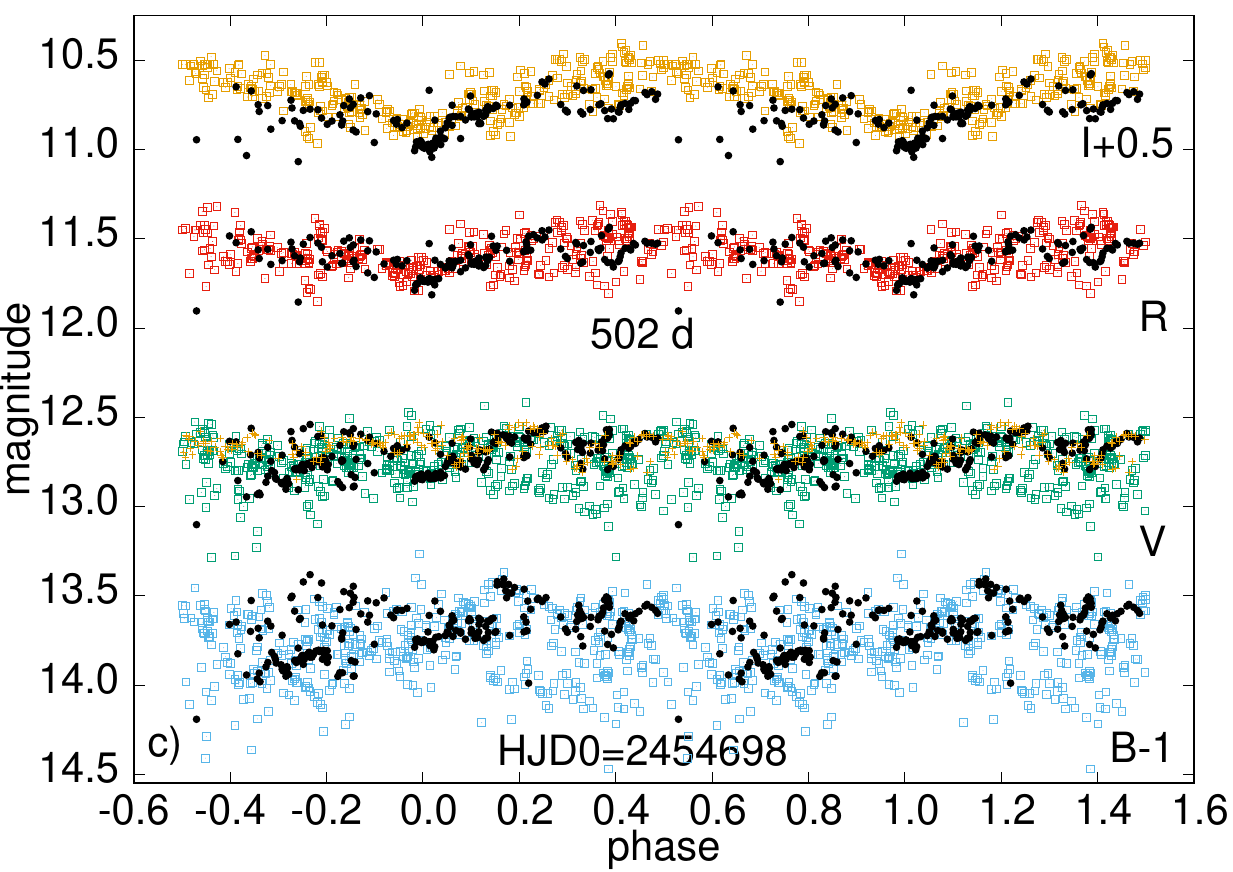}\\
\includegraphics[width=0.5\textwidth]{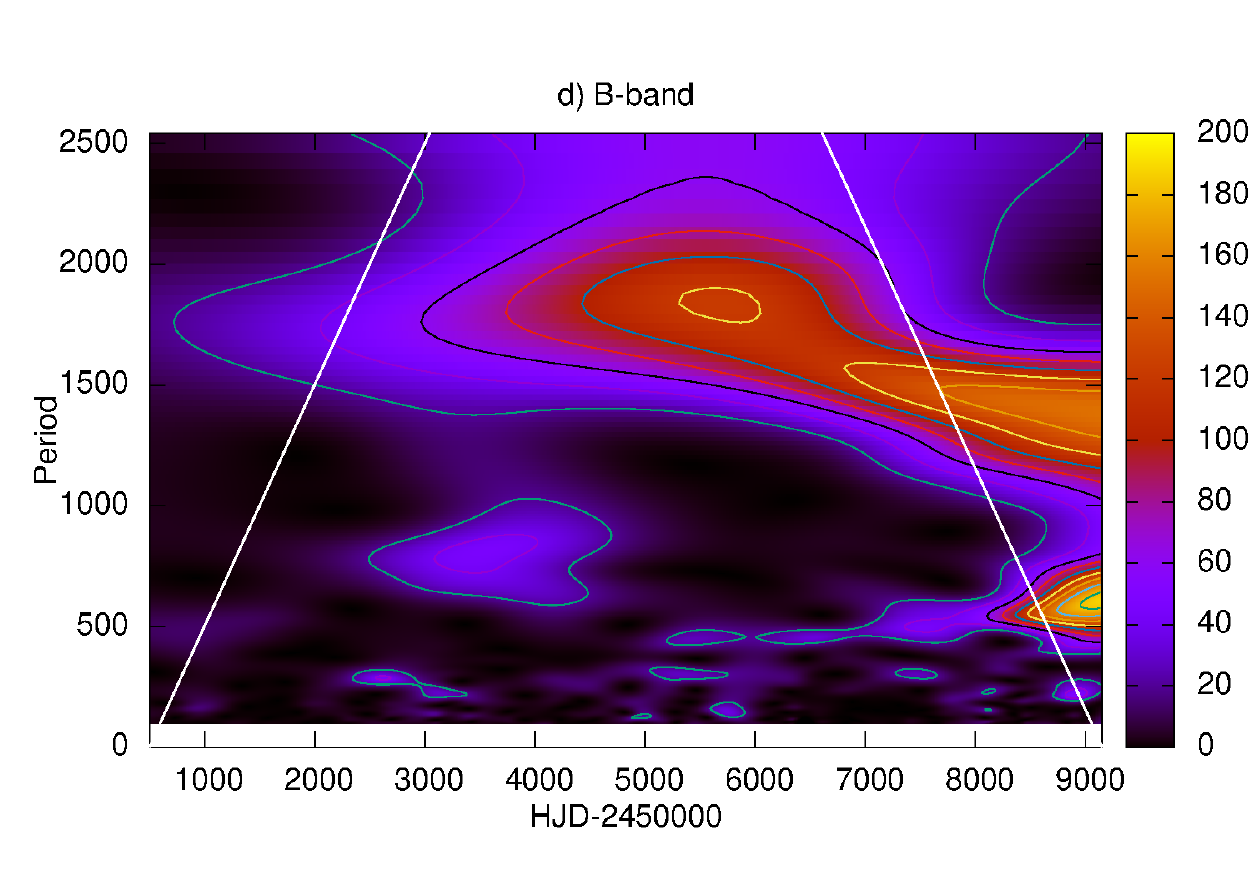}
\includegraphics[width=0.5\textwidth]{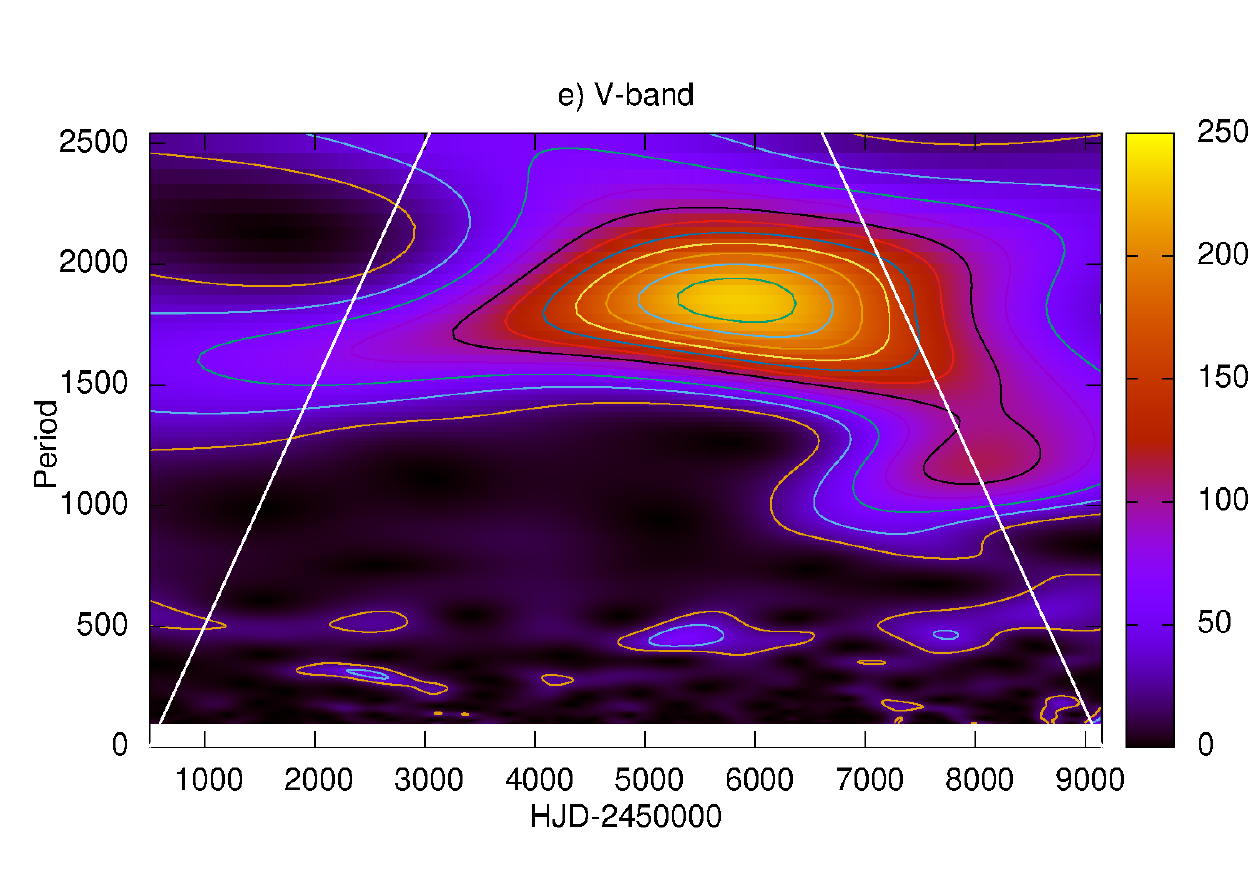}
\includegraphics[width=0.5\textwidth]{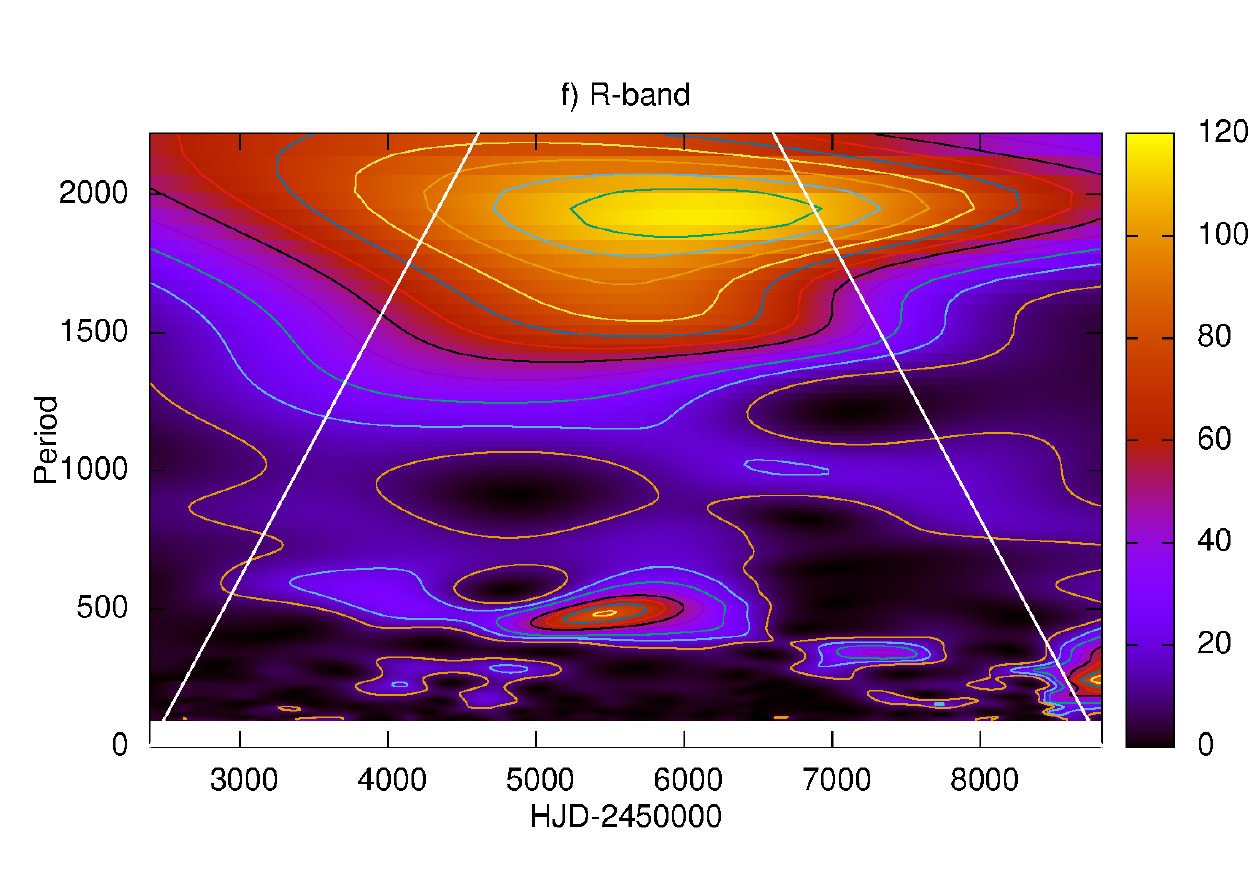}
\includegraphics[width=0.5\textwidth]{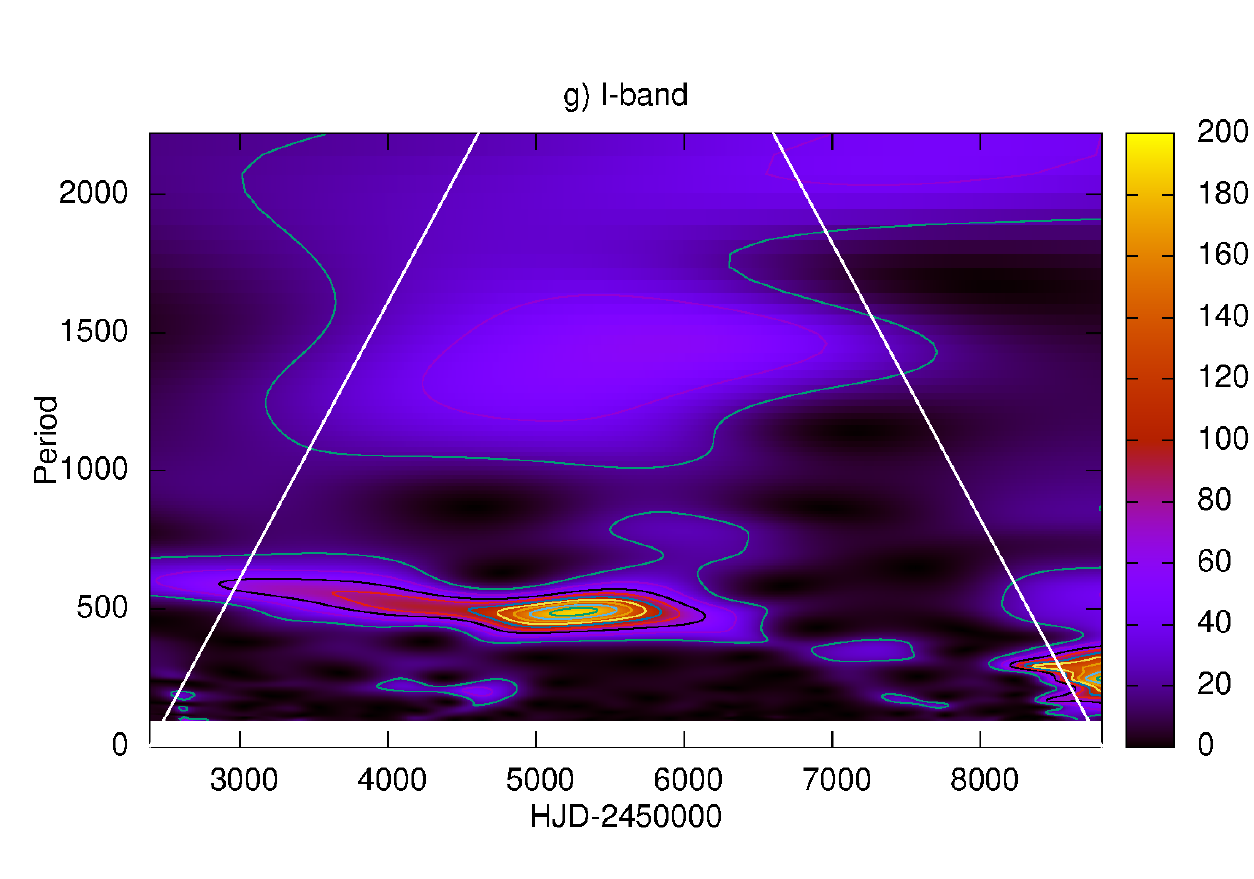}
\caption{Results obtained by means of PDM technique for the `second plateau' $BVR_{\rm C}I_{\rm C}$ data (panel a) and light curves phased with the longer (1707~d) and the shorter (502~d) period (panels b-c). Initial epochs used for phase calculations are indicated on the plots. Archival data are marked by colors, Piszkéstet\H{o} data are marked by dark dots, while ASAS-SN data by yellow crosses. Panels d-g show the WWZ spectra calculated for individual filters. Edge effects are contained outside the two white lines. The colors represent the Z-statistic values.
}
\label{fig:PdmWavGroundData}
\end{figure*}

\begin{figure*}
\includegraphics[width=0.33\textwidth]{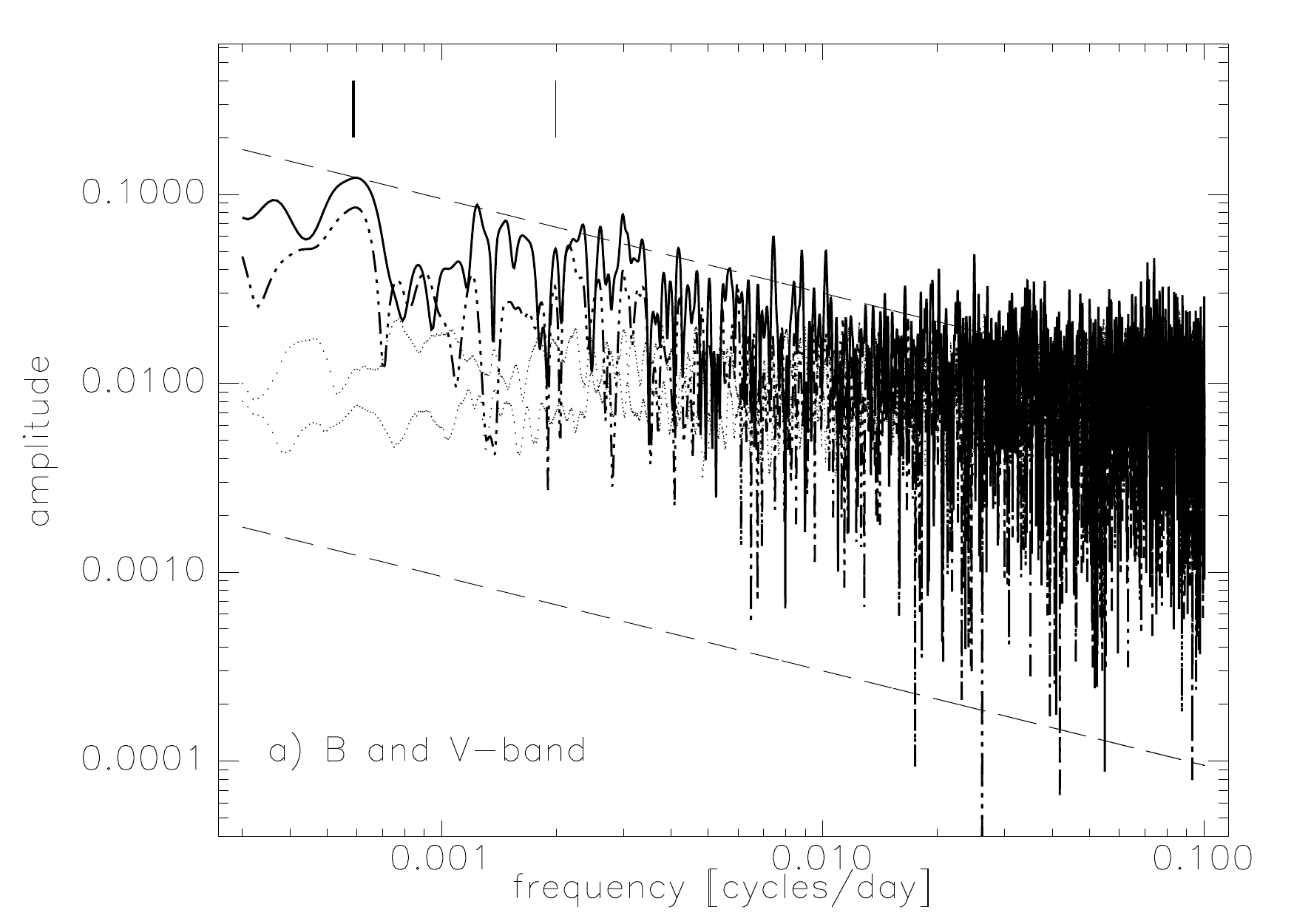}
\includegraphics[width=0.33\textwidth]{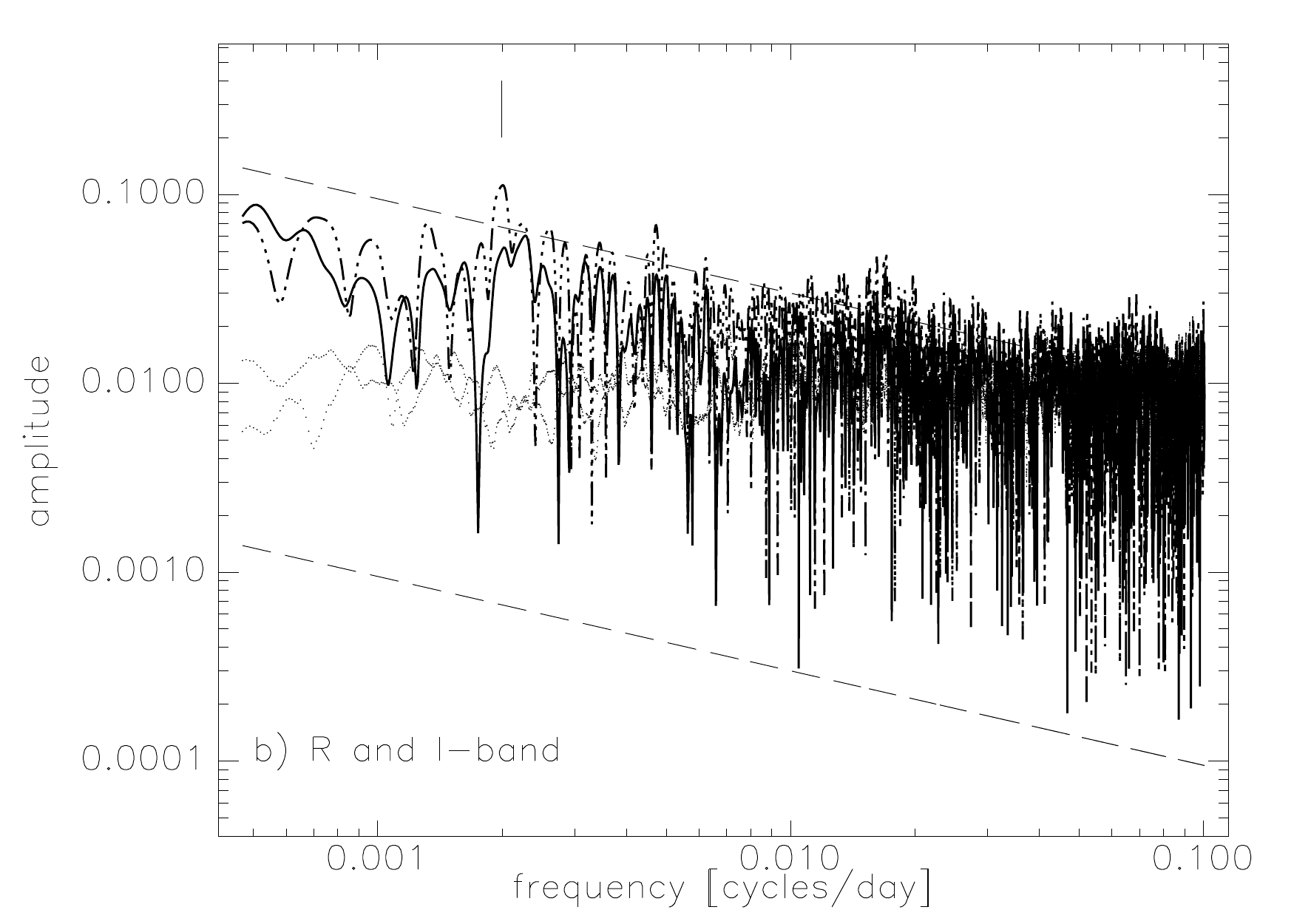}
\includegraphics[width=0.33\textwidth]{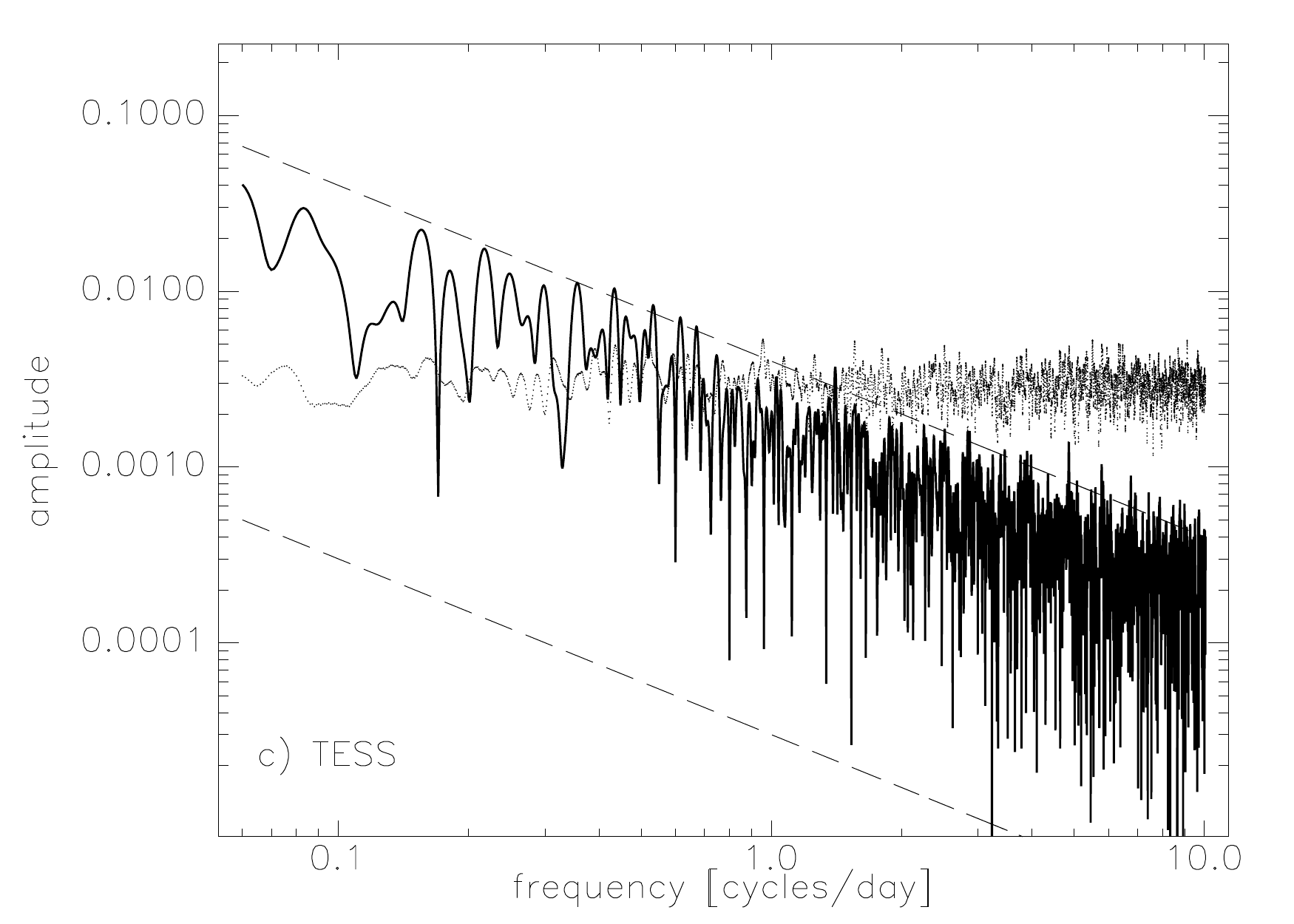}
\caption{Amplitude-frequency spectra in log-log scale represented by solid ($BR_{\rm C}$-filters) and dotted-dashed lines ($VI_{\rm C}$-filters), calculated from the `second plateau' data. The amplitude errors are marked by small dots; the significant peaks are located to the left from $\approx$0.01~c~d$^{-1}$. The short marks indicate the frequencies corresponding to the periods determined by means of the PDM method. No dominant period can be indicated in {\it TESS} spectrum (panel c). The red flicker-noise spectrum slope is indicated by two parallel dashed lines: they show $a_f\sim f^{-1/2}$ relation for the ground-based, and $a_f\sim f^{-1}$ for {\it TESS} data.
}
\label{fig:freqSP}
\end{figure*}

\begin{figure*}
\includegraphics[width=0.33\textwidth]{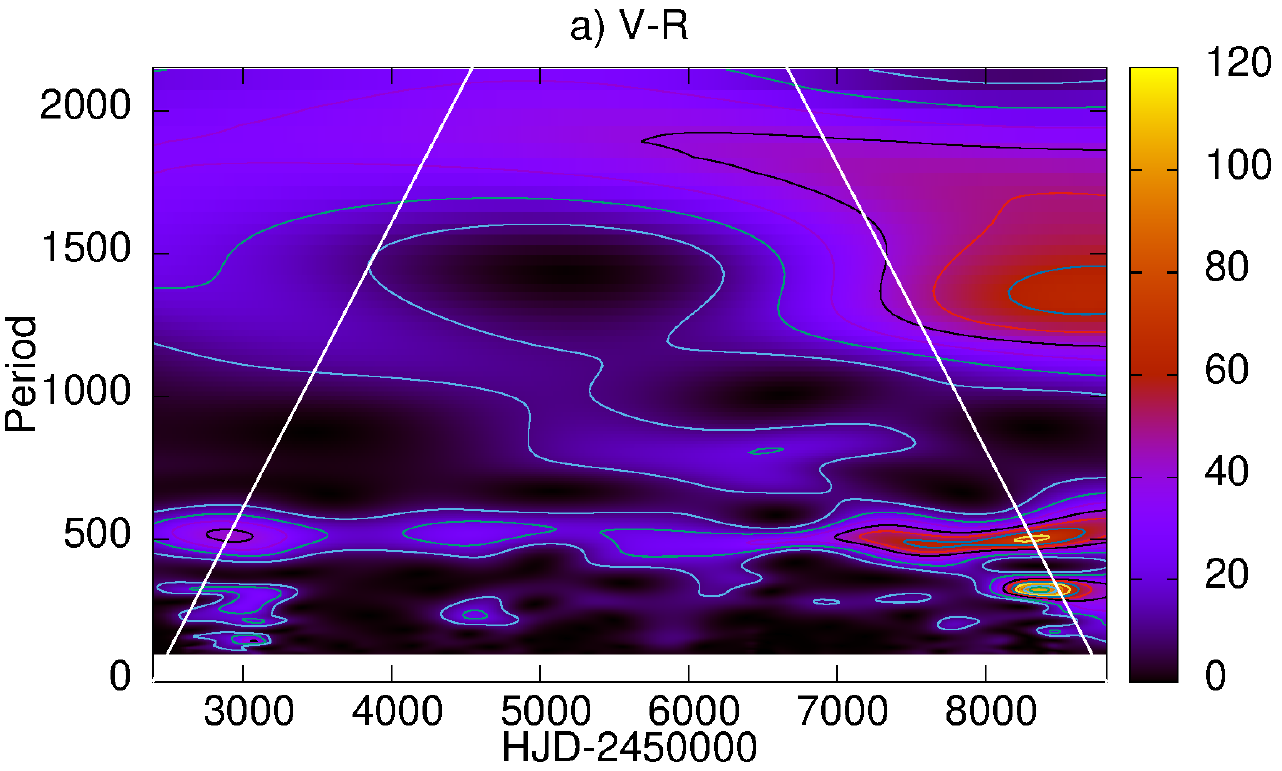}
\includegraphics[width=0.33\textwidth]{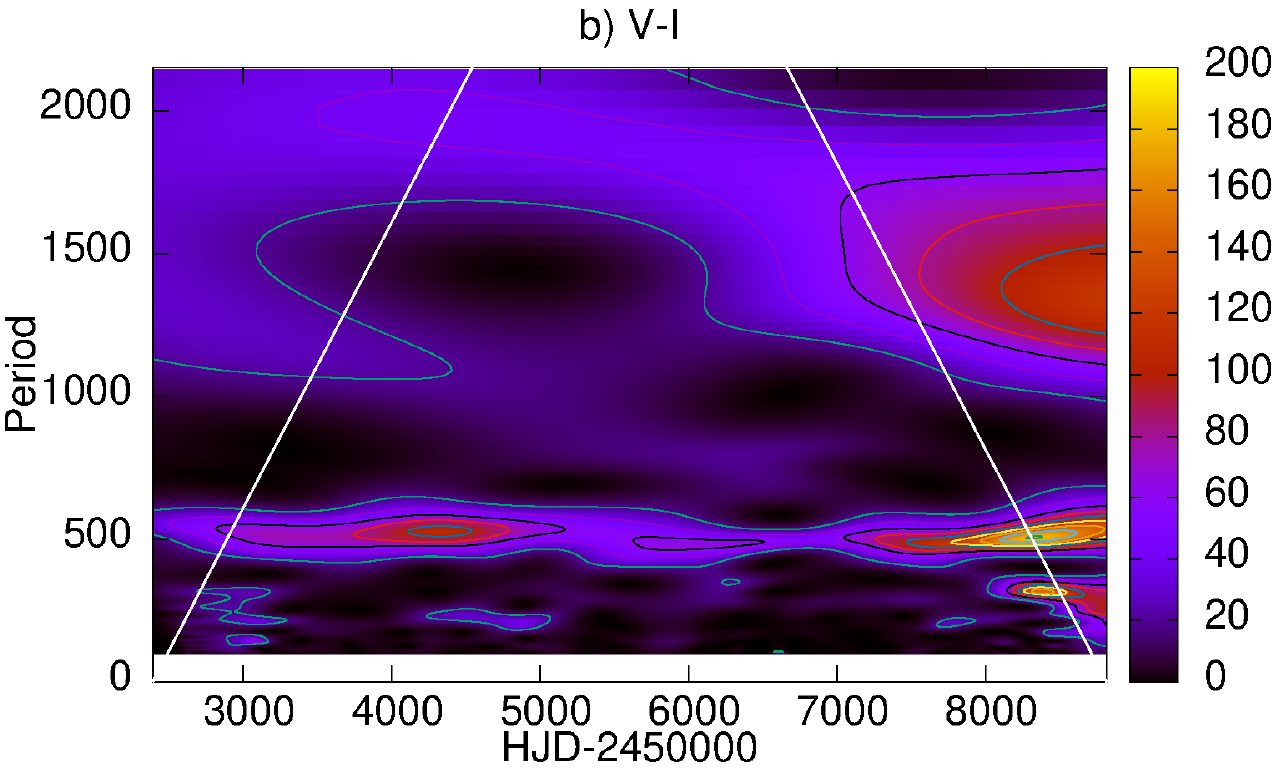}
\includegraphics[width=0.33\textwidth]{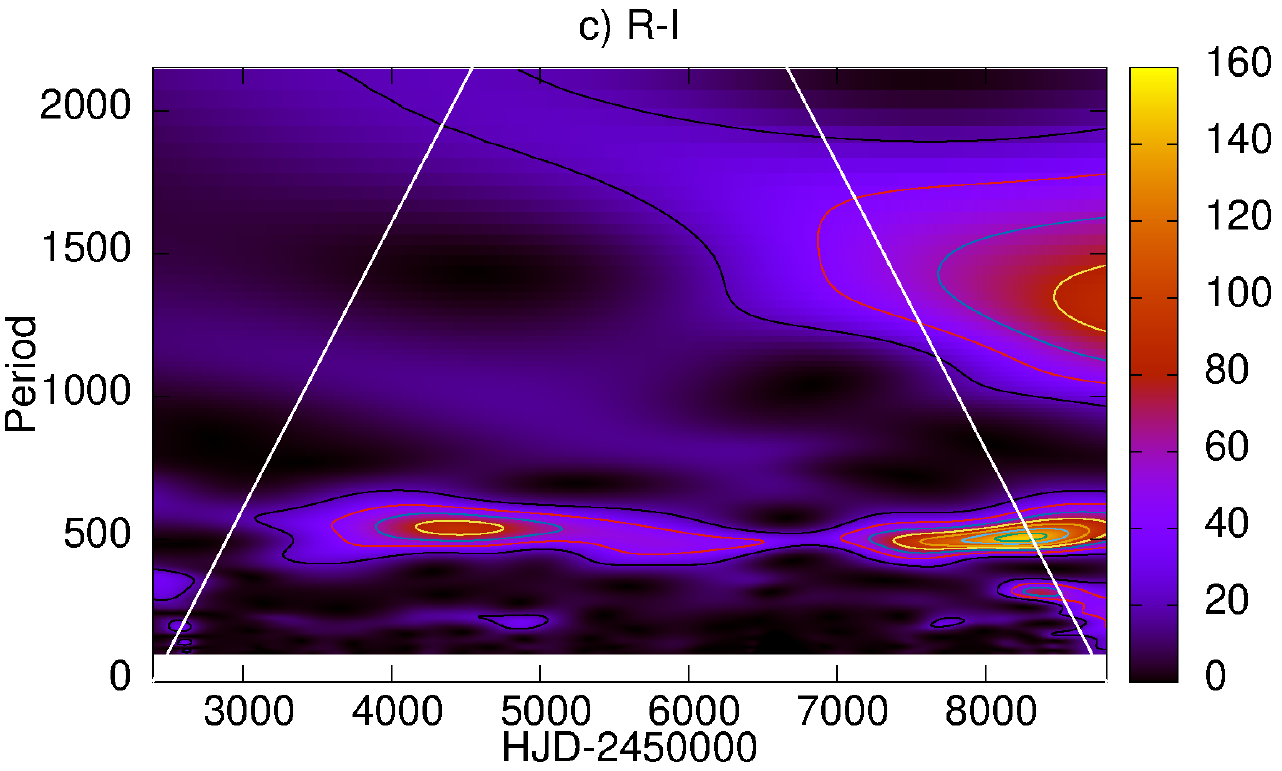}\\
\includegraphics[width=0.5\textwidth]{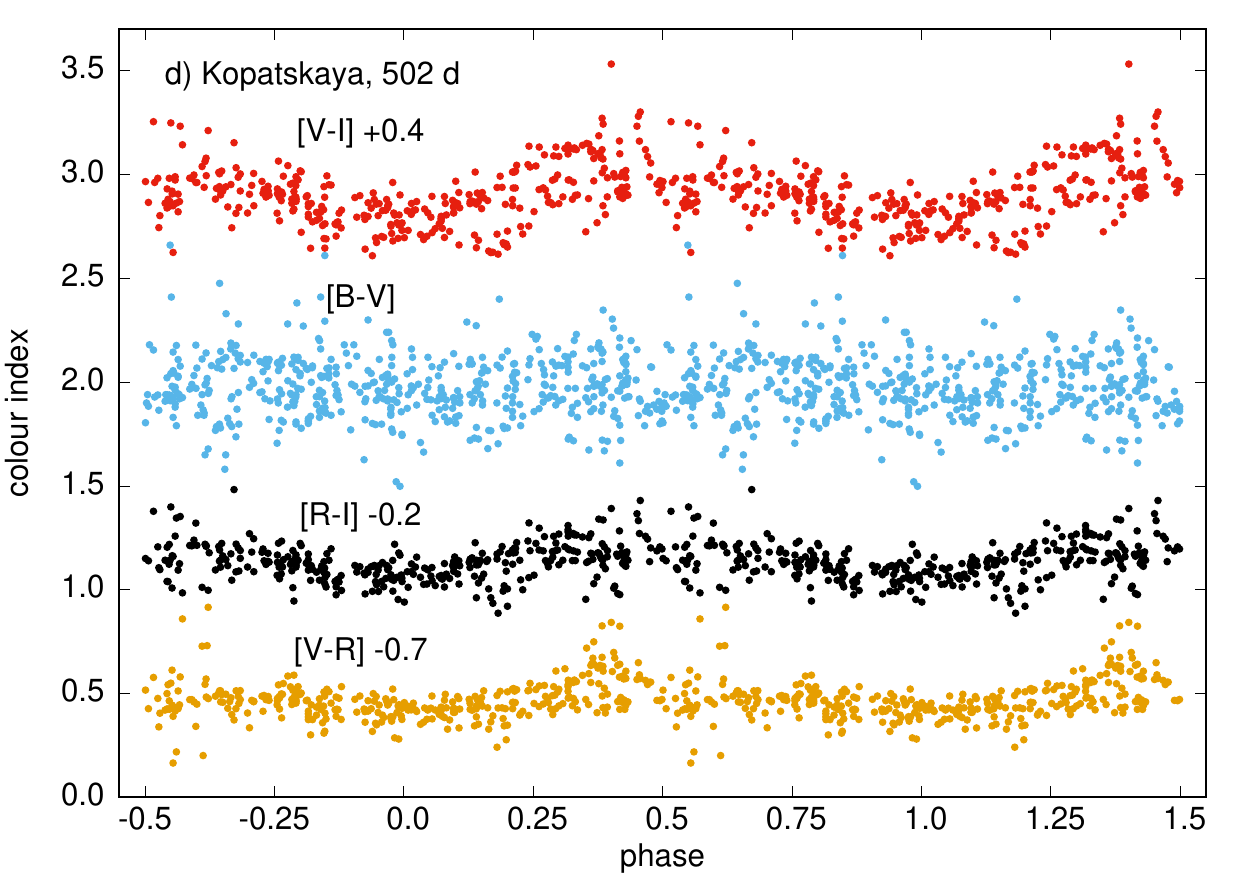}
\includegraphics[width=0.5\textwidth]{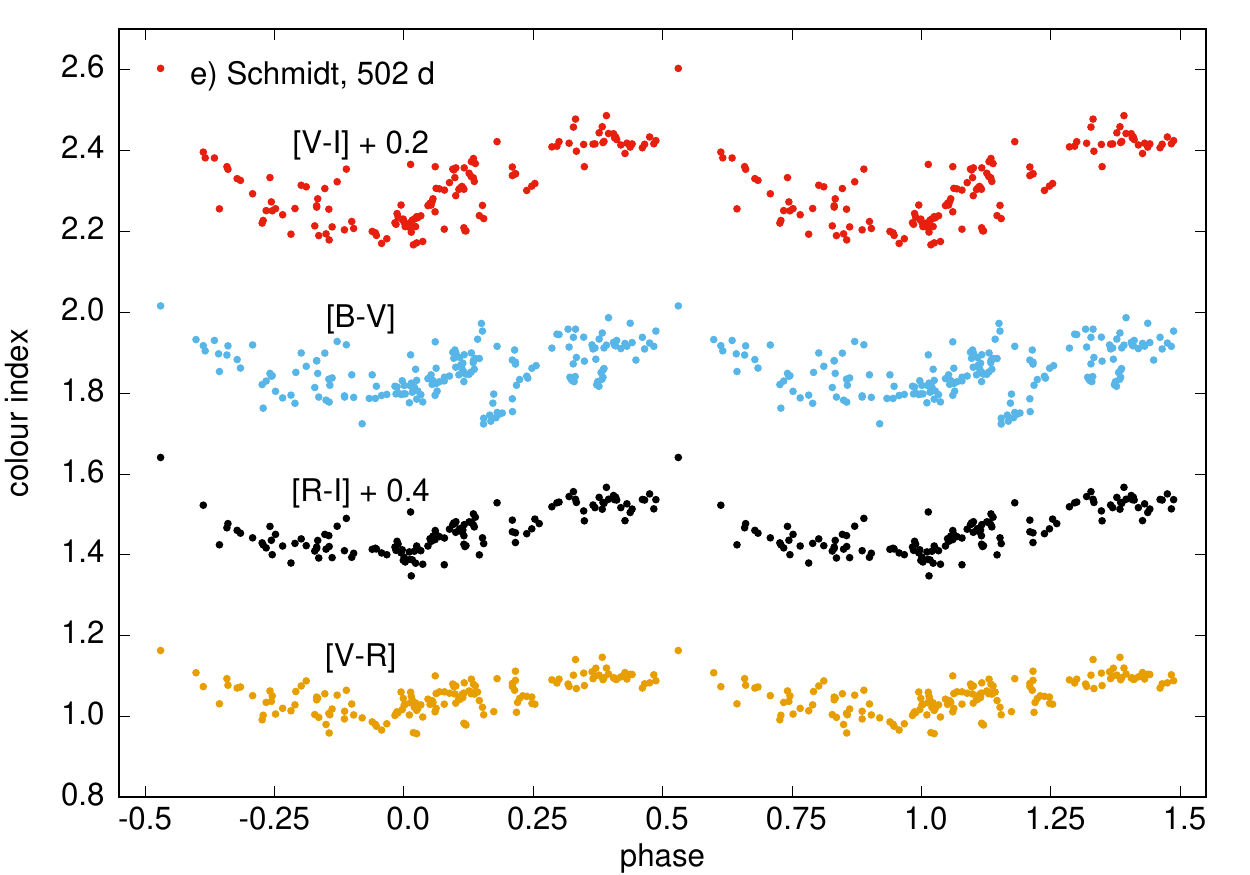}
\caption{WWZ results for CI variations (upper and middle panels), and phase-folded CI light curves (bottom panels). 
}
\label{fig:wwzCI}
\end{figure*}

\subsection{Period analysis}
\label{sec:periodanalysis}

\subsubsection{Long-term variability as seen from the ground}
\label{sec:PiszASASfreq}

As mentioned above, \citet{kopatskaya2013} discovered wavelength-dependent periodic components during the `second plateau' in all bands but $U$. The authors initially interpreted this finding as caused by the presence of a stellar companion or a forming planet, but strongly emphasized that future photometric observations will be essential to verify the driving mechanisms that they proposed. For this reason, we combined archival and new light curves to check if these oscillatory features are stable in time. In contrast to \citet{kopatskaya2013}, who for period analysis utilised detrended $UBV$ data collected since 1995, in this study we use their $BV$ data obtained since 1997 ($HJD=2450509$, i.e. when the brightness level rested on that typical for the `second plateau'), and the $R_{\rm C}I_{\rm C}$-filter data obtained since 2002. 
Afterwards, we included data gathered with the Schmidt ($BVR_{\rm C}I_{\rm C}$), RC80 ($BV$) and NSO ($VI$) telescopes, as well as the public-domain {\it ASAS--SN} Johnson--$V$ data. The new and the archival light curves were aligned to 0.002--0.005~mag by means of constant shifts to form uniform 19--23 years long time series. To ensure linearity during period analysis, the light curves were transformed from magnitudes to flux units, and were then normalised to unity at the mean brightness level of the complete 19--23 years light curve. 

Three period analysis techniques were used: as the light curves do not generally exhibit sine-like brightness variations, we decided to rely on the phase dispersion minimization (PDM) method \citep{stellingwerf1978}. We confronted these results with those obtained by means of 
the Fourier analysis, in which the mean standard errors of the amplitudes are conservatively evaluated using the bootstrap sampling technique \citep{ruc08}. Finally, in order to check for stability of these oscillatory features in time, we used the weighted wavelet Z-transform (WWZ, \citealt{foster1996}), designed for analysis of unevenly sampled time series and available within the {\sc Vartools} package \citep{hartman2016}.

Results obtained by means of the PDM technique are shown in Fig.~\ref{fig:PdmWavGroundData}a. Only the significant parts of the periodograms, showing periods covered at least three times and longer than 100 days, are presented. The most significant peaks for $BV$ filters are centered at 1707$\pm$70 days. In spite of the formally inconclusive value (0.6) of $\theta$ statistic, both the archival and the new $BV$-filter phased data clearly show periodic behaviour (Fig.~\ref{fig:PdmWavGroundData}b). $HJD_0^{BV}=2454410$ -- the best defined minimum in $BV$-filters that occurred at the end of 2007 -- was assumed during phase calculation.

At first sight the PDM diagram obtained from $R_{\rm C}$-filter data may appear to be inconclusive. First, because the primary $\sim$2000~d peak is poorly defined, the full extent of the long period lies outside the plotted portion of this periodogram. 
Second, the derived primary period is a multiple of the identified 497~d periodicity, which is also seen clearly in the $R_{\rm C}$-filter data. Note that within the measurement uncertainty, 497~d is indistinguishable from 502~d obtained from $I_{\rm C}I$-filter data (see below) and 523~d obtained from the archival $R_{\rm C}I_{\rm C}JHK$-filter data \citep{kopatskaya2013}. After rejecting the 2000~d peak, following the authors in Fig.~\ref{fig:PdmWavGroundData}b, we plot the $R_{\rm C}$-filter light curve phased with 1707~d period to examine the wavelength-amplitude evolution of this quasi-periodicity. We note that this peak is fairly well defined (although shifted to 1750~d) in the $R_{\rm C}$-band PDM diagram as well. 

For the same reasons as above, we adopted 1707~d for $I_{\rm C}$-filter data phasing (Fig.~\ref{fig:PdmWavGroundData}b). This peak is visible between the major ones at 1500 and 2000~days ($\theta=0.4$), which are the multiples of the dominant 502~d quasi-period ($\theta=0.6$). According to Fig.~1 in \citet{Czerny1997}, the false alarm probability of this 502~d quasi-period is $\leq$ few \%. 

In Fig.~\ref{fig:PdmWavGroundData}c we show the light curves phased with the 502~d quasi-period. $HJD_0^{I}=$2454698 -- the best defined minimum in the $I_{\rm C}$-filter that occurred in 2008 (288 days after the best-defined minimum in $BV$-filters) -- was assumed as the reference moment during phase calculation. In order to prepare these light curves, we applied a custom procedure to clear the original `second plateau' observations from the 1707~d QPO variability: the specific shape of each light curve shown in Fig.~\ref{fig:PdmWavGroundData}b was approximated by ordinary 7--9th order polynomial fit and then periodically subtracted. Thanks to our Piszkéstet\H{o} data being of a higher precision, the presence of the 502~d component was for the first time directly confirmed in the $V$-band light curve.

The wavelet analysis of the entire 19--23 long $BVR_{\rm C}I_{\rm C}$ light curves confirms the above results: the WWZ spectra indicate the broad, fairly stable in time, primary $\sim$1700--2000~d period for $BVR_{\rm C}$-filters (Fig.~\ref{fig:PdmWavGroundData}d-f) and the strong $\sim$502~d period for the $I_{\rm C}$-filter (Fig.~\ref{fig:PdmWavGroundData}g)\footnote{Note that the 1000, 1500 and 2000~d periods, the same that turned out 
to be the multiples of the dominant 502~d oscillation, are poorly defined in the $I_{\rm C}$-band spectrum.} Interestingly, signatures of the 502~d signal 
are noticeable in the form of a few isolated features in the archival and new $BVR_{\rm C}I_{\rm C}$ data, although surprisingly in certain bands this quasi-period appears to evolve or even to be suppressed. We stress that even though WWZ is designed for analysis of unevenly sampled data, the resulting spectrum does strongly depend on photometric quality, data density and a mixture of these effects makes existing quasi-periods impossible to disentangle at all times. This limitation allows us to firmly detect the 502~d QPO only in periods with good temporal sampling. However, these problems can be partially overcome, as  described in the discussion of the color index variations (Sec~\ref{sec:perreex}).

Finally we calculated Fourier spectra to check the PDM and WWZ results and to investigate the relationship of the amplitudes ($a$) in the frequency ($f$) space (Fig.~\ref{fig:freqSP}ab), which is carrying information about the nature of these small-scale oscillations.
Except for the rough confirmation of the PDM results, we found that the ground-based spectra that `feel' the longer family of quasi-periodic oscillations (QPOs) only, show the stochastic flicker-noise nature characterised by $a_f\sim f^{-1/2}$ \citep{press1978}. We will return to this issue in Sec.~\ref{sec:TESSperioddiscussion}.

\subsubsection{Periodic color index variations}
\label{sec:perreex}

The light curves themselves are affected by secular light changes, which in turn worsen 
the above obtained PDM, WWZ and Fourier results. Therefore we decided to reexamine the above obtained quasi-periods by means of the color index (CI) variations. In other words, analysis of light curves formed from the CIs can be treated as a counterpart of the usual whitening that is sensitive to the non-periodic and intrinsic to the disk's environment gray variability factors. The majority of these undesirable effects is expected to be removed, while the pure quasi-periodic variations driven by the not yet well-understood mechanisms should still be preserved.

We have performed time variability analysis of $B-V$, $V-R_{\rm C}$, $V-I_{\rm C}$ and $R_{\rm C}-I_{\rm C}$ CIs with PDM and WWZ technique. To ensure homogeneity, we decided to use only the archival and the Schmidt-telescope data. The PDM analysis confirmed the previous 502~d value. 
We also found that the long-periodic variations weakened to a large extent. This is most visible in the new wavelet spectra dominated by the 502~d oscillation, which now appear as persistent and stable for the entire `second plateau' (Fig.~\ref{fig:wwzCI}abc). In Fig.~\ref{fig:wwzCI}d-e we show associated archival and new CI light curves phased with the 502~d quasi-period and assuming $HJD_0^I$. Note that this approach reveals 502~d variations in the $B-V$ light curve, but only in the part composed of the precise 
Schmidt data (Fig.~\ref{fig:wwzCI}e, see also in Fig.~\ref{fig:PdmWavGroundData}de). This analysis clearly shows that in the zero phase, all CI's are significantly bluer than during the light maximum. This is in line with the CIs that can be directly inferred from light curves alone (Fig.~\ref{fig:PdmWavGroundData}c), but opposite compared with the 1707~d period (Fig.~\ref{fig:PdmWavGroundData}b), in which the respective CIs are redder at times when the disk is fainter.

\subsubsection{Short-term variability as seen by TESS} 
\label{sec:tess-period}

To gain insight into the variability occurring on the time scales of hours and days, we performed Fourier analysis of {\it TESS} data obtained with 30~min sampling. In accordance with the visual inspection of the light curve itself (Fig.~\ref{fig:tess}), the amplitude-frequency spectrum does not show any dominant peaks (Fig.~\ref{fig:freqSP}c). We also found that this spectrum exhibits a Brownian random-walk, described by $a_f\sim f^{-1}$  \citep{press1978}. 

We also performed wavelet analysis of the {\it TESS} data. We do not report these results here, as the analysis is strongly affected by the previously mentioned (Sec.~\ref{sec:obs}) six breaks in the data acquisition, which have a duration comparable to the characteristic time scale of observed light changes.

\subsection{Amplitude-wavelength dependency of the QPOs}
\label{sec:ampl-wav_dep}

As already shown by \citet{kopatskaya2013} and confirmed in Fig.~\ref{fig:PdmWavGroundData}bc, the two QPOs observed in the `second plateau' show very different amplitude-wavelength dependencies. We also noted that these amplitudes evolve in time in our observations. To characterize this effect more profoundly, for each Johnson filter we determined the amplitudes by sinusoidal-fits to the phase-folded (Fig.~\ref{fig:PdmWavGroundData}bc) light curves constructed from the archival (1997--2011) and from the new (2011--2020) data only. This approach minimizes the non-periodic overlapping effects.

In the case of 1707~d period, the amplitudes decrease with increasing effective wavelength of a filter: for the archival data we obtained 0.143(9), 0.122(6), 0.088(6) and 0.068(7)~mag for $BVR_{\rm C}I_{\rm C}$ filters, respectively. The errors shown in parentheses represent the $1\sigma$ uncertainty obtained from the least-square fits. New data show the same well-defined amplitude-wavelength trend, but the resulting 0.066(9), 0.033(6) and 0.021(6)~mag for $BVR_{\rm C}$ filters, respectively, clearly indicate (within 3$\sigma$) that the amplitudes are systematically becoming smaller in all bands, to the point that no variability has recently been detected in $I_{\rm C}$-band.

The amplitudes associated with the 502~d period are gradually increasing with the wavelength: no variability has been detected in $B$-band, both in the archival and the new light curves. There is no evidence of variability in the archival $V$-band data, and only archival red and near-infrared  data show significant variation -- 0.068(5) ($R_{\rm C}$), 0.131(5) ($I_{\rm C}$),  0.161(31) ($J$)  0.146(33) ($H$)  0.130(33)~mag ($K$), respectively. We find variation of 0.026(7), 0.057(6) and 0.080(7)~mag for $VR_{\rm C}I_{\rm C}$ filters, respectively, in the accurate Schmidt-telescope data. Recent $JHK$ data are too sparse to estimate the current amplitudes. We conclude that unlike the 1707~d, there is no obvious sign of time-evolution of the amplitudes associated with the 502~d QPO.

\subsection{Color-magnitude diagrams}
\label{sec:cc}

Evolution of the color indices during the first post-outburst stages has already been investigated by \citet{kopatskaya2013}. The authors found that after the gradual colour evolution along the extinction path in the phase of the exponential decay (1971--1985), during the `first plateau' (1985--1995), when the source became fainter than $V\sim11.5$\,mag, the color index showed a `blueing effect', which can be observed in the young UX~Ori type objects. According to the authors, this effect has no longer been obviously present in the `second plateau'.

\begin{figure*}
\centering
\includegraphics[width=0.32\textwidth]{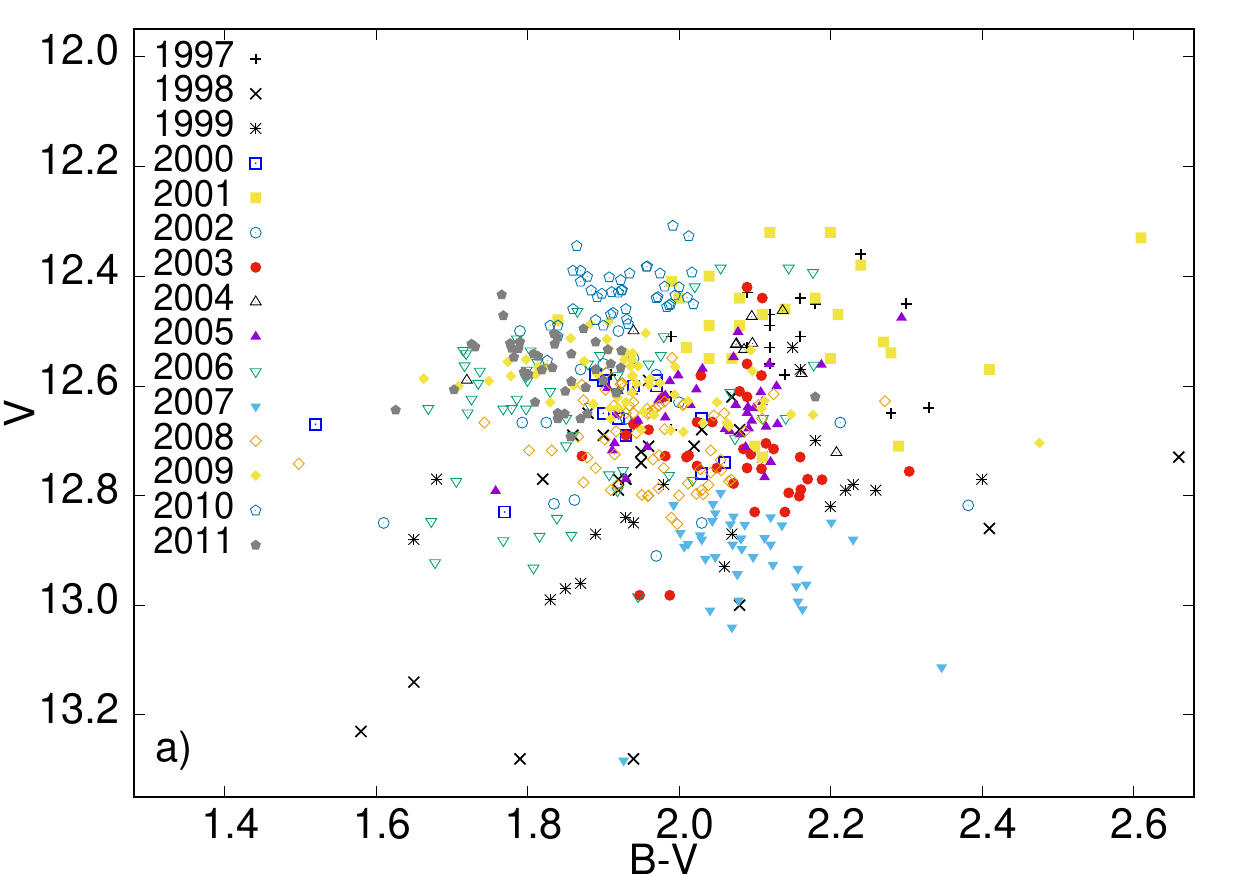}
\includegraphics[width=0.32\textwidth]{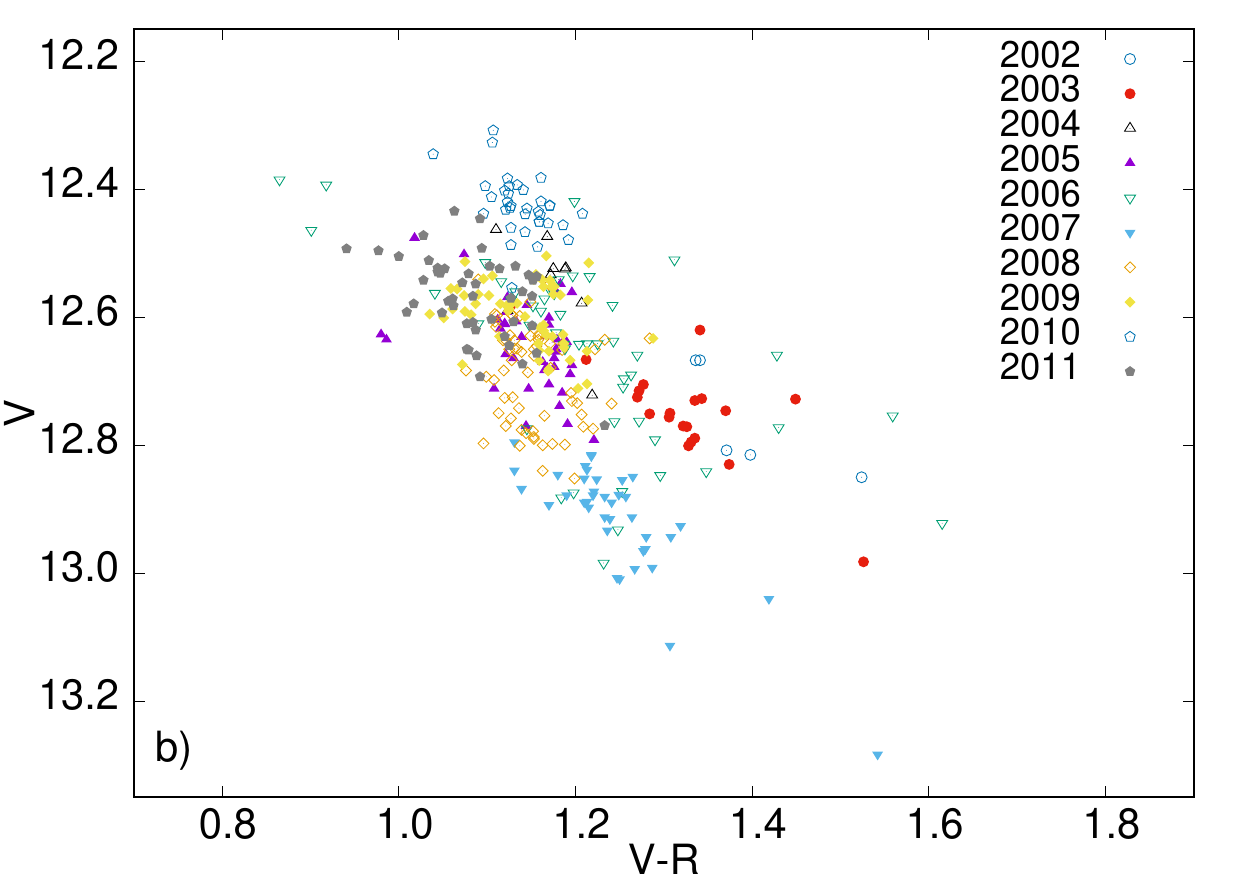}
\includegraphics[width=0.32\textwidth]{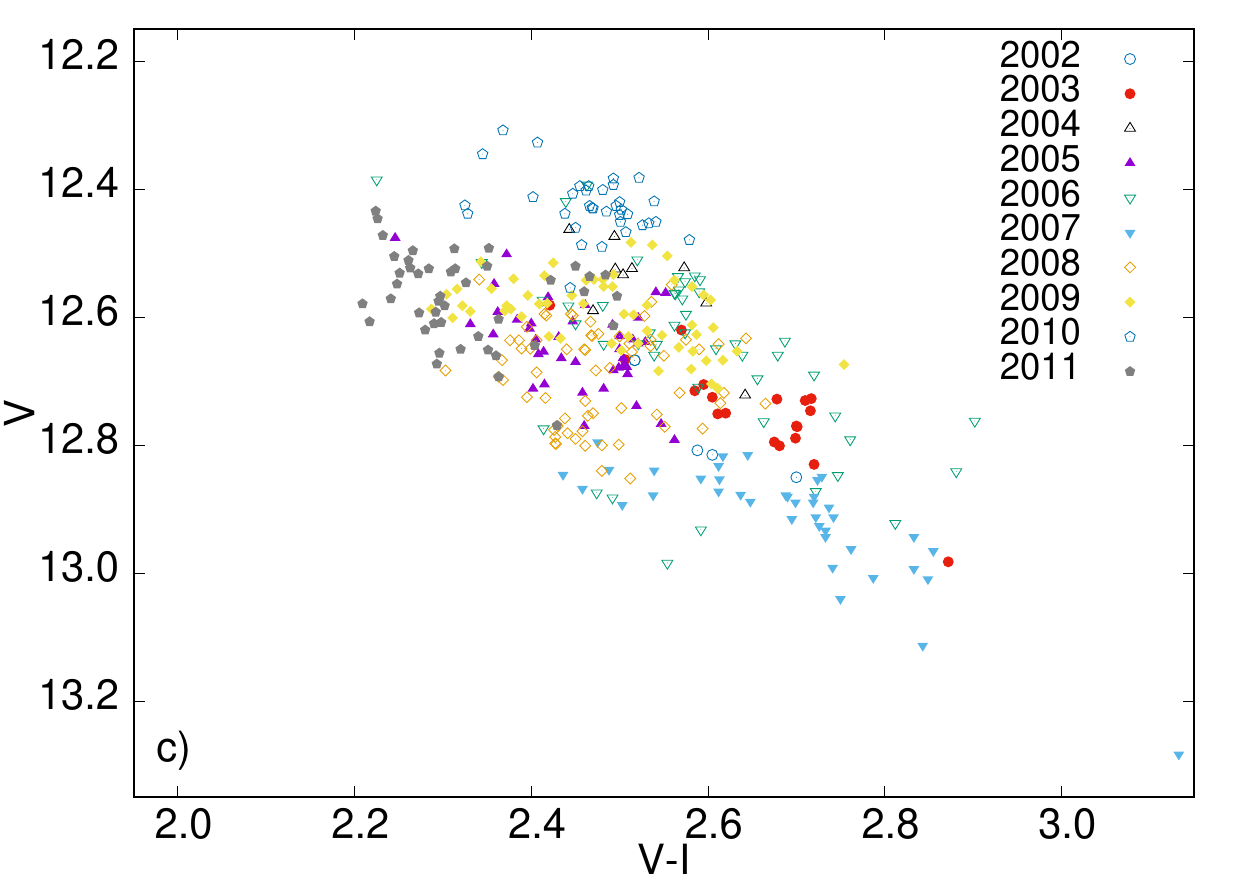}\\
\includegraphics[width=0.32\textwidth]{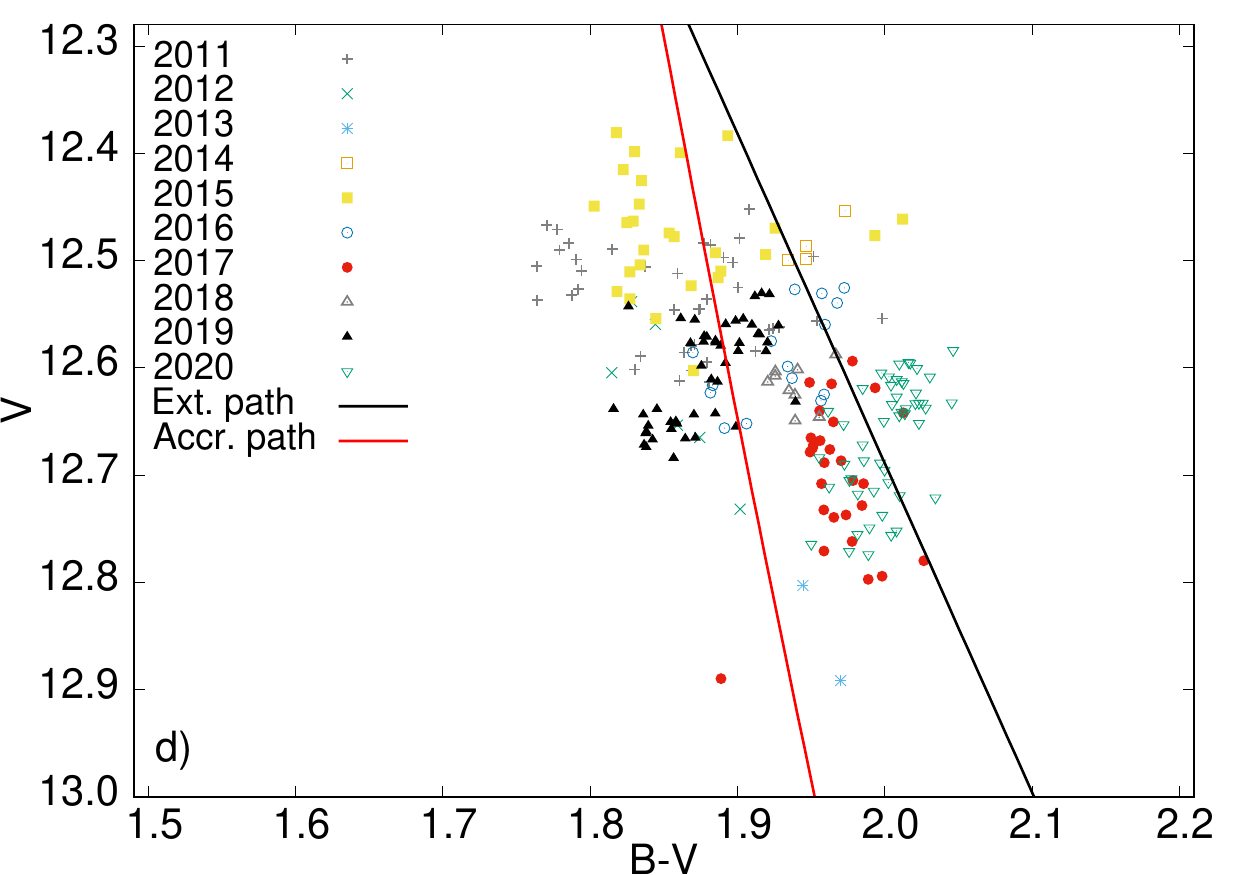}
\includegraphics[width=0.32\textwidth]{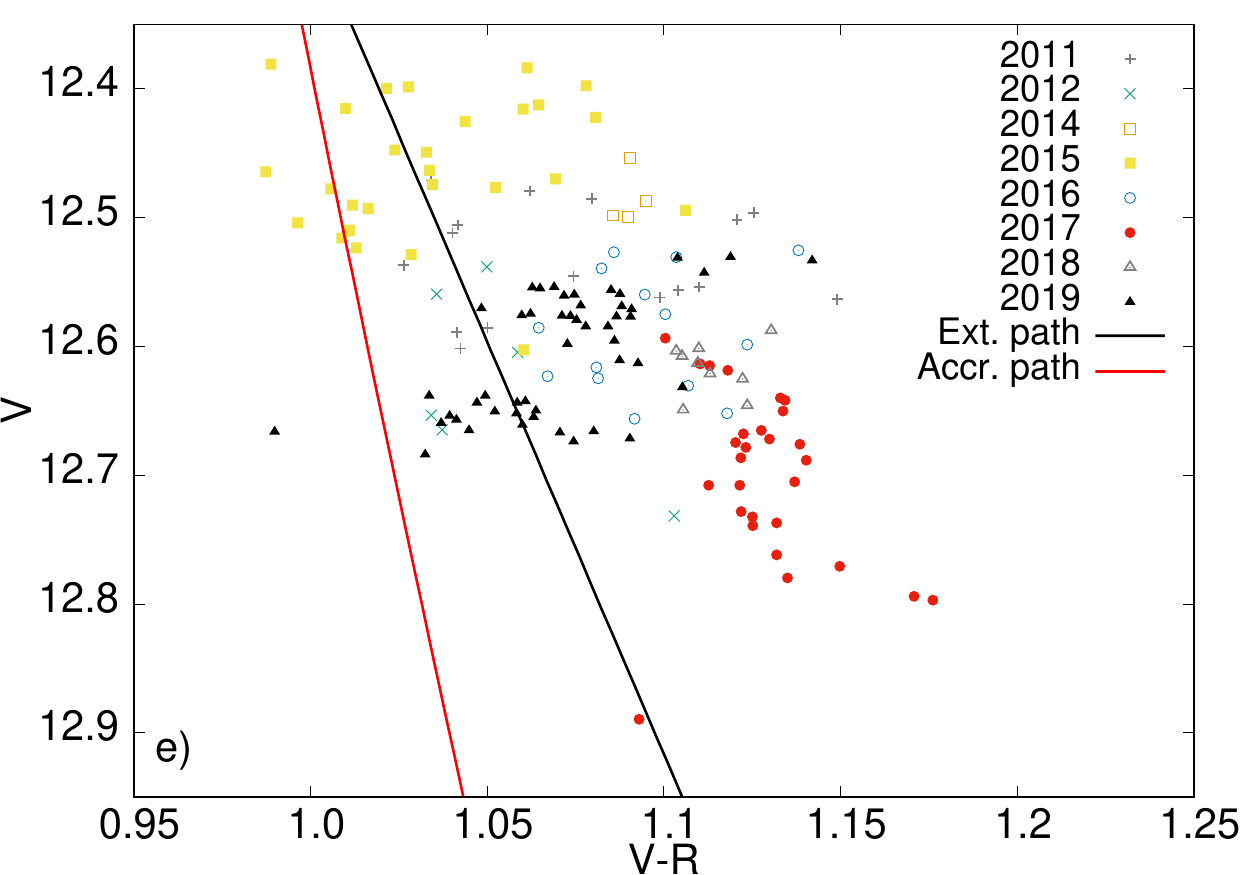}
\includegraphics[width=0.32\textwidth]{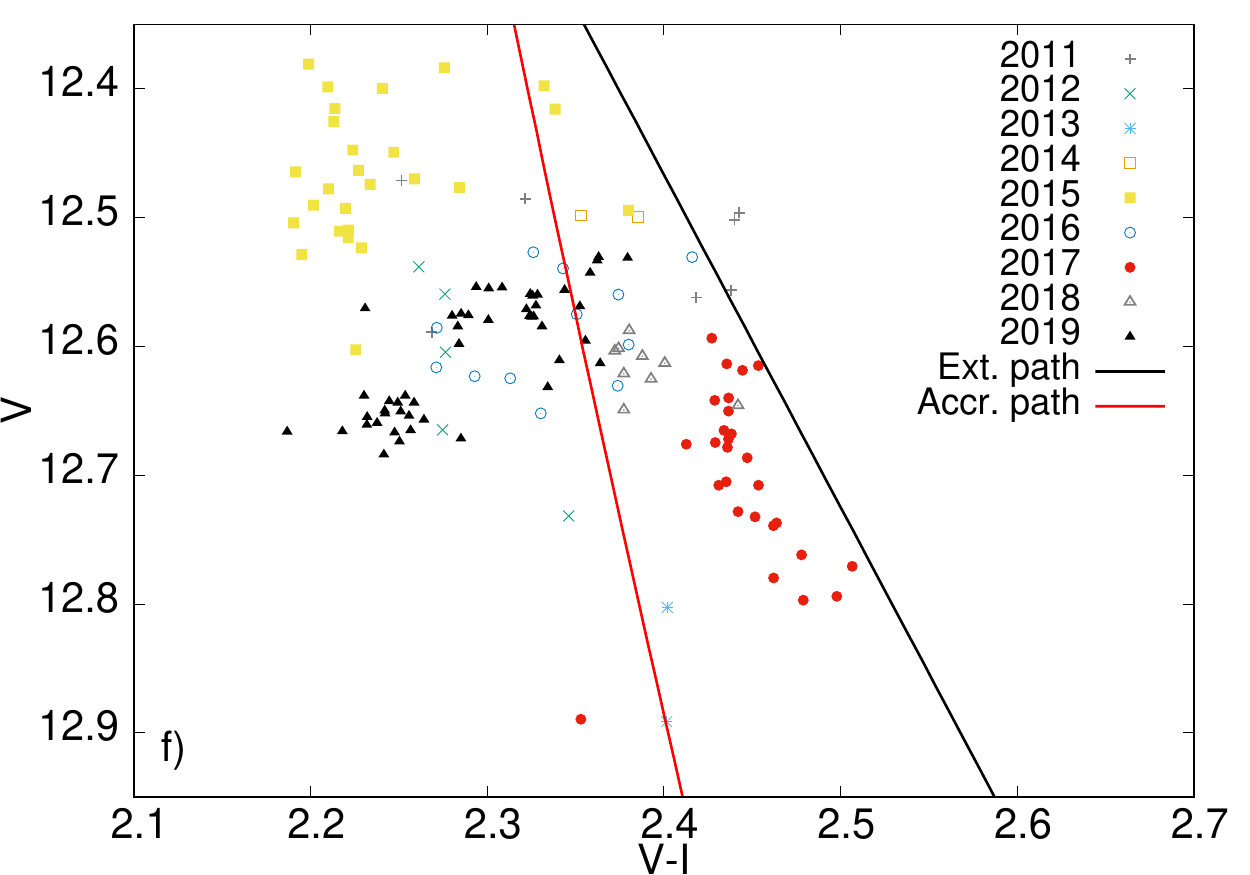}
\caption{Color-magnitude diagrams prepared from the archival (upper panels) and 
Schmidt and RC80 data (bottom panels). Theoretical color index variations caused by variable extinction and variable accretion obtained by our model (see in Sec.~\ref{sec:accdisc}) are also shown in the last three panels. 
}
\label{fig:CIall}
\end{figure*}

Here we continue investigation of the colour index evolution during the `second plateau'. We utilize the archival data combined with the new one obtained in $BVR_{\rm C}I_{\rm C}$ and $BV$-filters with the Schmidt and RC80 telescopes, respectively.~We show obtained results in Fig.~\ref{fig:CIall}. Data obtained during individual years are marked by different colors and symbols.

\begin{figure}
\centering
\includegraphics[width=\columnwidth]{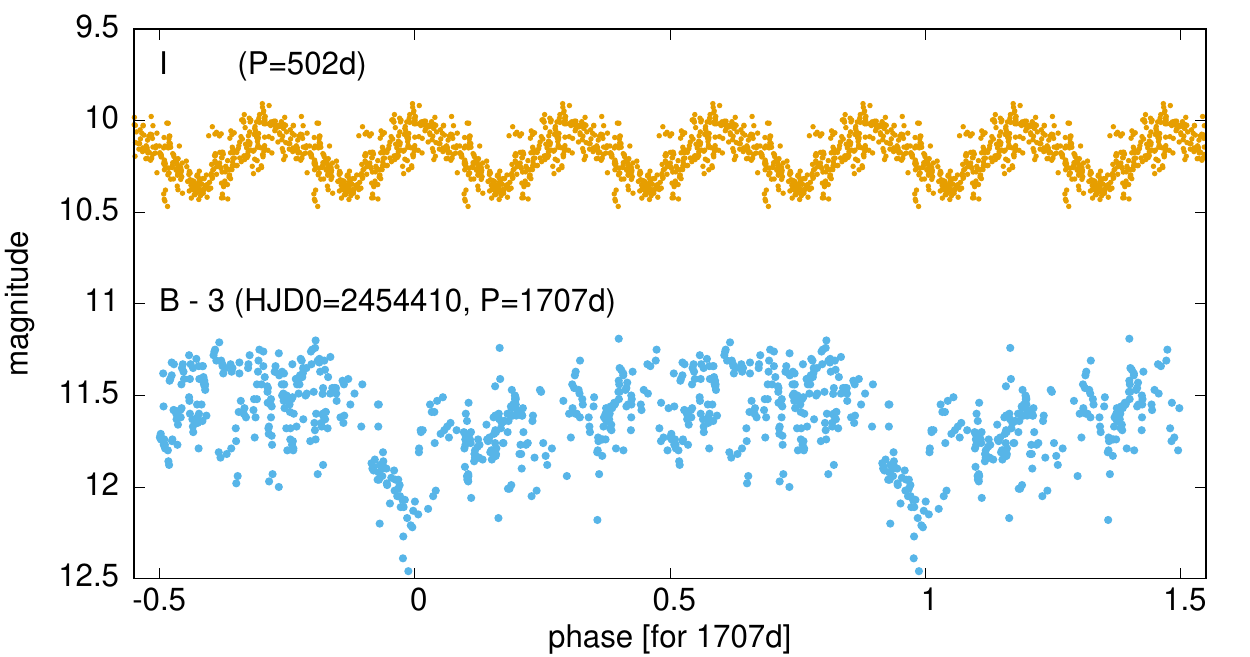}
\caption{Relation between the longer and the shorter period, shown in function of phase calculated for the longer period. The $B$-filter light curve is the same as shown in Fig.~\ref{fig:PdmWavGroundData}b, the $I$-filter light curve is the same as shown in Fig.~\ref{fig:PdmWavGroundData}c. Only archival data are plotted.
} 
\label{fig:compBIphase}
\end{figure}

In our figures, the majority of the CI variations most closely follow the extinction path (dark continuous line), which is calculated by our accretion disc model assuming the mean extinction law ($R_V=3.1$, see also Sec.~\ref{sec:accdisc} for more details). Both the uncertainty related to the true level of $I_{\rm C}$-band photometry, and simplified assumptions about the disk photosphere radiation function, are potential sources of the differences between the observed and synthetic color-magnitude diagrams. However, a more detailed look into the 2019 data (panels b, d, f) does reveal different relationship. The same is valid for the 2020 $(B-V)-V$ diagram (Fig.~\ref{fig:CIall}d) and for the associated $(V-r)-V$ and $(V-i)-V$ diagrams. These trends cannot be explained only by variable accretion (represented by continuous red line) - these brief `blueing events' (similar to those observed during the first plateau) are currently observed when the target is at the minimum brightness during the 502~d quasi-period and the CI variations related with the 1707~d QPO are relatively constant (see Fig.~\ref{fig:compBIphase} for illustration of mutual relations between both QPOs during 1997--2011).
Searching for similar events, we also examined archival 2002--2011 data. Only the $(B-V) -V$ diagrams obtained in 2005, 2006 and 2010 exhibit signatures of the CI reversal, but they are absent in the associated $(V-R_{\rm C})-V$ and $(V-I_{\rm C})-V$ diagrams.

We also investigated color-magnitude diagrams from the 2019 data gathered simultaneously with {\it TESS}. The spacecraft coincidentally observed V1057 Cyg during the major brightness increase (phases 0.98--0.1 according to ephemeris adopted for the 502~d QPO).  Thus the associated diagrams show the same well-defined CI reversal evidence characteristic of the entire 2019 dataset. In addition, we performed analysis of two specific color-magnitude diagrams, constructed from data obtained during the fainter and the brighter stages. Several (but not all) diagrams indicated 
variations along the extinction path, suggesting that the small-scale light changes noticed by {\it TESS} are just scaled-down counterparts of the major ones observed from the ground. However, given the limited precision of ground-based data and these relatively small brightness changes, respective correlation rank numbers are not high enough to confirm this behavior with high certainty.

\subsection{Spectroscopy}
\label{sec:spectroscopy}

We detected several emission and absorption lines in the spectra of V1057~Cyg. We used the NIST Atomic Spectra Database for the line identification (Kramida, A., Ralchenko, Yu., Reader, J. and NIST ASD Team 2020, NIST Atomic Spectra Database (version 5.8)\footnote{https://www.nist.gov/pml/atomic-spectra-database}). 
The lines detected in the spectra of V1057~Cyg are also listed in Appendix B, Tab.~\ref{tab:lines}, and Tab.~\ref{tab:lines_IR}.

\subsubsection{Optical Spectroscopy}
\label{sec:optical_spectroscopy}

Classical FUors show several common optical spectroscopic characteristics: P~Cygni profile of H$\alpha$, strongly blueshifted absorption lines, Li\,{\footnotesize I} absorption, and spectra similar to F/G supergiants/giants \citep{kenyon&hartmann1996, audard2014}. These spectroscopic features are also seen in our observations and most of the features vary with time.

\begin{figure}
\includegraphics[width=\columnwidth]{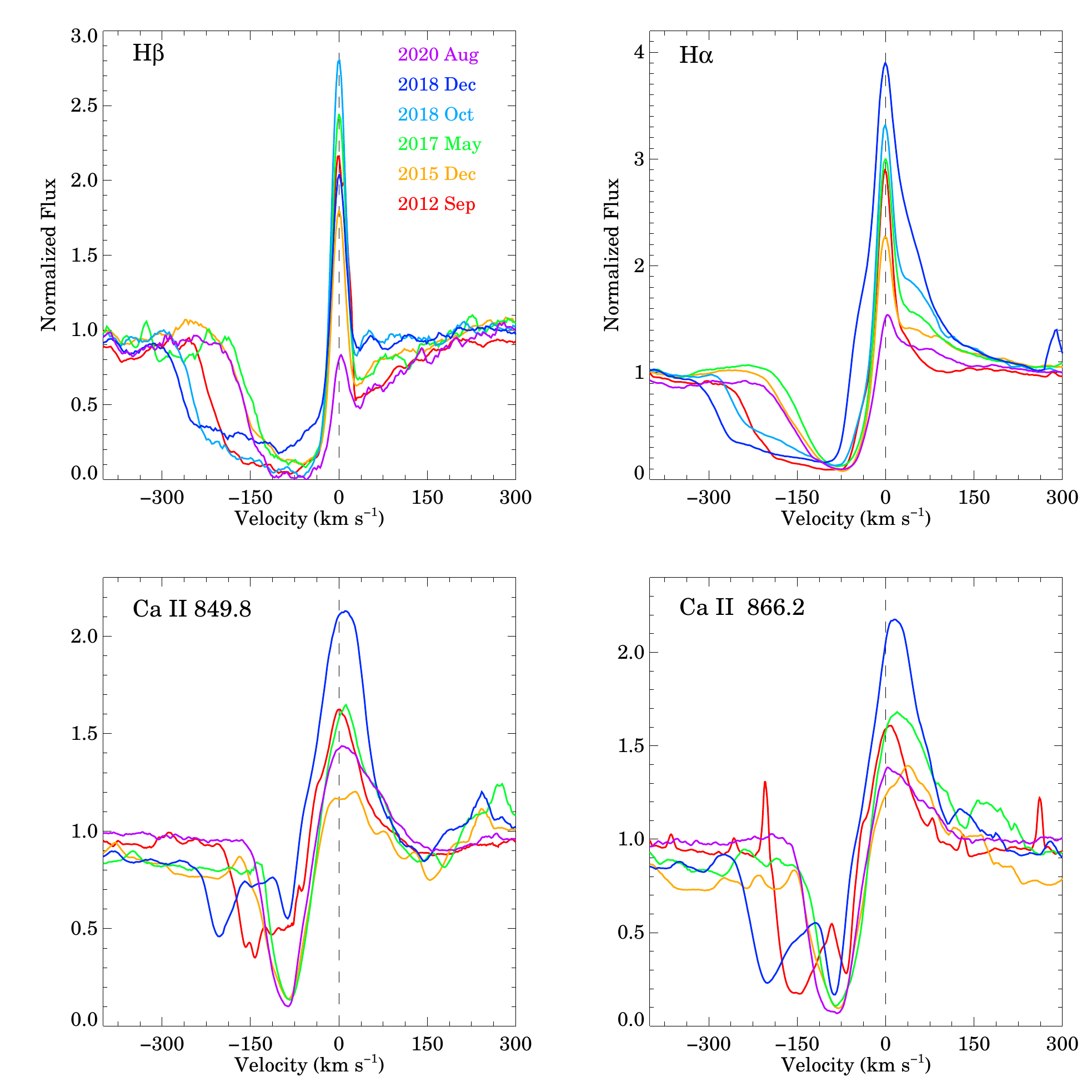}
\caption{Lines showing strong P~Cygni profile in the BOES and the NOT spectra of V1057~Cyg between 2012 and 2020: H$\beta$ 486.2\,nm, H$\alpha$ 656.3\,nm, Ca\,{\footnotesize II} 849.8\,nm, and Ca\,{\footnotesize II} 866.2\,nm. The BOES spectrum from 2018 October (sky blue) only covered wavelengths below 822.5\,nm. Different colors indicate different observation dates.}
\label{fig:pcygprofiles}
\end{figure}

P~Cygni profiles of several lines of hydrogen and metallic lines are found in the spectra of V1057~Cyg.
 The blueshifted absorption component of these profiles is formed by an outflowing wind \citep{kenyon&hartmann1996, hartmann2009, herbig2009, reipurth2010}. The strength of the blueshifted absorption component in the P~Cygni profile of the H$\alpha$ line is related to mass-loss in the wind \citep{herbig2003}.
 In our observations, P Cygni profiles of H$\beta$ 486.2, H$\alpha$ 656.3, and the Ca II infrared triplet (849.8, 854.2, and 866.2 nm) lines are identified. Fig.~\ref{fig:pcygprofiles} shows examples for the observed P~Cygni profiles.
The blueshifted absorption component of all P~Cygni profiles strongly varies with time. The high velocity component of the wind was observed in all P~Cygni profiles, and the highest velocity component was extended to about $-$300 $\sim$ $-$350\,km\,s$^{-1}$ in 2018 December.

The strength of the emission component of the lines with P~Cygni profiles also varies with time. Although there is no tight correlation between the variation of the absorption and emission components in most lines, they show similar trends in the case of H$\alpha$: when the blueshifted absorption component of the H$\alpha$ P~Cygni profile was at the highest velocity (2018 December), the strength of the redshifted emission component was also the strongest, and vice versa (the weakest in 2020 August). 

\begin{figure}
\includegraphics[width=\columnwidth]{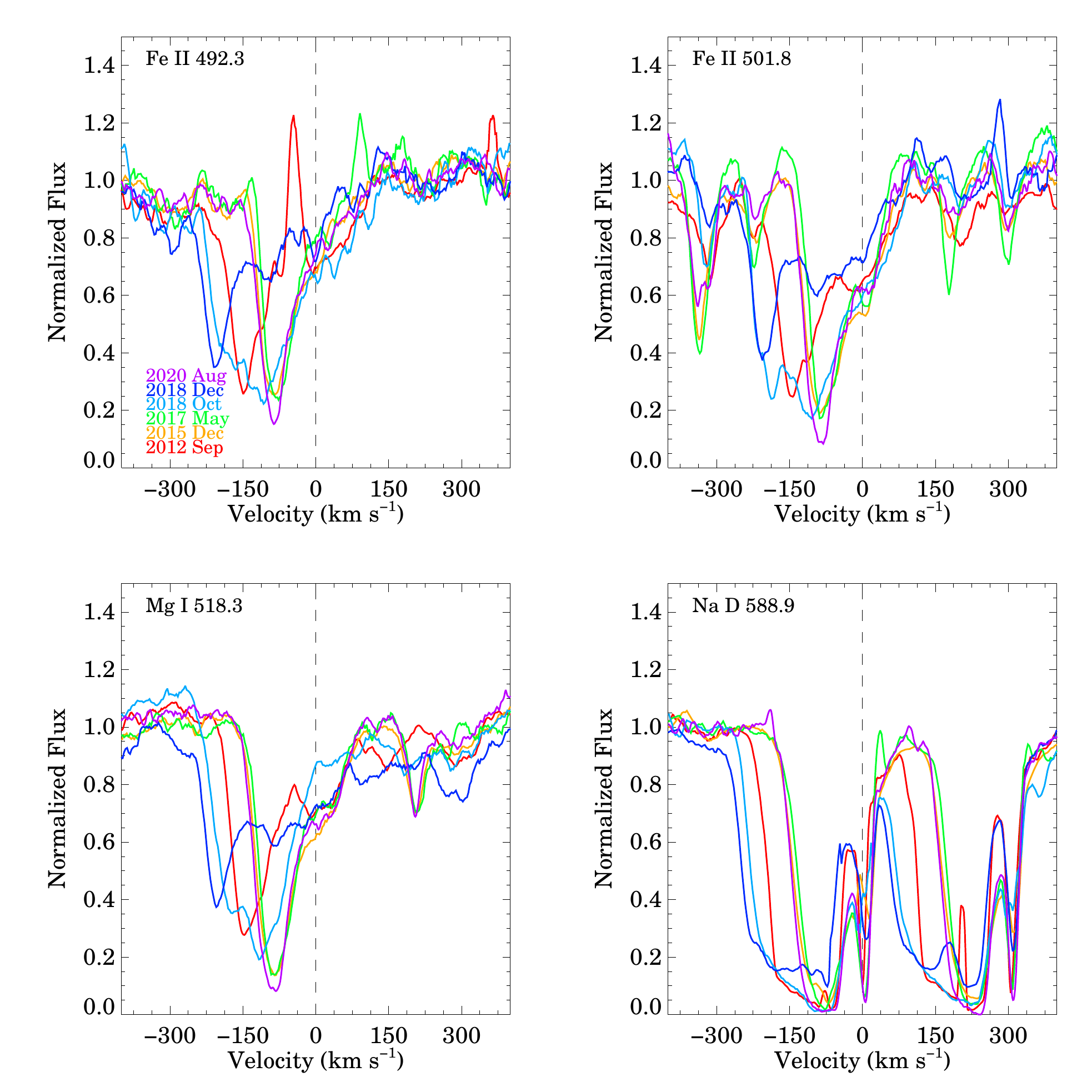}
\caption{Variation of the strong blueshifted absorption lines (Fe\,{\footnotesize II} 492.3\,nm, Fe\,{\footnotesize II} 501.8\,nm, Mg\,{\footnotesize I} 518.3\,nm, and Na D 588.9 and 589.5\,nm) detected in the spectrum of V1057~Cyg between 2012 and 2020. Different colors indicate different observation dates.}
\label{fig:absorption features}
\end{figure}

Strongly blueshifted absorption profiles caused by wind \citep{bastian1985, herbig2003, hartmann2009, miller2011} are also observed in V1057~Cyg. Some of the strongest examples are Fe\,{\footnotesize II} 492.3\,nm, Fe\,{\footnotesize II} 501.8\,nm, Mg\,{\footnotesize I} 518.3\,nm, and the Na D doublet (588.9 and 589.5\,nm), and these are plotted in Fig.~\ref{fig:absorption features}. All of the observed blueshifted absorption lines vary with time, and the variation trend is similar to that of the blueshifted absorption component of the P~Cygni profiles (Fig.~\ref{fig:pcygprofiles}). 
Among the observed blueshifted absorption lines, the Fe\,{\footnotesize II} 501.8\,nm and the Mg\,{\footnotesize I} 518.3\,nm lines show the same velocity variation over time, and thus likely originate from the same location in the structure.

\begin{figure}
\includegraphics[width=\columnwidth]{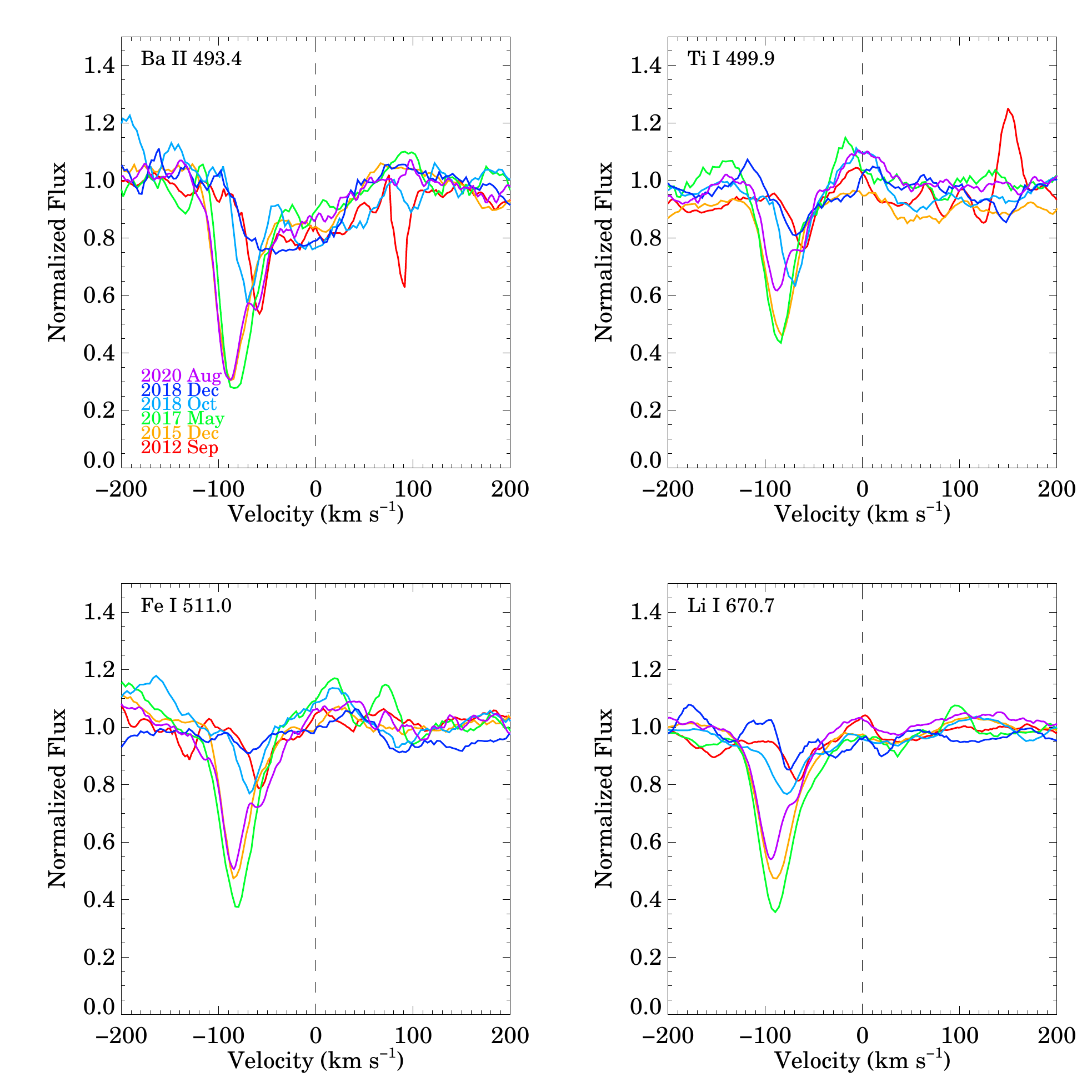}
\caption{Observed shell features of Ba~{\footnotesize II}~493.4 nm,  Ti~{\footnotesize I}~499.9 nm,
Fe~{\footnotesize I}~511.0 nm, and Li~{\footnotesize I}~670.7 nm lines. Different colors indicate different observation dates.}
\label{fig:ti_lines}
\end{figure}

Several shell features are also found in the spectra of V1057 Cyg. A total of eight shell features in the range of 493 -- 671\,nm, showing similar velocity variations with time as the Li\,{\footnotesize I} 670.7\,nm line, are selected. Four representative lines which show clear spectral profiles are presented in Fig.~\ref{fig:ti_lines}: Ba\,{\footnotesize II} 493.4\,nm, Ti\,{\footnotesize I} 499.9\,nm, Fe\,{\footnotesize I} 511.0\,nm, and Li\,{\footnotesize I} 670.7\,nm.
Since various atomic lines show the same velocity distribution, the correlation between atomic properties (lower energy level E$_{i}$, upper energy level E$_{k}$, and transition probability A$_{ki}$) and line profiles of shell features was investigated. However, no correlations between the line profiles and the different atomic parameters were found. All the detected shell features also vary with time during our observations. The highest velocity and strongest absorption profile is detected in 2017 May (green line) when the width of the blueshifted absorption component of the P~Cygni profile and the wind features are the narrowest (lower velocity).
As noted in previous studies \citep{herbig2003, herbig2009, kopatskaya2013}, a weak emission component of the Li \,{\footnotesize I} 670.7\,nm line was also observed in September 2012.

We also detected several forbidden emission lines in the spectra of V1057~Cyg, such as [N\,{\footnotesize II}] 654.8, 658.3, [S\,{\footnotesize II}] 671.6, 673.1, [O\,{\footnotesize I}] 630.0, 636.3, [O\,{\footnotesize III}] 495.9, 500.7\,nm, and [Fe\,{\footnotesize II}] 715.5\,nm, which are rarely detected in classical FUors. Among the several forbidden emission lines, the relatively weak [N\,{\footnotesize II}] 654.8, 658.3, [S\,{\footnotesize II}] 671.6, 673.1\,nm, and [O\,{\footnotesize III}] 495.9, 500.7\,nm lines are detected for the first time in the spectra of V1057~Cyg. The [S\,{\footnotesize II}] emission line was previously found in only three known FUors: V2494~Cyg \citep{magakian2013}, V960~Mon \citep{takagi18, park2020}, and V346~Nor \citep{kospal2020}, and the [N\,{\footnotesize II}] emission line was only found in V346~Nor \citep{kospal2020}.

\begin{figure}
\includegraphics[width=\columnwidth]{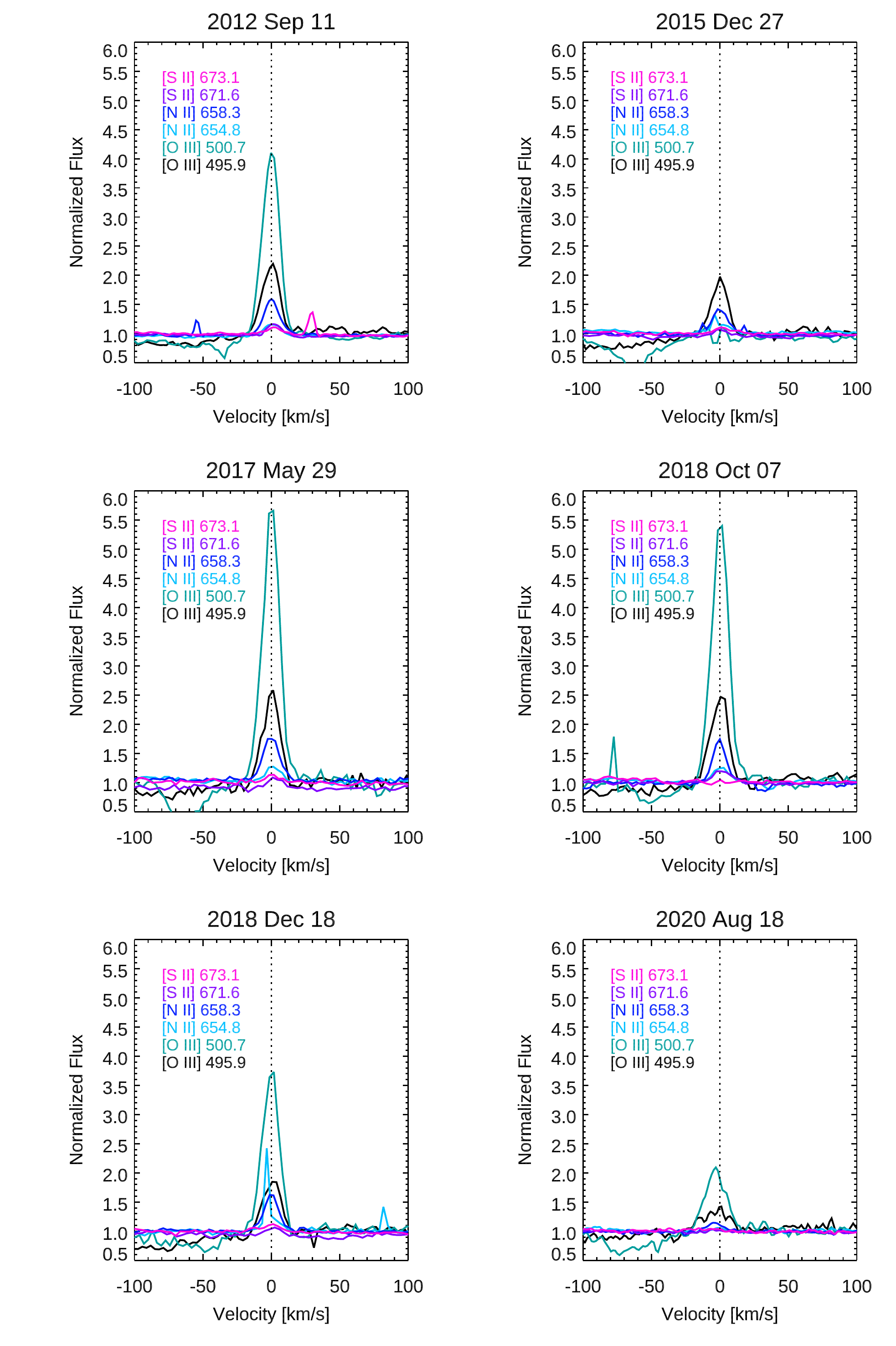}
\caption{Relatively narrow forbidden emission lines detected in the spectra of V1057~Cyg: [O\,{\footnotesize III}] 495.9\,nm (black), [O\,{\footnotesize III}] 500.7\,nm (green), [N\,{\footnotesize II}] 654.8\,nm (sky blue), [N\,{\footnotesize II}] 658.3\,nm (blue), [S\,{\footnotesize II}] 671.6\,nm (purple), and [S\,{\footnotesize II}] 673.1\,nm (pink). The narrow emission close to the [S\,{\footnotesize II}] 671.6\,nm at the upper left panel (2012 September 11) is a sky emission line \citep{hanuschik03}. The other narrow emission lines are cosmic rays, and also the [N\,{\footnotesize II}] 654.8\,nm observed on 2018 December~18 is affected by cosmic rays.}
\label{fig:sii}
\end{figure}

These forbidden emission lines are generally associated with spatially resolved jets or outflows in Class II YSOs \citep{cabrit1990, hartmann2009}. The [N\,{\footnotesize II}], [S\,{\footnotesize II}], and [O\,{\footnotesize III}] emission lines are relatively narrow, and the peak velocity is located around systemic velocity (Fig.~\ref{fig:sii}). These emission lines are detected in most epochs, and their strengths also changed. The [O\,{\footnotesize III}] emission lines are relatively stronger than the [N\,{\footnotesize II}] and [S\,{\footnotesize II}] lines. Most of the [O\,{\footnotesize III}] lines are detected during our observations, except [O\,{\footnotesize III}] 500.7 nm in 2015 December. The strength of the [S\,{\footnotesize II}] emission lines is weaker than those of the [N\,{\footnotesize II}] emission lines. The strength of [N\,{\footnotesize II}] emission lines is very weak in 2020 August, and the [S\,{\footnotesize II}] 673.1\,nm line is not detected in 2018 October, and neither of the [S\,{\footnotesize II}] emission lines was observed in 2020 August, indicating that the jet/outflow is also showing variability in time.

\begin{figure}
\includegraphics[width=\columnwidth]{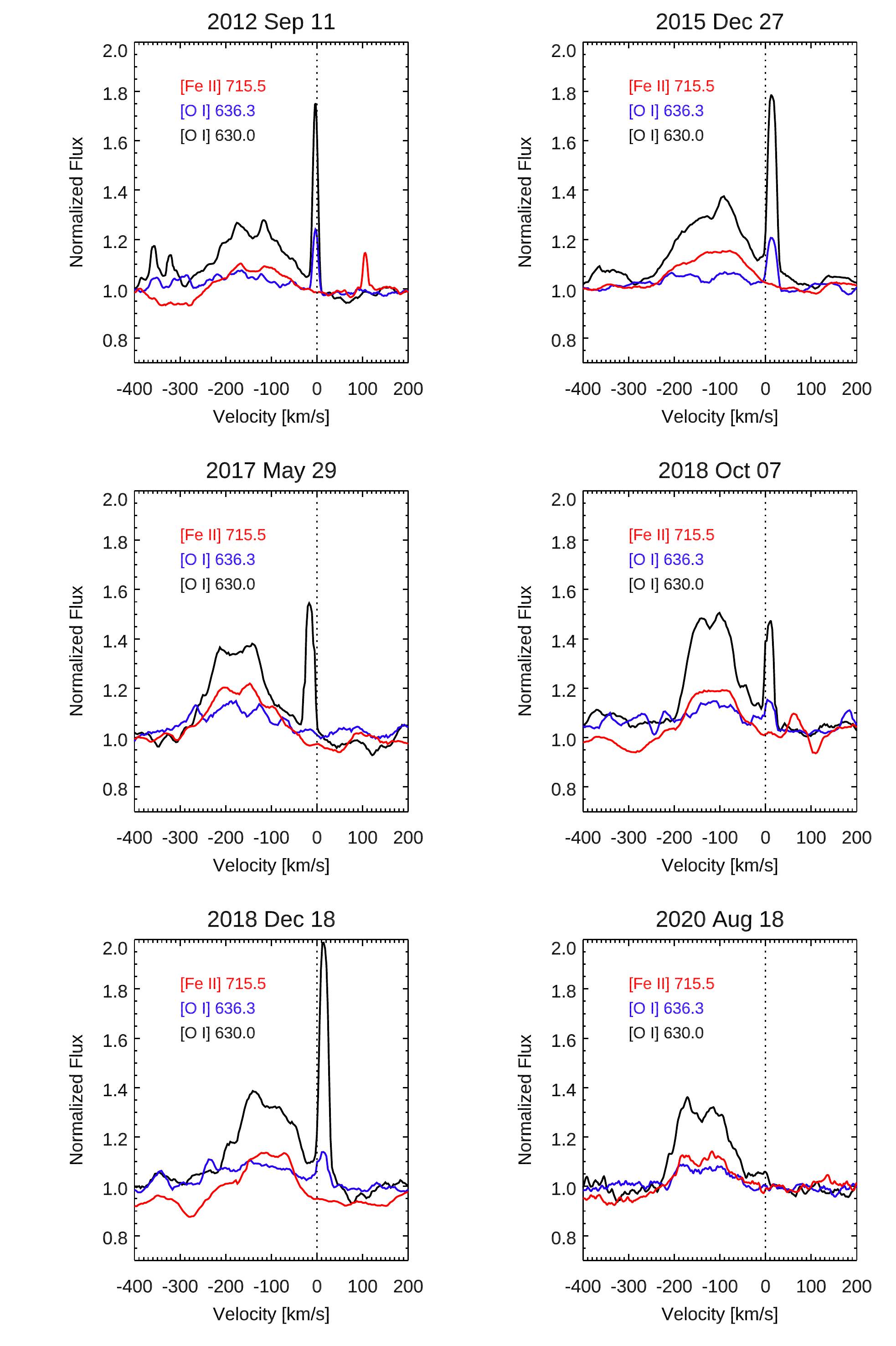}
\caption{Strong forbidden emission lines detected in V1057 Cyg: [O\,{\footnotesize I}] 630.0\,nm (black), [O\,{\footnotesize I}] 636.3\,nm (blue), and [Fe\,{\footnotesize II}] 715.5\,nm (red). These three forbidden emission lines are strongly blue-shifted, and the emission peaks are at around $-$150 km/s. The narrow emission component at around the systemic velocity in the spectra taken between 2012 and 2018 is sky emission line \citep{hanuschik03}.}
\label{fig:emission}
\end{figure}

In the case of T Tauri stars, the [O\,{\footnotesize I}] 630.0\,nm emission line is often observed as two components: high-velocity (a few hundred km\,s$^{-1}$) and low-velocity (a few tens of km\,s$^{-1}$) \citep{hartigan1995, hartmann2009}. 
In our observations, [O\,{\footnotesize I}] 630.0\,nm line shows two velocity components which are both greater than 91 km\,s$^{-1}$ wide. The high-velocity component can be formed by the outflowing wind \citep{hartmann2009}.
The relatively higher velocity peaks are at around $-$140 to $-$213 km\,s$^{-1}$, and the relatively lower velocity peaks are at around $-$91 to $-$117 km\,s$^{-1}$ (Fig.~\ref{fig:emission}).
The velocity variation of these components is similar to those of lower velocity components of shell, wind, and P~Cygni profiles. Therefore, the lower velocity component of these lines can be formed at the same place of the structure. The strength of the forbidden emission lines varies slightly, but less than that of the wind features.

In contrast with previous studies \citep{kenyon-and-hartmann1988, kenyon&hartmann1996, hartmann2009}, we did not detect double-peaked line profiles in our observations. 

\subsubsection{Near-infrared Spectroscopy}

\begin{figure*}
\centering
\includegraphics[width=\textwidth]{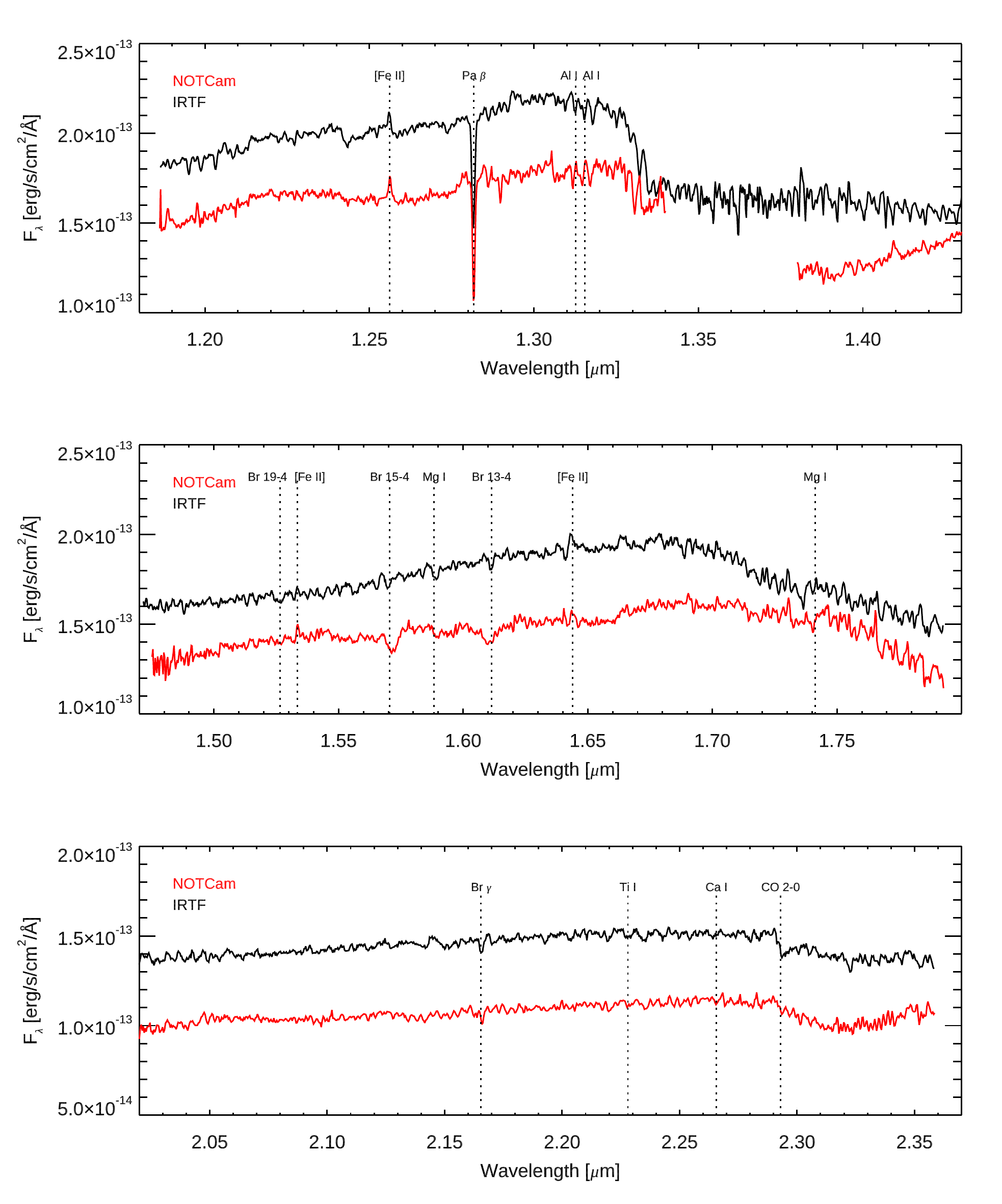}
\caption{The near-infrared $J$, $H$, and $K$ spectrum of V1057 Cyg observed with NOTCam (red) and IRTF \citep[black;][]{connelley2018}. The region of 1.34--1.38$\,\mu$m at the $J$ band was removed due to strong telluric absorption features.} 
\label{fig:nir_spec}
\end{figure*}

We detected several absorption and emission lines in the near-infrared spectrum.
Fig.~\ref{fig:nir_spec} shows the comparison between our NOTCam spectrum observed in 2020 August 29 (red) and that of IRTF \citep{connelley2018} observed in 2015 June 26 (black). Similarly to \citet{connelley2018}, we also detected Pa$\beta$ 1.281 $\mu$m, Al\,{\footnotesize I} 1.312, 1.315$\,\mu$m, and strong water absorption bands, although our spectrum is not corrected well around 1.35$\,\mu$m in the $J$-band because of strong telluric absorption features. In the $H$-band, the 19--4, 15--4, and 13--4 lines of the Br series are detected in broad absorption, and the [Fe\,{\footnotesize II}] 1.533, 1.644$\,\mu$m lines are detected in emission. The Mg\,{\footnotesize I} 1.588, 1.741$\,\mu$m absorption lines are also detected. The Br$\gamma$ 2.165, Ti\,{\footnotesize I} 2.228, Ca\,{\footnotesize I} 2.265$\,\mu$m lines are detected in absorption in the $K$-band. The Br$\gamma$ appeared as a weak P~Cygni profile in the previous study \citep{connelley2018}, but it appeared as an absorption line in this study. The difference between the two spectra is the detection of the [Fe\,{\footnotesize II}] emission lines and the shape of the CO overtone bandhead features. Emission lines are rarely detected in classical FUors.  However, as in the optical spectra, we also detected a few [Fe\,{\footnotesize II}] emission lines in the near-infrared spectrum. Compared to earlier observations of V1057~Cyg, the strength of the CO first overtone bandhead feature appears to be the weakest in 2020 (see Sec.~\ref{sec:discussion_co}).

\section{Discussion} \label{sec:discussion}

\subsection{Accretion disk modeling}
\label{sec:accdisc}

While the long-term light curve of V1057~Cyg suggests a general decay of the accretion rate after the outburst peak in 1971, changing extinction towards the source might also play a role. In this section we attempt to separate the effects of variable accretion and extinction, and study their long term evolution quantitatively. Following the method we have successfully applied on several young eruptive stars \citep{kospal2016,kospal2017a,abraham2018,kun2019,szegedi-elek2020}, we model the inner part of the system with a steady, optically thick and geometrically thin viscous accretion disk, whose mass-accretion rate is constant in the radial direction \citep[Eq.~1 in ][]{kospal2016}. We neglect any contribution from the star itself, assuming that all optical and near-infrared emission in the outburst originates from the hot accretion disk. We calculated synthetic SEDs of the  disk by integrating the blackbody emission of concentric annuli between the stellar radius and $R_{\rm out}$. 

A fundamental input parameter of the model is the inclination of the accretion disk. Estimates in the literature, mainly based on SED fitting, range between 0$^{\circ}$ (pole-on) and 30$^{\circ}$ \citep[for a review, see][]{gramajo2014}. In order to derive a value based on observations, we analysed the 1.3 mm continuum observations of \citet{liu2018} obtained with the Submillimeter Array (SMA). Deconvolving the measured size of their continuum source (1$\farcs$00$\times$0$\farcs$59, PA=84$^{\circ}$) by a beam of 0$\farcs$87$\times$0$\farcs$50, PA=76$^{\circ}$, the resulting ratio of the minor and major axes implies an inclination of $i$=62$^{\circ}$. This result indicates a more edge-on view of the disk than previously thought. While this inclination value was derived from measurements of the whole disk, including both the outer cold regions and the hot inner disk, we will adopt it for the subsequent modeling of the accretion disk. 
This assumption is independently confirmed based on comparison of our Na~I doublet spectra with those obtained from disk wind models by \citet{milliner2019}.  

The outer radius of the accretion disk, another input parameter, mainly affects the mid-IR emission. We fixed it to $R_{\rm out} = 1$~au, which matches the early L-band observations of V1057~Cyg in the 1970's. The inner radius of the disk, equal to the stellar radius, mainly influences the optical bands. However, we cannot discriminate between the cases of smaller stellar radius with higher line-of-sight extinction as opposed to larger radius with lower extinction using our broad-band optical photometry. In order to circumvent this problem, we prescribed that the A$_V$ value computed for 2020 August must comply with the $A_V$=3.9$\pm$1.6 mag proposed by \citet{connelley2018} based on an infrared spectrum taken in 2015. This constraint set $R_{\rm in}$ = 4.6~R$_{\odot}$. 


With this model setup, only two free parameters remain: the product of the accretion rate $\times$ stellar mass $M\dot{M}$, and the line-of-sight extinction $A_V$. 
We calculated disk SEDs for a large range of $M\dot{M}$, and at each step the fluxes were reddened using a large grid of $A_V$ values assuming the standard extinction law from \citet{cardelli1989} with $R_V = 3.1$. Finding the best $M\dot{M}$ -- $A_V$ combination was performed with ${\chi}^2$ minimization, by taking into account all measured flux values between 0.4 and 2.5$\,\mu$m. Preferentially we performed our modeling when both optical and infrared data were available for the same night, but we also included epochs when only $JHK$ photometry was taken but 
optical data were available within 10 days, thus interpolation in the optical fluxes was acceptable. 
The formal uncertainties of the data points were set to a homogeneous 5\% of the measured flux value, which also accounted for possible differences among photometric systems.
The model fits usually reproduced the measurements reasonably well, with typical reduced $\chi^2$ values below four. 

\begin{figure}
\includegraphics[width=0.9\columnwidth]{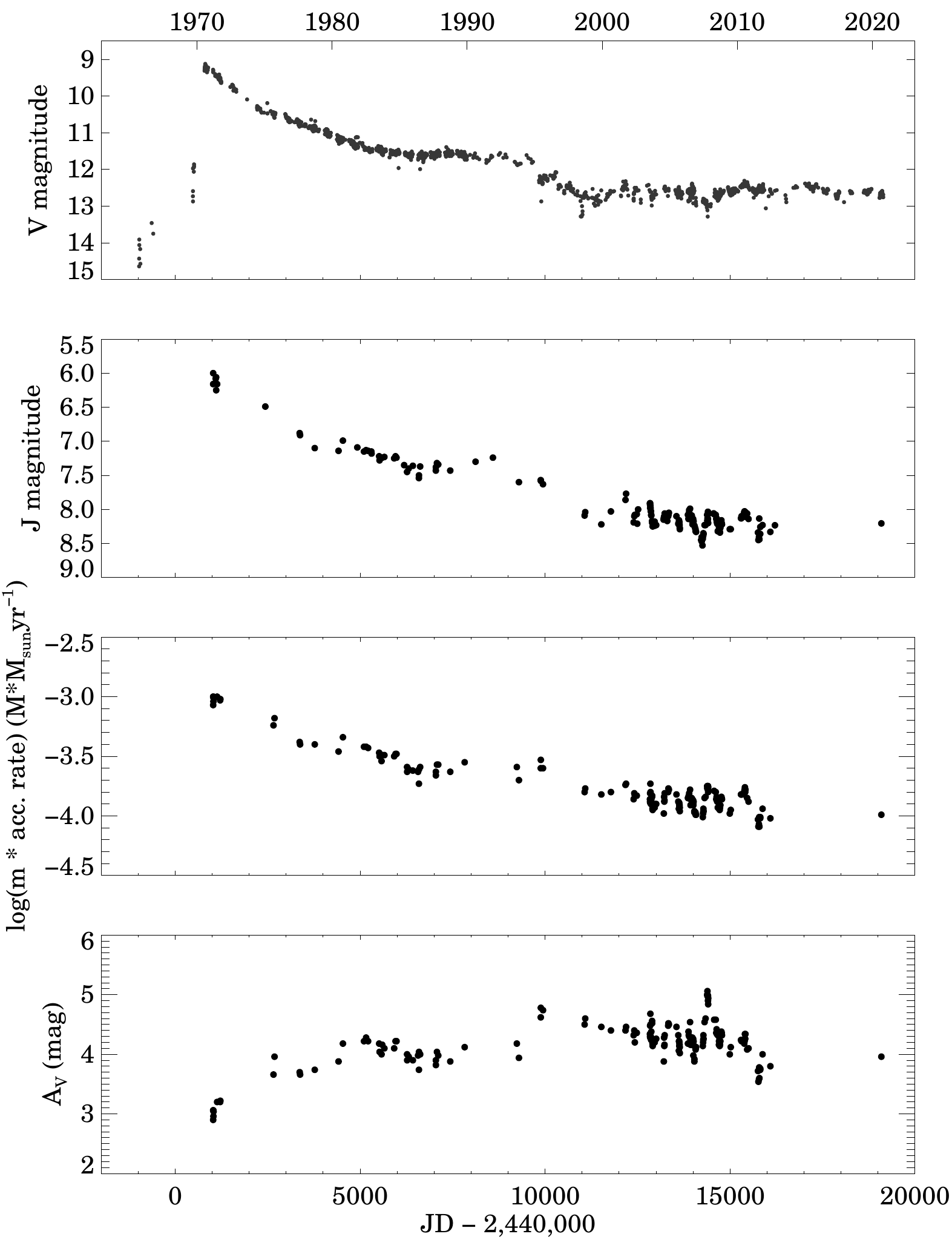}
\caption{$V$ and $J$-band light curves of V1057~Cyg for reference (first and second panel), temporal evolution of the accretion rate (third panel) and line-of-sight extinction (bottom panel) derived from our accretion disk modeling described in Sect.~\ref{sec:accdisc}.}
\label{fig:accdisk}
\end{figure}

The resulting temporal evolution of the accretion rate and extinction values, together with the $V$ and $J$-band light curves, are plotted in Fig.~\ref{fig:accdisk}. The initial decay of the source, between the outburst maximum and ~1987, can be explained by an exponential drop of the accretion rate from 10$^{-3}$~M$_{\odot}$M$_{\odot}$yr$^{-1}$ to $\sim$2.5$\times$10$^{-4}$~M$_{\odot}$M$_{\odot}$yr$^{-1}$, with an e-folding time of 4300 days ($\sim$12\,yr). During this fading phase (1971--1987)
the extinction first slowly increased by $\sim$1\,mag, while after 1983 
slightly decreased again, suggesting a rearrangement of the circumstellar structure, and/or a change in the dust size distribution in the line-of-sight, leading to a different extinction law. Then between 1987 and 1993 
both the accretion rate and the extinction stayed constant. In 1994--95 $A_V$ suddenly rose by $\sim$0.6\,mag (probably causing the sudden drop of optical brightness at the same time). In the `second plateau' phase no long-term trend can be seen in $M\dot{M}$, and only a weak initial decay in $A_V$. On top of this relatively constant behavior in the `second plateau', correlated oscillations can be seen in the $M\dot{M}$ and $A_V$ curves. These are probably due to the fact that the unusual optical-infrared color variations, caused by the superposition of two periodic processes of very different wavelength dependencies (Fig.~\ref{fig:compBIphase}), cannot be simply reproduced by our simple analytical model, and thus these variations should not be overinterpreted. 
The current luminosity of the accretion disk is about 330\,$L_{\odot}$, but its value depends on the disk inclination value. Since we adopted a more edge-on orientation than before in the literature, our inferred luminosity also increased. The current accretion rate of V1057~Cyg in our model, also slightly dependent on the inclination and the stellar radius, is about 10$^{-4}$\,M$_{\odot}$M$_{\odot}$yr$^{-1}$.

\subsection{Spectral energy distribution}
\label{sec:sed}
\begin{figure}
\includegraphics[width=\columnwidth]{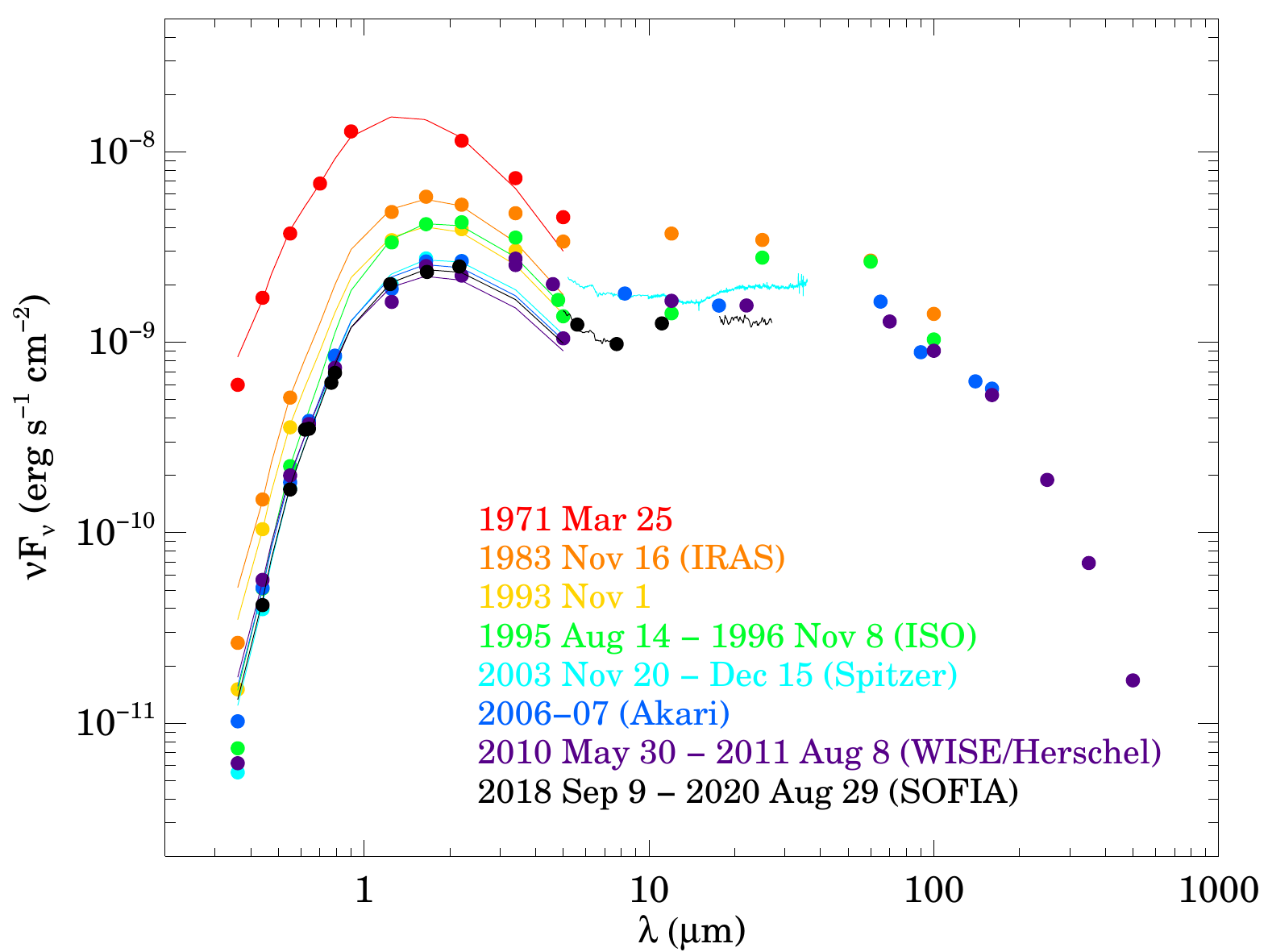}
\caption{Spectral energy distribution of V1057~Cyg at different representative epochs. 
The data points are from Fig.~\ref{fig:lc}, as well as from space-based (\textit{IRAS}, \textit{ISO}, \textit{Spitzer},  \textit{Akari}, \textit{WISE},  \textit{Herschel}) and airborne (SOFIA) missions, as indicated in the legend. Solid curves show the results of our accretion disk models for the individual epochs.
}
\label{fig:sed}
\end{figure}

In Fig.~\ref{fig:sed} we plot the spectral energy distribution of V1057~Cyg at several epochs since the outburst. The optical and near-infrared points are from Fig.~\ref{fig:lc}, while longer wavelength photometry was collected from different space-borne (\textit{IRAS}, \textit{ISO}, \textit{Spitzer}, \textit{WISE}, \textit{Akari}, \textit{Herschel}) or airborne (SOFIA) missions. The data points from Herschel and the AllWISE catalog were taken within a year, thus we combined the two data sets into a single SED. The SOFIA spectra were smoothed and scaled to simultaneous SOFIA photometry. 

The gradual decrease in the short wavelength part reflects the evolution of the hot inner accretion disk as modeled in Sect.~\ref{sec:accdisc}. The difference between the 1993 and 1995 SEDs displays how the fading in 1995 became apparent first in the optical regime, while the near-infrared part stayed constant. The SEDs after 2003 (`second plateau' phase) were very similar; their slight differences reflect only the periodic behavior described in Sect.~\ref{sec:periodanalysis}.
Between 5$\,\mu$m and 100$\,\mu$m, V1057~Cyg also faded, although significantly less than at optical wavelengths (part of this flux drop might be related to the improving spatial resolution, and thus smaller aperture size of the subsequent telescopes). 

Based on a comparison of \textit{IRAS} and \textit{ISO} measurements, \citet{abraham2004b} claimed that below 25$\,\mu$m the flux was variable while at longer wavelength it remained constant. 
Extending the temporal baseline of this study with subsequent \textit{Spitzer}, \textit{Akari}, \textit{WISE}, \textit{Herschel}, and SOFIA measurements, almost a factor of 3 systematic fading was observed between \textit{IRAS} and SOFIA at $\sim$25~$\mu$ma. This fading was also seen at far-infrared wavelengths by comparing the \textit{Akari} and \textit{Herschel} photometric points to earlier \textit{IRAS} and \textit{ISO}, although the fading was less pronounced (less than a factor of 2).
\citet{abraham2004b} concluded that the outer part of the system, responsible for the long wavelength SED, has an energy source different from the central star. However our new results imply that the circumstellar medium does react to the changing irradiation by the central source, and thus the origin of the energy emitted by the envelope is more likely the outbursting star than an external source.


\subsection{Optical spectroscopy}
As described above in Sec.~\ref{sec:spectroscopy}, we detected several wind features in the spectra of V1057~Cyg. The velocity of the blueshifted absorption component and the strength of the redshifted emission component of P Cygni profiles vary from year to year.
The highest velocity of the blueshifted absorption component was observed in 2018 December in all P~Cygni profiles (Fig.~\ref{fig:pcygprofiles}) and wind features (Fig.\ref{fig:absorption features}), and blueshifted velocity components of P Cygni profiles and wind features change similarly with time. From our observations, we can confirm that the year-to-year variability of
strongly blueshifted absorption components of P~Cygni profiles and wind features are similar to those observed by \citet{herbig2003}, suggesting variability over time in the strength of the wind.


The emission components of the H$\alpha$ and Ca {\footnotesize II} IRT P~Cygni profiles are the strongest in 2018 December, while the other lines behave differently. All of the absorption and emission component of P~Cygni profiles change with time, but there is no tight correlation between the two components, except for H$\alpha$ (Sec.~\ref{sec:optical_spectroscopy}).

The shell features were variable during our observations, but the data do not suggest a well-defined trend. The shell features were the strongest in 2017 May, when the blueshifted components of wind and P Cygni profiles were the lowest velocity, and also when the system was close to the minimum light in that year (Fig.~\ref{fig:lc-piszkes}).
Since both depth and velocity change over time discontinuously, this variation over time can be interpreted as a rapidly-changing wind or rotation of non-axisymmetric components \citep{powell2012, Sicilia-Aguilar2020}.

We also detected several forbidden emission lines: [N\,{\footnotesize II}] 654.8, 658.3, [S\,{\footnotesize II}] 671.6, 673.1, [O\,{\footnotesize I}]~630.0, 636.3, 
[O\,{\footnotesize III}] 495.9, 500.7\,nm, and [Fe\,{\footnotesize II}] 715.5\,nm. 
These are rarely found in FUors: 
so far, only three known FUors show these properties \citep{magakian2013, takagi18, kospal2020, park2020}. In contrast, these lines are generally found in classical T~Tauri stars as tracers of outflow or jets \citep{cabrit1990, hartmann2009}. 
The lack of forbidden emission lines in FUors could be due to
the lack of detailed spectroscopic studies and the small number of FUors known at this point.
In addition, typically, the continuum of FUors is very bright, which makes it hard to detect the forbidden emission lines due to the contrast.
On several epochs, [O\,{\footnotesize III}] 495.9, 500.7, [N\,{\footnotesize II}] 654.8, 658.3, and [S\,{\footnotesize II}] 671.6, 673.1\,nm lines are detected for the first time in the spectra of V1057~Cyg. All of the detected forbidden emission lines also vary with time, but less than the wind features. However, the variation of these emission lines suggests that any jets/outflows in the system also change with time.

\subsection{Variation of the CO first overtone bandhead}
\label{sec:discussion_co}
\begin{figure}
\centering
\includegraphics[width=\columnwidth]{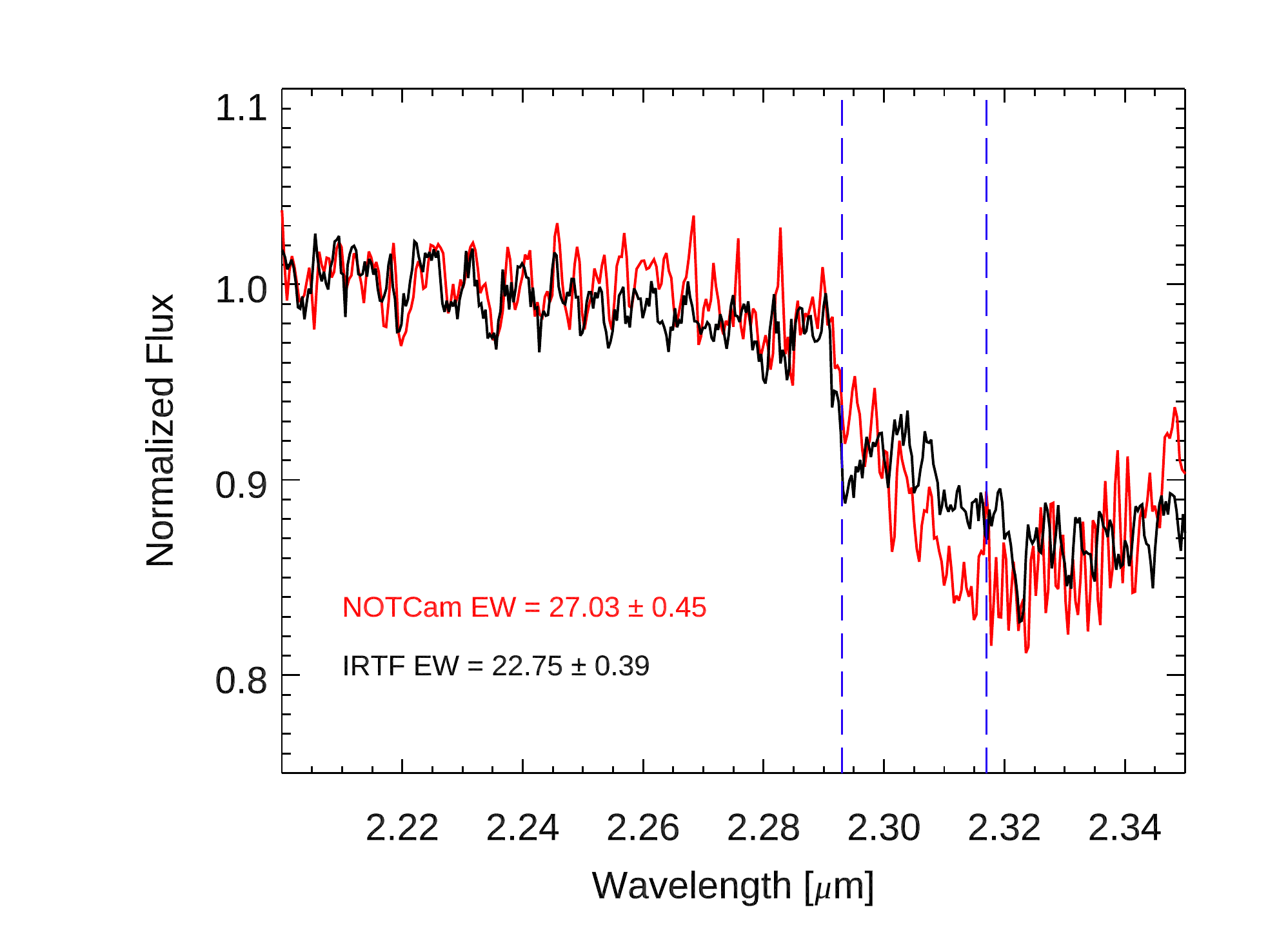}
\caption{CO first overtone bandhead features observed in NOTCam (red) and IRTF \citep[black;][]{connelley2018}. The EW was measured between the blue dashed lines (from 2.293 to 2.317$\,\mu$m). 
} 
\label{fig:ew_co}
\end{figure}

The strength of the CO bandhead feature in V1057~Cyg decreased and the equivalent width (EW) increased in these epochs, according to the original studies \citep{mould1978, hartmann1987, biscaya1997}, and this trend continued in recent observations. Fig.~\ref{fig:ew_co} shows the recent observations of the CO v=2--0 2.293$\,\mu$m first overtone bandhead with the NOTCam (red) and the IRTF (black; connelley).
We measured the EW of the CO feature from 2.293 to 2.317$\,\mu$m (blue dashed line), which is the same region used by \citet{biscaya1997} (see their Table 3). The EW of the NOTCam (27.03 $\pm$ 0.45 \AA{}) and the IRTF (22.75 $\pm$ 0.39 \AA{}) data were estimated with a Monte Carlo method. The EWs were measured 1000 times with random Gaussian errors multiplied by the observation errors. The standard deviation derived from all 1000 EW measurements was adopted as the uncertainty of the EW. Our results, together with values from the literature, are listed in Tab.~\ref{tab:co}. The measured EW is stronger in 2015 and 2020 than in 1986 (from 2.293 to 2.305$\,\mu$m) and 1996, as the K-band magnitude decreases (Fig.~\ref{fig:lc}). We suggest that the weakened strength of the CO overtone bandhead features in our observation of V1057 Cyg is also caused by the decrease in brightness \citep{biscaya1997, connelley2018}, which can then be related to decreasing mass accretion rate and disk midplane temperature. Our modeling of the disk (Fig.~\ref{fig:accdisk}) confirms the proposed explanations of decreasing brightness and therefore likely decreasing mass accretion rate and midplane temperature.

\begin{deluxetable}{lll}
\tabletypesize{\scriptsize}
\tablecaption{EW of CO overtone bandhead \label{tbl_co}}
\tablewidth{0pt}
\tablehead{
\colhead{Observation Date} & \colhead{EW} & \colhead{Reference} \\[-3mm]
\colhead{[UT]} & \colhead{[\AA{]}} & \colhead{}}
\startdata
1986 & 14.6$^{a}$ $\pm$ 0.7 & \citet{carr1989} \\
1996 June & 18.3 $\pm$ 1 & \citet{biscaya1997} \\
2015 June$^{b}$ & 22.75  $\pm$ 0.36 & This work \\
2020 August & 27.03 $\pm$ 0.45 & This work \\
\enddata
\tablenotetext{a}{Measured EW range: 2.293 -- 2.305$\,\mu$m}
\tablenotetext{b}{Spectrum from \citet{connelley2018}}
\label{tab:co}
\end{deluxetable}

\subsection{About the nature of the two quasi-periodic components in the `second plateau'}
\label{sec:perioddiscussion}

\begin{figure}
\centering
\includegraphics[width=\columnwidth]{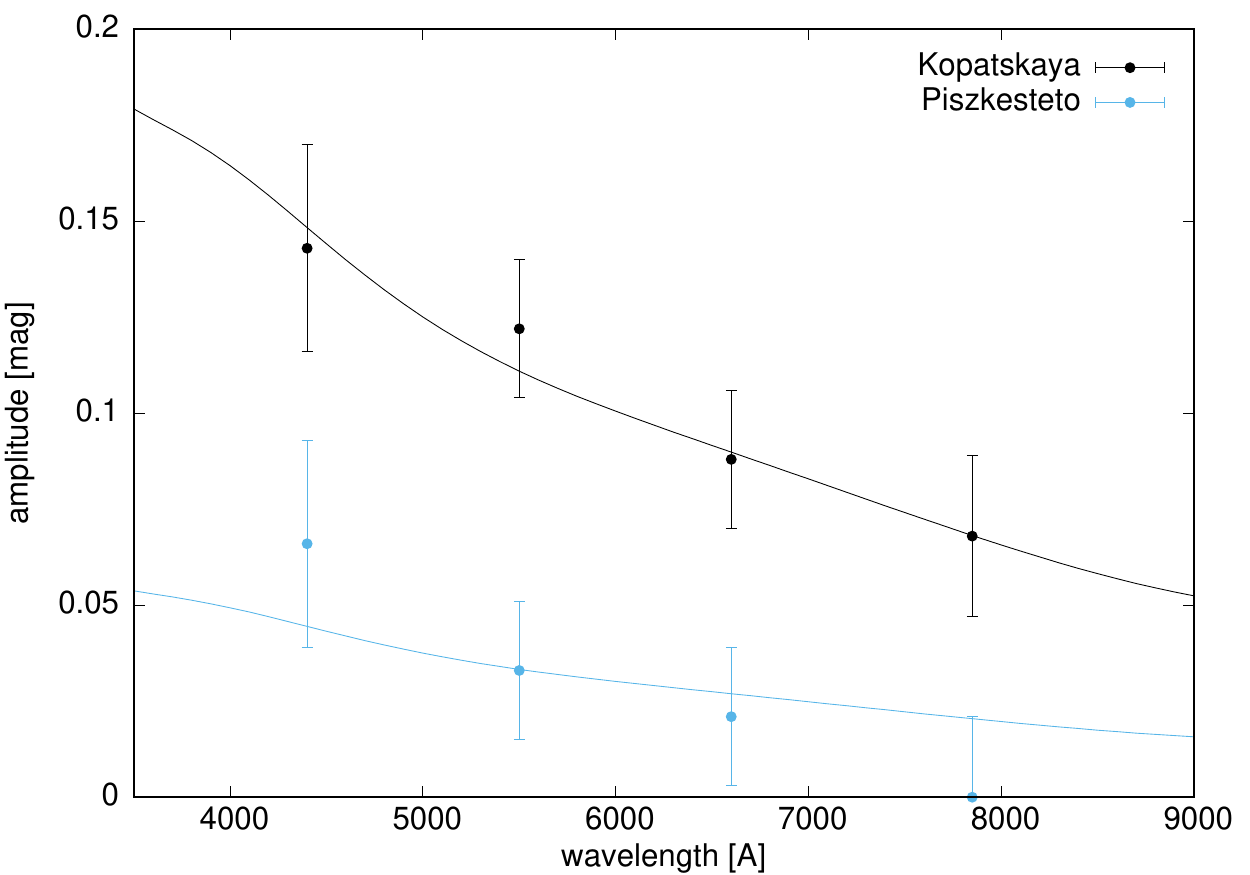}
\caption{Amplitudes of 1707~d light variations obtained from sinusoidal fits to the archival and new data (Sec~\ref{sec:ampl-wav_dep}), with the arbitrarily scaled reddening curve for $R_V=3.1$. The formal 1-$\sigma$ errors were multiplied by three to show realistic uncertainties.} 
\label{fig:ampl-evol}
\end{figure}

\begin{figure}
\includegraphics[width=0.49\columnwidth]{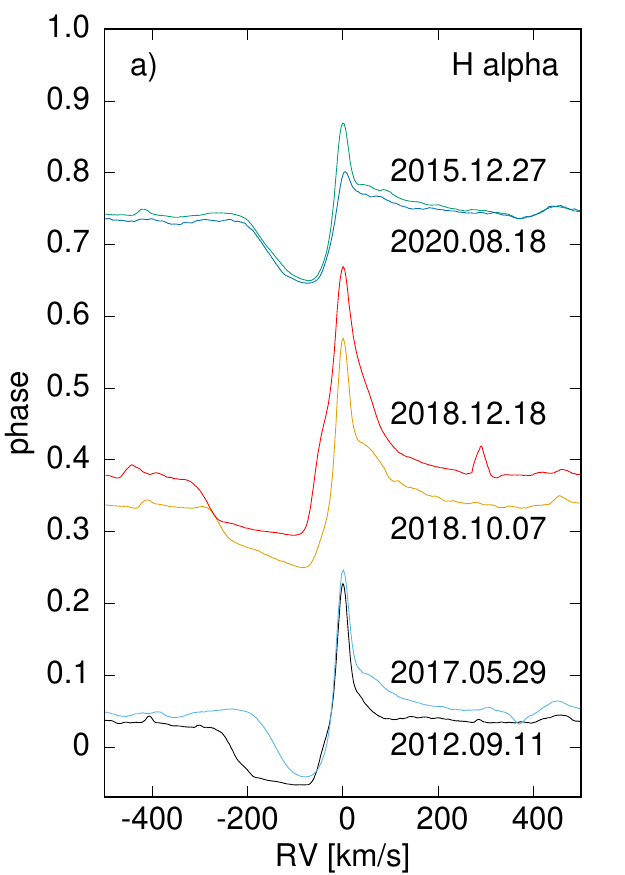}
\includegraphics[width=0.49\columnwidth]{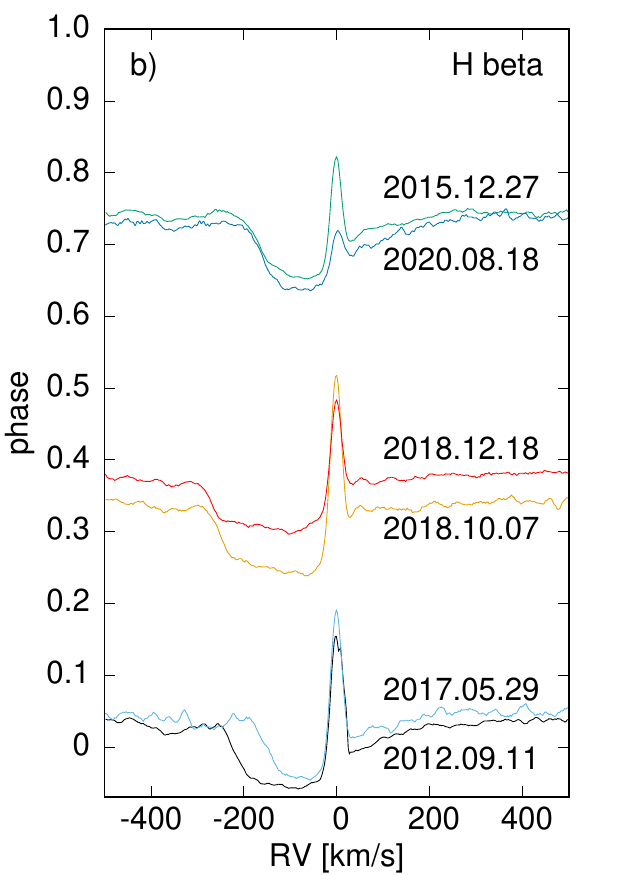}
    \caption{Variations of the H$\alpha$ an H$\beta$ lines in phase, calculated for the 1707~d quasi-period. 
}
\label{fig:PCyg-phase}      
\end{figure}

\begin{figure}
\centering
\includegraphics[width=\columnwidth]{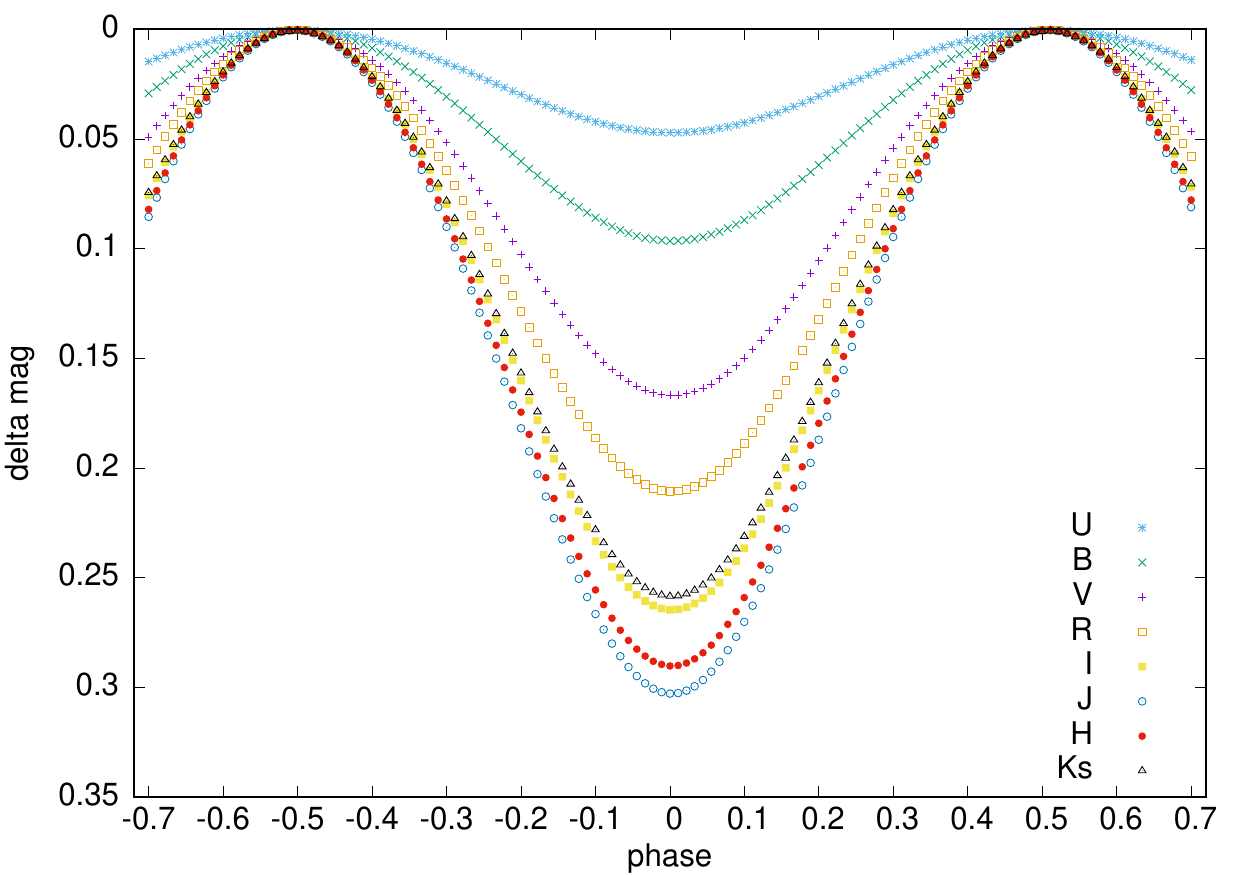}
\caption{Synthetic light curves caused by rotation of hot spot of 30$^{\circ}$ azimuthal width contained between 200-240~R$_{\odot}$. The brightness minimum is at phase 0, when the hypothetical hot spot is located in front of the star and its projected surface area is the smallest. As the $U-V$ bands are strongly affected by 1707~d QPO, only $R_{\rm C}-K$ bands were used to refine the final modelling result.}
\label{fig:hotspot}
\end{figure}

With 23 years of coverage of the `second plateau', in Sec.~\ref{sec:periodanalysis} we refined the values of the associated quasi-periods to 1707$\pm$70~d and 502$\pm$20~d. Precise Schmidt and RC80 telescope data enabled detection of the shorter period in the $B$ and $V$-band for the first time. Similarly, we obtained the first marginal detection of the longer period in the $I_{\rm C}$-band. 
Furthermore we obtained that during the light minimum associated with the 502~d period all CIs are becoming bluer, but found the opposite (redder) for the 1707~d quasi-period. The amplitudes related to the longer period decrease, while the amplitudes related to the shorter period increase with increasing effective wavelength ($BVR_{\rm C}I_{\rm C}$), respectively (Sec.~\ref{sec:ampl-wav_dep}). Moreover, our precise Schmidt-telescope data revealed that as the time progresses, the amplitudes related with the 1707~d period are becoming smaller in all these filters. No significant evolution of amplitudes related to the 502~d period is observed.

In order to scrutinize possible mechanisms driving these quasi-periods, we searched for their signatures during the earlier post-peak epochs. We calculated the residual $BVR_{\rm C}$ light curves obtained by subtraction of the general trend during the exponential decay and the `first plateau'. No significant peaks other than those closely related to the breaks in the data acquisition (340--410~d) were found (see also \citealt{clarke2005, kopatskaya2013} for similar attempts). This suggests that both quasi-periods are not permanent features of V1057~Cyg, but have a close relationship with the mechanism that led to the brightness drop in 1995-1996.

Two mechanisms driving these periodic variations have been proposed. \citet{clarke2005} concluded that the erratic photometric variability observed in V1057~Cyg between 1996--2003 is associated with the fall back of dusty material to small radii and the subsequent passage of dust condensations across the line of sight to the inner accretion disk (Sec.~4.5.2 in their paper). However, at the time of this study the periodic behavior was unknown. Enriched with this knowledge, we conclude that this event can be uniquely associated with the 1707~d period. We base this result on respective amplitudes obtained in Sec.~\ref{sec:ampl-wav_dep}, which are decreasing with increasing effective wavelengths of consecutive filters roughly in line with the mean extinction law (Fig.~\ref{fig:ampl-evol}). Assuming Keplerian rotation of this inhomogeneous region, this dust condensation scene could be located 1.9--2.8~au from the 0.3--1~M$_{\odot}$ mass star, respectively. In these circumstances the recently observed amplitude decrease could also be explained either by dispersion of this dusty region in time, and/or lower dust production rates as the disk wind is the subject of weakening due to the decreasing accretion rate. 
On the other hand, this dimming and its subsequent evolution could also be caused by the disk warp localized at 1.9--2.8~au, as the disk is seen more edge-on (Sec.~\ref{sec:accdisc}) than previously thought. According to the unified models of innermost disk warps \citep{McGinnis2015}, the maximum warp height is 20--30\% of the disk radius at which it originates (i.e.~up to 0.5--0.8~au assuming that this model scales to these distances) and it may vary by 10–20\% during a single rotation.

According to \citet{guver2009}, $N_H$~[g cm$^{-2}$]$=0.00367 A_V$. Assuming that whatever feature localized at 2.2~au (for the 0.5~M$_{\odot}$ star) causing the extinction changes occupies a quarter orbit (based on Fig.~\ref{fig:PdmWavGroundData}b), its linear size measured along the disk plane is about 3.5~au. Using an arbitrary height of 0.8~au, the mass of this structure would be 0.0004~M$_{\oplus}$ for $\Delta A_V=1$~mag during the `second plateau', at most (see the bottom panel in Fig.~\ref{fig:accdisk}). Considering that V1057~Cyg is accreting about 33 Earth masses per year, this represents a negligible fraction of the total disk mass and an order of magnitude less than estimated for an analogous phenomenon in V582~Aur by \citet{abraham2018}.

In spite of the very limited number of spectra, we decided to search for correlations between the spectroscopic and the known photometric variability. We tentatively find correlation between the 1707~d QPO and the absorption components of the wind lines. In Fig.~\ref{fig:PCyg-phase}ab we present H$\alpha$ and H$\beta$ P~Cygni profiles in function of phase calculated for this QPO (Fig.~\ref{fig:PdmWavGroundData}b). 
The radial velocities (RV) measured consistently at a depth of $2/3$ of the virtual line bisector of the first (usually the deepest) absorption peak \citep{herbig2009} are possibly periodic (i.e. 1707~d), with mean amplitude of 
$\Delta$RV$=5.4\pm1.6$~km~s$^{-1}$ and mean velocity of $\gamma = -81.4\pm1.2$~km~s$^{-1}$ 
The reason for measuring this peak is that it represents the last interaction of the inner disk light with the dusty environment that modifies it and is probably the most important factor in determining what kind of photometric variations will ultimately be seen by the observer. With a limited number of spectra, the uncertainties are too large for meaningful considerations of the distinct lines. If the above spectroscopic-photometric connection is real, the obtained $\gamma$~value weighs against the hypothesis of occultations caused by a disk warp, as those should produce RV variations around the mean systemic velocity.

Similar analysis performed for shell lines and the broad and narrow emission peaks in forbidden lines do not reveal variations correlated with phase. However, the lower and higher velocity components of the double-peaked broad emission peaks appear with change in accordance to the 1707~d quasi-period.


With regard to the shorter period, the most plausible interpretations invoked binarity or even planet(s) forming and/or tidally disrupting in the very inner disk \citep{lodato2004, clarke2005}. This possibility has been considered more seriously once new observations revealed that the 502~d period correlates with radial velocity changes of the emission component of the lithium line \citep{kopatskaya2013}. 
The reversed color index behavior associated with the 502~d period is similar both to the `blueing effect' observed in UXors, and also to the effect caused by the rotation of a locally warmer plasma bubble, as proposed for FU~Ori's inner disk \citep{siwak2018}. If the latter scenario is valid and the disk rotation is Keplerian, the `hot spot' in V1057~Cyg's disk would have to be localised at the distance of 1~au from the 0.5~M$_{\odot}$ star \citep{gramajo2014} to be responsible for the 502~d QPO. Using the model presented in \citet{siwak2018}, but assuming blackbody instead of supergiant spectral intensities\footnote{This was done to let this particular model deal with $JHK$ data.}, $R_{\rm in}$ = 4.6~R$_{\odot}$ and i=62$^{\circ}$ (Sec.~\ref{sec:accdisc}), we were able to reproduce the amplitudes observed in $R_{\rm C}I_{\rm C}JHK$ filters at the same time by \citet{kopatskaya2013} and listed in Sec.~\ref{sec:ampl-wav_dep}. A hot spot of $T_{\rm eff}\approx 3500-4000$~K (vs. $\approx 800\pm100$~K predicted for these disk annuli assuming steady accretion) with 30$^{\circ}$ azimuthal extent and 200--240~R$_{\odot}$ (0.93--1.12~au), is necessary to qualitatively reproduce these observations (Fig.~\ref{fig:hotspot}). The obtained hot spot size and temperature is in principle consistent with that numerically obtained for luminous optically thick shocks on the circumplanetary disks around giant forming planets \citep{Szulagyi2017}. It is debatable whether such a shock could produce the associated RV variations observed in the Li line emission component by \citet{herbig2009} and \citet{kopatskaya2013}. The single piece of evidence for it can be found in our 2012 spectrum at $+5$~km~s$^{-1}$, which is very different from those reported by the authors (from $-$9 to $-$19~km~s$^{-1}$). We found no correlations between the 502~d quasi-period and shell or P~Cyg or forbidden line profiles, but this may be due to the poor temporal coverage. The weakness of the hot spot scenario is that it fails to explain why the shorter QPO became observable simultaneously with the longer one. Disk instabilities ignited during the enhanced accretion are not well-understood and could be the cause, or possibly the disk fragmentations that led to the FUor phenomenon \citep{vorobyov2018}. Hot spot structures often attributed to forming planets have been directly imaged in disks of young stars at distances from a dozen to several dozen of astronomical units \citep{Reggiani2014, Biller2014}.

The 502~d quasi-periodic phenomenon can also be caused by obscuration of certain disk annuli, i.e.~those mostly emitting in the red and near-infrared bands, by dust cloud. In order to reproduce the observed amplitudes, we first calculated magnitudes for an unshaded disk assuming steady accretion, then compared the results with the synthetic magnitudes calculated assuming total eclipsing of certain disk annuli, contained within the {\it a priori} chosen azimuthal angle of 120$^{\circ}$. This parameter has negligible effect on the final result. We approximated this phenomenon by setting $T_{\rm eff}=0$ in the shaded area. We obtained that a substantial part of the inner disk contained roughly between $20-70$~R$_{\odot}$ must be periodically obscured to produce the observed effect. As the Keplerian period at 45~R$_{\odot}$ is about 50 days, a cold spot in a disk photosphere or a dusty cloud corotating with the disk on circular orbit can immediately be excluded. Occultation of the innermost disk annuli as in the 1707~d scenario proposed above does not apply because it would produce a strong signal at $UBV$ wavelengths (in fact 
only barely detected in $BV$-bands). The remaining possibilities are a dusty cloud -- a remnant of the envelope -- rotating on inclined and eccentric orbit, and a `dust streamer' structure caused by interaction of forming planets and elevating a  substantial amount of dusty material high above the disk midplane \citep{Loomis2017}.
However, the first possibility seems to be unlikely, as this cloud would likely be absorbed after the first attempt to break through the protoplanetary disk midplane. A dust streamer could act as an occulting screen of certain disk annuli only, although it is not clear the mechanism leading to periodic and continuous brightness variations. 
Thus, based on the data available so far, we conclude that obscuration scenarios considered above fail to explain the 502~d period and its spectral properties in a manner as self-consistent as a locally heated disk at 1~au.

\subsection{About the intra-day and weekly variability observed by {\it TESS}}
\label{sec:TESSperioddiscussion}

While the nature of the quasi-periodic light variations observed from the ground since 1997 seem to be now better understood, the nature of these observed from space is still not clear to us. We obtained different Fourier spectrum slopes for the ground-based ($a_f\sim f^{-1/2}$) and {\it TESS} data ($a_f \sim f^{-1}$). These different relationships could be initially interpreted by combination of extinction, accretion and light-scattering processes. We arrive at this conclusion due to the lack of time-coherent QPOs in the {\it TESS} light curve of V1057~Cyg, similar to those observed in the disk-only FU~Ori and V646~Pup \citep{siwak2018, siwak2020}. This lack could be due to strong reprocessing of time-coherent inner disk light variations on their way to the observer. On the other hand, analysis of the accretion-dominated public-domain light curves of FU~Ori itself and highly-accreting CTTSs obtained during several seasons by several spacecraft revealed that the steeper, random-walk Brownian nature observed by {\it TESS} during its 56~day run, can also be an effect of a particular realization of stochastic accretion-related processes in the inner disk. Longer or more frequent observing runs with photometric precision provided by space telescopes are required to clarify these assumptions.

\section{Conclusions} \label{sec:conclusions}

In this paper, we reported on a multi-epoch, multi-wavelength study of V1057~Cyg, a classical FU~Orionis-type object. We arrived at the following conclusions: 
\begin{itemize}
    \item The Gaia DR2 distance of V1057~Cyg of 897\,pc is significantly larger than previous estimates in the literature, making this object more luminous than previously thought. We constructed multi-epoch SEDs, which we modeled with a simple geometrically thin, optically thick accretion disk model, with $A_V$ and $M\dot{M}$ as the two free parameters. Our results show that the accretion rate reached $1\times10^{-3} M_{\odot}M_{\odot}$\,yr$^{-1}$ at the peak of the outburst in 1971 and is still about $1\times10^{-4} M_{\odot}M_{\odot}$\,yr$^{-1}$. This makes it the most heavily accreting FUor ever discovered.

    \item Our long-term photometric monitoring shows the continuation of the second photometric plateau, showing a year to year variability in the optical bands limited to only a few tenths of a magnitude. A  period analysis reveals a longer 1707$\pm$70~d and a shorter 502$\pm$20~d period. Our study detected the shorter period in the $BV$-filters for the first time. The amplitudes related to the longer period decrease with wavelength, while the amplitudes related to the shorter period increase with wavelength. Our data revealed that the amplitudes related to the 1707~d period have decreased during the last two decades. No clear evidence for evolution of the amplitudes related to the 502~d period was observed. 
   
   \item Based on optical color-magnitude diagrams, we conclude that during the `second plateau' the color index variations generally follow the extinction path. Due to mutual overlap of the two periodic components having different spectral characteristics, color index variations observed at certain times (especially in 2019) show different relationship, which cannot be explained either by changing extinction along the line of sight, or variable accretion, or combination of both.
 
    \item The origin of the 1707~d periodicity might be related to an orbiting dust structure periodically eclipsing the central part of the disk. This may likely arise from a fall back of dusty material from the envelope to small disk radii. For the 502~d period, the most plausible interpretation is a $0.3\times0.2$\,au `bubble' heated to 3500--4000\,K, located at 1~au from the star. 
    
    \item We conducted optical spectroscopic monitoring of V1057~Cyg between 2012--2020. We detected high velocity wind features in the form of the absorption components of P-Cygni profile lines and pure absorption lines that are blueshifted by up to 100--300\,km\,s$^{-1}$. All the detected features in 2020 show lower blueshifted velocity profiles both in the P~Cygni profiles and pure absorption lines than in spectra observed at previous epochs. The emission component of the P~Cygni profiles and the forbidden emission lines show the weakest line strength in 2020.
    The variation of velocity and strength of absorption profiles vary strongly over time, indicating that the strength of the wind also varies with time.
    The wind features and the absorption component of the lines with P~Cygni profiles show the same velocity variation trend in time. 
    Shell features, blueshifted by about 100\,km\,s$^{-1}$, are also detected and exhibit strong variation in velocity and strength with time.
    All the detected absorption components of the P~Cygni profile, wind features, and shell features vary with time but not with a well-defined long-term trend. The variation trend of the lower velocity component (between $-$50 to $-$150 km\,s$^{-1}$) of those absorption profiles shows the same tendency.
    In addition, the lower velocity component of the [O~\,{\footnotesize I}] 630\,nm shows the same variation trend. Therefore, it is suspected that the absorption components of the P~Cygni profile, wind features, and the lower velocity component of the [O~\,{\footnotesize I}] 630.0\,nm line are formed by the same phenomenon, most likely in the dusty envelope leading to the 1707~d QPO. 

    \item In several epochs, the relatively narrow [O~\,{\footnotesize III}] 495.9, 500.7\,nm, [N~\,{\footnotesize II}] 654.8, 658.3\,nm, and [S~\,{\footnotesize II}] 671.6, 673.1\,nm lines are detected for the first time in the spectra of V1057~Cyg. In classical T Tauri stars, these lines are jets/outflows tracers, but they are not common in FUors. This is the first detection ever of these lines in a classical FUor. These lines are also variable in time, suggesting that the jet/outflow activity is not constant in V1057~Cyg.
    \vspace{0mm}
    \item We obtained a new NIR spectrum of V1057~Cyg in 2020 and compared it with previous spectra from the literature. In this wavelength range we also detected various absorption and emission features. The strength of the CO first overtone bandhead absorption has been decreasing since its first observation, weakest in 2020. This can be interpreted as a sign of decreasing mass accretion rate.
    \item Our spectroscopic analysis shows that the properties of V1057~Cyg still mostly resemble those of a classical FUor. The photometric monitoring also indicates that it has not yet returned to quiescence, therefore, the FUor outburst of V1057~Cyg is still ongoing. 

\end{itemize}


There remain several open questions related to the origin of variability in the photometric and spectroscopic data sets. Continuing long-term photometric observations as well as snapshot spectroscopic observations at least once a year will be essential to better understand the change in the accretion and outflow activity of V1057~Cyg. This study highlights the importance of expanding our knowledge on classical FUors, since studying them is crucial to identify new members of this class.
    
\section*{acknowledgements}
We kindly thank Michael Connelley for handing us over the 2015 IRTF spectra of V1057~Cyg in order to carry out a more accurate analysis.
ZsMSz is supported by the ÚNKP-20-2 New National Excellence Program of the Ministry for Innovation and Technology from the source of the National Research, Development and Innovation Fund.
This project has received funding from the European Research Council (ERC) under the European Union's Horizon 2020 research and innovation programme under grant agreement No 716155 (SACCRED). 
The work was also supported by the Hungarian NKFIH grant K-132406, K-131508, KH-130526 and by the NKFIH grant 2019-2.1.11-TÉT-2019-00056. Authors acknowledge the financial support of the Austrian-Hungarian Action Foundation (101\"ou13, 104\"ou2). LK acknowledges the financial support of the Hungarian National Research, Development and Innovation Office grant NKFIH PD-134784.
Our study is supported by the project ''Transient Astrophysical Objects'' GINOP 2.3.2-15-2016-00033 of the National Research, Development and Innovation Office (NKFIH), Hungary, funded by the European Union. 
This project has been supported by the GINOP-2.3.2-15-2016-00003 grant of the Hungarian National Research, Development and Innovation Office (NKFIH).  
This project has been supported by the Lendület grant LP2012-31 of the Hungarian Academy of Sciences.
J-E. Lee was supported by the Basic Science Research Program through the National Research Foundation of Korea (grant no. NRF-2018R1A2B6003423).
This paper includes data collected by the TESS mission. Funding for the TESS mission is provided by the NASA Explorer Program. This project has been supported by the GINOP-2.3.2-15-2016-00003 grant of the Hungarian National Research, Development and Innovation Office (NKFIH). 
L. Kriskovics is supported by the Bolyai János Research Scholarship of the Hungarian Academy of Sciences.
Based on observations made with the Nordic Optical Telescope, operated by the Nordic Optical Telescope Scientific Association at the Observatorio del Roque de los Muchachos, La Palma, Spain, of the Instituto de Astrofisica de Canarias.
The data presented here were obtained (in part) with ALFOSC, which is provided by the Instituto de Astrofisica de Andalucia (IAA) under a joint agreement with the University of Copenhagen and NOTSA. 
Based [in part] on observations made with the NASA/DLR Stratospheric Observatory for Infrared Astronomy (SOFIA). SOFIA is jointly operated by the Universities Space Research Association, Inc. (USRA), under NASA contract NNA17BF53C, and the Deutsches SOFIA Institut (DSI) under DLR contract 50 OK 0901 to the University of Stuttgart.

\facilities{NOT, BOAO, SOFIA, TESS}
\software{FITSH package \citep{pal2012}, Vartools \citep{hartman2016}, molecfit \citep{smette2015,kausch2015}, IRAF \citep{tody1986,tody1993}}
\appendix
\section{Photometry of V1057~Cyg}
Tab.~\ref{tab:phot} contains our original optical and near-infrared photometry of V1057~Cyg before the shifts discussed in Sec.~\ref{sec:obs}, while Tab.~\ref{tbl_phot_wise} contains the saturation corrected \textit{WISE} data that we use in Fig.~\ref{fig:lc}.
\startlongtable
\begin{deluxetable*}{ccccccccccccccc}
\tabletypesize{\footnotesize}
\tablecaption{Optical and near-IR photometry of V1057~Cyg.\label{tab:phot}}
\tablehead{%
\colhead{Date} & %
\colhead{MJD} &%
\colhead{$B$} &%
\colhead{$V$} &%
\colhead{$g'$} &%
\colhead{$R_{\rm C}$} &%
\colhead{$I_{\rm C}$} &%
\colhead{$R_{\rm J}$} &%
\colhead{$I_{\rm J}$} &%
\colhead{$r'$} &%
\colhead{$i'$} &%
\colhead{$J$} & \colhead{$H$} & \colhead{$K_{\rm s}$} & \colhead{Instrument}} 
\startdata 
2005-10-29 & 53672.75 & $14.43$ & $12.53$ &\dots & $11.52$ & $10.54$ &\dots&\dots&\dots&\dots&\dots&\dots&\dots& RCC\\
2005-10-31 & 53674.75 & $14.43$ & $12.51$ &\dots & $11.51$ & $10.51$ &\dots&\dots&\dots&\dots&\dots&\dots&\dots& RCC\\    
2005-11-17 & 53691.74 & $14.51$ & $12.66$ &\dots & $11.61$ & $10.59$ &\dots&\dots&\dots&\dots&\dots&\dots&\dots& RCC \\    
2005-11-18 & 53692.82 & $14.26$ & $12.61$ &\dots & $11.60$ & $10.50$ &\dots&\dots&\dots&\dots&\dots&\dots&\dots& RCC \\
2006-07-15 & 53932.12 & \dots   & \dots   &\dots & \dots   & \dots   &\dots&\dots&\dots&\dots&$8.08$& $7.27$ & $6.25$ & TCS \\
2006-07-20 & 53936.92 & $14.53$ & $12.51$ & \dots  & \dots & \dots   &$11.39$ & $10.42$ &\dots&\dots&\dots&\dots&\dots& IAC80 \\
2007-06-02 & 54254.22 &\dots    &\dots & \dots   &\dots    &\dots    &\dots  & \dots&\dots&\dots& $8.52$ & $7.19$ & $7.29$ & TCS \\
2007-07-23 & 54304.92 & $14.56$ & $12.71$ &\dots & $11.62$ & $10.49$ &\dots&\dots&\dots&\dots&\dots&\dots&\dots& RCC \\
\enddata
\label{tab:photometry}
\end{deluxetable*}

\startlongtable
\begin{deluxetable*}{ccc}
\tabletypesize{\footnotesize} 
\tablecaption{Saturation corrected \textit{WISE} data \label{tbl_phot_wise}}
\tablewidth{0pt}
\tablehead{\colhead{MJD} & \colhead{W1 mag} & \colhead{W2 mag}}
\startdata
55346.5    & 4.993$\pm$0.006 & 4.360$\pm$0.017   \\
55527.1    & 5.851$\pm$0.020 & 4.937$\pm$0.070   \\
56811.0    & 5.280$\pm$0.072 & 4.492$\pm$0.074   \\
56990.3    & 5.304$\pm$0.049 & 4.549$\pm$0.066   \\
57172.9    & 5.493$\pm$0.058 & 4.542$\pm$0.092    \\
57351.0    & 5.231$\pm$0.105 & 4.399$\pm$0.075   \\
57538.4    & 5.344$\pm$0.054 & 4.704$\pm$0.056    \\
57710.0    & 5.492$\pm$0.044 & 4.715$\pm$0.094    \\
57904.0    & 5.202$\pm$0.075 & 4.548$\pm$0.078   \\
58070.8    & 5.418$\pm$0.061 & 4.609$\pm$0.087   \\
58268.2    & 5.303$\pm$0.086 & 4.559$\pm$0.137  \\
58431.3    & 5.382$\pm$0.043 & 4.810$\pm$0.102  \\
58634.9    & 5.214$\pm$0.125 & 4.653$\pm$0.169  \\
58798.5    & 5.289$\pm$0.085 & 4.633$\pm$0.102  \\
58999.8    & 5.262$\pm$0.102 & 4.625$\pm$0.104  \\
59162.7    & 5.428$\pm$0.080 & 4.722$\pm$0.114  \\
\enddata
\vspace{1cm}
\end{deluxetable*}

\section{Spectral lines in V1057~Cyg}
For line identification, the NIST Atomic Spectra Database\footnote{\url{https://www.nist.gov/pml/atomic-spectra-database}} was used. The line information of transition probabilities and energy levels are collected from the NIST database. Tab.~\ref{tab:lines} and~\ref{tab:lines_IR} list the detected spectral lines in the optical and NIR spectrum of V1057~Cyg, respectively. The observed wavelength of absorption and emission profiles in Tab.~\ref{tab:lines} and~\ref{tab:lines_IR} were found as the wavelength where the intensity is the strongest. For line identification, the latest observation of optical spectrum (FIES, 2020 August) is used, and NOTCam spectrum is used for the NIR line identification. The laboratory and observed wavelengths are given in air.


\startlongtable
\begin{deluxetable*}{cccccccc}
\tablecaption{Atomic lines in the optical spectrum of V1057~Cyg in 2020.\label{tab:lines}}
\tablehead{
\colhead{Species} & \colhead{Lab. $\lambda$} & \colhead{Obs. $\lambda$} & 
\colhead{Lower level} & \colhead{Upper level} & \colhead{$A_{ki}$} & \colhead{$E_i - E_k$} & \colhead{Notes}\\
\colhead{} & \colhead{(nm)} & \colhead{(nm)}  & \colhead{Conf., term, J} & 
\colhead{Conf., term, J} & \colhead{(s$^{-1}$)}  & \colhead{(eV)} & }
\startdata
Fe I &  414.387 & 414.387 &    3d$^{7}$($^{4}$F)4s a $^{3}$F 3 &               3d$^{7}$($^{4}$F)4p y $^{3}$F$^{\circ}$ 4 &     1.33$\cdot10^{7}$ &     1.557 -- 4.548 &  A \\
Fe I &  420.203 & 420.047 &    3d$^{7}$($^{4}$F)4s a $^{3}$F 4 &               3d$^{7}$($^{4}$F)4p z $^{3}$G$^{\circ}$ 4 &     8.22$\cdot10^{6}$ &     1.484 -- 4.434 &  A \\
Fe I &  421.597 & 421.417 &    3d$^{7}$($^{2}$G)4s a $^{3}$G 5 &               3d$^{6}$($^{3}$H)4s4p($^{3}$P$^{\circ}$) y $^{3}$G$^{\circ}$ 4 &         &      2.692 -- 5.632 &  A \\
V I &   422.662 & 422.531 &    3d$^{4}$($^{3}$F2)4s b $^{4}$F $\sfrac{5}{2}$ &         3d$^{4}$($^{3}$F2)4p u $^{4}$F$^{\circ}$ $\sfrac{7}{2}$ &       1.64$\cdot10^{7}$ &     1.945 -- 4.877 &  A \\
Fe II & 423.316 & 423.155 &    3d$^{6}$($^{3}$P2)4s b $^{4}$P $\sfrac{5}{2}$ &         3d$^{6}$($^{5}$D)4p z $^{4}$D$^{\circ}$ $\sfrac{7}{2}$ &        7.2$\cdot10^{5}$ &      2.582 -- 5.510 &  A \\
V I &   424.671 & 424.549 &    3d$^{3}$4s$^{2}$ a $^{2}$P $\sfrac{3}{2}$ & 3d$^{3}$($^{2}$D2)4s4p($^{3}$P$^{\circ}$) w $^{2}$F$^{\circ}$ $\sfrac{5}{2}$ &   &   1.711 -- 4.629 &  A \\
V I &   425.443 & 425.285 &    3d$^{3}$($^{4}$F)4s4p($^{3}$P$^{\circ}$) z $^{4}$D$^{\circ}$ $\sfrac{5}{2}$ &   3d$^{3}$4s($^{5}$F)4d $^{4}$F $\sfrac{7}{2}$ &   &  2.582 -- 5.495 &  A \\
Fe I &  427.176 & 427.064 &    3d$^{7}$($^{4}$F)4s a $^{3}$F 4 &               3d$^{7}$($^{4}$F)4p z $^{3}$G$^{\circ}$ 5 &     2.28$\cdot10^{7}$ &     1.484 -- 4.386 &  A \\
\enddata
\tablenotetext{}{A: absorption profile; E: emission profile; PA: absorption part of a P~Cyg profile; PE: emission part of a P~Cyg profile; SA: absorption shell feature; SE: emission shell feature}
\end{deluxetable*}

\startlongtable
\begin{deluxetable*}{cccccccc}
\tablecaption{Atomic lines in the infrared spectrum of V1057~Cyg in 2020.\label{tab:lines_IR}}
\tablehead{
\colhead{Species} & \colhead{Lab. $\lambda$} & \colhead{Obs. $\lambda$} & 
\colhead{Lower level} & \colhead{Upper level} & \colhead{$A_{ki}$} & \colhead{$E_i - E_k$} & \colhead{Notes}\\
\colhead{} & \colhead{($\mu$m)} & \colhead{($\mu$m)}  & \colhead{Conf., term, J} & 
\colhead{Conf., term, J} & \colhead{(s$^{-1}$)}  & \colhead{(eV)} & 
\colhead{}}
\startdata
[Fe II] &  1.256 & 1.256 & 3d$^{6}$($^{5}$D)4s a $^{6}$D  $\sfrac{9}{2}$ & 3d$^{6}$($^{5}$D)4s a $^{4}$D  $\sfrac{7}{2}$ & 4.74$\cdot10^{-3}$ & 0.000 -- 0.986  &  E \\
Pa$\beta$ &  1.281 & 1.281 & 3 & 5 &    2.20$\cdot10^{6}$ & 12.087	-- 	13.0545017   &  A \\
Al I &  1.312 & 1.311 &    3s$^{2}$4s $^{2}$S $\sfrac{1}{2}$ & 3s$^{2}$4p $^{2}$P$^{\circ}$ $\sfrac{3}{2}$  & 1.60$\cdot10^{7}$  & 3.142  -- 4.087  &  A \\
Al I &  1.315 & 1.318 &  3s$^{2}$4s $^{2}$S $\sfrac{1}{2}$ & 3s$^{2}$4p $^{2}$P$^{\circ}$ $\sfrac{1}{2}$  &  1.59$\cdot10^{7}$ &    	3.142  -- 4.085 &  A \\
Br 19-4 &  1.526 & 1.525 & 4 & 19 &     1.53$\cdot10^{3}$ & 12.748 -- {}[13.560] &  A \\
{}[Fe~II] &  1.533 & 1.533 &    3d$^{7}$ a $^{4}$F $\sfrac{9}{2}$ & 3d$^{6}$($^{5}$D)4s a $^{4}$D $\sfrac{5}{2}$ &   3.12$\cdot10^{-3}$ &    0.232  -- 1.040 &  E \\
Br 15-4 &  1.556 & 1.525 & 4 & 15 &     5.11$\cdot10^{35}$ & 12.7485393  -- {}[13.537] &  A \\
Mg I &  1.588 & 1.533 &    3s3d $^{3}$D 2 &  3s5p $^{3}$P$^{\circ}$ 1 &     2.65$\cdot10^{5}$ & 5.945 -- 6.726  &  A \\
\enddata
\tablenotetext{}{A: absorption profile; E: emission profile}
\end{deluxetable*}


\bibliography{paper}{}

\begin{thebibliography}{}
\expandafter\ifx\csname natexlab\endcsname\relax\def\natexlab#1{#1}\fi
\providecommand{\url}[1]{\href{#1}{#1}}
\providecommand{\dodoi}[1]{doi:~\href{http://doi.org/#1}{\nolinkurl{#1}}}
\providecommand{\doeprint}[1]{\href{http://ascl.net/#1}{\nolinkurl{http://ascl.net/#1}}}
\providecommand{\doarXiv}[1]{\href{https://arxiv.org/abs/#1}{\nolinkurl{https://arxiv.org/abs/#1}}}

\bibitem[{{{\'A}brah{\'a}m} {et~al.}(2004){{\'A}brah{\'a}m}, {K{\'o}sp{\'a}l},
  {Csizmadia}, {Kun}, {Mo{\'o}r}, \& {Prusti}}]{abraham2004b}
{{\'A}brah{\'a}m}, P., {K{\'o}sp{\'a}l}, {\'A}., {Csizmadia}, S., {et~al.}
  2004, \aap, 428, 89, \dodoi{10.1051/0004-6361:20040315}

\bibitem[{{{\'A}brah{\'a}m} {et~al.}(2018){{\'A}brah{\'a}m}, {K{\'o}sp{\'a}l},
  {Kun}, {Feh{\'e}r}, {Zsidi}, {Acosta-Pulido}, {Carnerero},
  {Garc{\'\i}a-{\'A}lvarez}, {Mo{\'o}r}, {Cseh}, {Hajdu}, {Hanyecz}, {Kelemen},
  {Kriskovics}, {Marton}, {Mez{\H{o}}}, {Moln{\'a}r}, {Ordasi},
  {Rodr{\'\i}guez-Coira}, {S{\'a}rneczky}, {S{\'o}dor}, {Szak{\'a}ts},
  {Szegedi-Elek}, {Szing}, {Farkas-Tak{\'a}cs}, {Vida}, \&
  {Vink{\'o}}}]{abraham2018}
{{\'A}brah{\'a}m}, P., {K{\'o}sp{\'a}l}, {\'A}., {Kun}, M., {et~al.} 2018,
  \apj, 853, 28, \dodoi{10.3847/1538-4357/aaa242}

\bibitem[{{Audard} {et~al.}(2014){Audard}, {{\'A}brah{\'a}m}, {Dunham},
  {Green}, {Grosso}, {Hamaguchi}, {Kastner}, {K{\'o}sp{\'a}l}, {Lodato},
  {Romanova}, {Skinner}, {Vorobyov}, \& {Zhu}}]{audard2014}
{Audard}, M., {{\'A}brah{\'a}m}, P., {Dunham}, M.~M., {et~al.} 2014, in
  Protostars and Planets VI, ed. H.~{Beuther}, R.~S. {Klessen}, C.~P.
  {Dullemond}, \& T.~{Henning}, 387,
  \dodoi{10.2458/azu_uapress_9780816531240-ch017}

\bibitem[{{Bailer-Jones} {et~al.}(2018{\natexlab{a}}){Bailer-Jones}, {Rybizki},
  {Fouesneau}, {Mantelet}, \& {Andrae}}]{bailerjones1}
{Bailer-Jones}, C.~A.~L., {Rybizki}, J., {Fouesneau}, M., {Mantelet}, G., \&
  {Andrae}, R. 2018{\natexlab{a}}, \aj, 156, 58,
  \dodoi{10.3847/1538-3881/aacb21}

\bibitem[{{Bailer-Jones} {et~al.}(2018{\natexlab{b}}){Bailer-Jones}, {Rybizki},
  {Fouesneau}, {Mantelet}, \& {Andrae}}]{bailerjones2}
---. 2018{\natexlab{b}}, VizieR Online Data Catalog, I/347

\bibitem[{{Bastian} \& {Mundt}(1985)}]{bastian1985}
{Bastian}, U., \& {Mundt}, R. 1985, \aap, 144, 57

\bibitem[{{Bell} {et~al.}(1995){Bell}, {Lin}, {Hartmann}, \&
  {Kenyon}}]{bell1995}
{Bell}, K.~R., {Lin}, D.~N.~C., {Hartmann}, L.~W., \& {Kenyon}, S.~J. 1995,
  \apj, 444, 376, \dodoi{10.1086/175612}

\bibitem[{{Biller} {et~al.}(2014){Biller}, {Males}, {Rodigas}, {Morzinski},
  {Close}, {Juhász}, {Follette}, Sylvestre, {Benisty}, {Sicilia-Aguilar}, \&
  et~al.}]{Biller2014}
{Biller}, B.~A., {Males}, J., {Rodigas}, T., {et~al.} 2014, \apjl, 792, 6,
  \dodoi{10.3847/1538-4357/aa6c63}

\bibitem[{{Biscaya} {et~al.}(1997){Biscaya}, {Rieke}, {Narayanan}, {Luhman}, \&
  {Young}}]{biscaya1997}
{Biscaya}, A.~M., {Rieke}, G.~H., {Narayanan}, G., {Luhman}, K.~L., \& {Young},
  E.~T. 1997, \apj, 491, 359, \dodoi{10.1086/304935}

\bibitem[{{Bonnell} \& {Bastien}(1992)}]{bonnell1992}
{Bonnell}, I., \& {Bastien}, P. 1992, \apjl, 401, L31, \dodoi{10.1086/186663}

\bibitem[{{Cabrit} {et~al.}(1990){Cabrit}, {Edwards}, {Strom}, \&
  {Strom}}]{cabrit1990}
{Cabrit}, S., {Edwards}, S., {Strom}, S.~E., \& {Strom}, K.~M. 1990, \apj, 354,
  687, \dodoi{10.1086/168725}

\bibitem[{{Cardelli} {et~al.}(1989){Cardelli}, {Clayton}, \&
  {Mathis}}]{cardelli1989}
{Cardelli}, J.~A., {Clayton}, G.~C., \& {Mathis}, J.~S. 1989, \apj, 345, 245,
  \dodoi{10.1086/167900}

\bibitem[{{Carr}(1989)}]{carr1989}
{Carr}, J.~S. 1989, \apj, 345, 522, \dodoi{10.1086/167927}

\bibitem[{{Clarke} {et~al.}(2005){Clarke}, {Lodato}, {Melnikov}, \&
  {Ibrahimov}}]{clarke2005}
{Clarke}, C.~J., {Lodato}, G., {Melnikov}, S.~Y., \& {Ibrahimov}, M.~A. 2005,
  \mnras, 361, 942, \dodoi{10.1111/j.1365-2966.2005.09231.x}

\bibitem[{{Connelley} \& {Reipurth}(2018)}]{connelley2018}
{Connelley}, M.~S., \& {Reipurth}, B. 2018, \apj, 861, 145,
  \dodoi{10.3847/1538-4357/aaba7b}

\bibitem[{{Cutri} \& {et al.}(2012)}]{cutri2012}
{Cutri}, R.~M., \& {et al.} 2012, VizieR Online Data Catalog, II/311

\bibitem[{{Cutri} \& {et al.}(2014)}]{cutri2014}
---. 2014, VizieR Online Data Catalog, II/328

\bibitem[{{Cutri} {et~al.}(2003){Cutri}, {Skrutskie}, {van Dyk}, {Beichman},
  {Carpenter}, {Chester}, {Cambresy}, {Evans}, {Fowler}, {Gizis}, {Howard},
  {Huchra}, {Jarrett}, {Kopan}, {Kirkpatrick}, {Light}, {Marsh}, {McCallon},
  {Schneider}, {Stiening}, {Sykes}, {Weinberg}, {Wheaton}, {Wheelock}, \&
  {Zacarias}}]{cutri2003}
{Cutri}, R.~M., {Skrutskie}, M.~F., {van Dyk}, S., {et~al.} 2003, {2MASS All
  Sky Catalog of point sources.} (NASA/IPAC Infrared Science Archive)

\bibitem[{{Feh{\'e}r} {et~al.}(2017){Feh{\'e}r}, {K{\'o}sp{\'a}l},
  {{\'A}brah{\'a}m}, {Hogerheijde}, \& {Brinch}}]{feher2017}
{Feh{\'e}r}, O., {K{\'o}sp{\'a}l}, {\'A}., {{\'A}brah{\'a}m}, P.,
  {Hogerheijde}, M.~R., \& {Brinch}, C. 2017, \aap, 607, A39,
  \dodoi{10.1051/0004-6361/201731446}

\bibitem[{{Fischer} {et~al.}(2012){Fischer}, {Megeath}, {Tobin}, {Stutz},
  {Ali}, {Remming}, {Kounkel}, {Stanke}, {Osorio}, {Henning}, {Manoj}, \&
  {Wilson}}]{fischer2012}
{Fischer}, W.~J., {Megeath}, S.~T., {Tobin}, J.~J., {et~al.} 2012, \apj, 756,
  99, \dodoi{10.1088/0004-637X/756/1/99}

\bibitem[{{Foster}(1996)}]{foster1996}
{Foster}, G. 1996, \aj, 112, 1709, \dodoi{10.1086/118137}

\bibitem[{{Gaia Collaboration} {et~al.}(2018){Gaia Collaboration}, {Brown},
  {Vallenari}, {Prusti}, {de Bruijne}, {Babusiaux}, {Bailer-Jones}, {Biermann},
  {Evans}, {Eyer}, {Jansen}, {Jordi}, {Klioner}, {Lammers}, {Lindegren},
  {Luri}, {Mignard}, {Panem}, {Pourbaix}, {Randich}, {Sartoretti}, {Siddiqui},
  {Soubiran}, {van Leeuwen}, {Walton}, {Arenou}, {Bastian}, {Cropper},
  {Drimmel}, {Katz}, {Lattanzi}, {Bakker}, {Cacciari}, {Casta{\~n}eda},
  {Chaoul}, {Cheek}, {De Angeli}, {Fabricius}, {Guerra}, {Holl}, {Masana},
  {Messineo}, {Mowlavi}, {Nienartowicz}, {Panuzzo}, {Portell}, {Riello},
  {Seabroke}, {Tanga}, {Th{\'e}venin}, {Gracia-Abril}, {Comoretto},
  {Garcia-Reinaldos}, {Teyssier}, {Altmann}, {Andrae}, {Audard},
  {Bellas-Velidis}, {Benson}, {Berthier}, {Blomme}, {Burgess}, {Busso},
  {Carry}, {Cellino}, {Clementini}, {Clotet}, {Creevey}, {Davidson}, {De
  Ridder}, {Delchambre}, {Dell'Oro}, {Ducourant},
  {Fern{\'a}ndez-Hern{\'a}ndez}, {Fouesneau}, {Fr{\'e}mat}, {Galluccio},
  {Garc{\'\i}a-Torres}, {Gonz{\'a}lez-N{\'u}{\~n}ez}, {Gonz{\'a}lez-Vidal},
  {Gosset}, {Guy}, {Halbwachs}, {Hambly}, {Harrison}, {Hern{\'a}ndez},
  {Hestroffer}, {Hodgkin}, {Hutton}, {Jasniewicz}, {Jean-Antoine-Piccolo},
  {Jordan}, {Korn}, {Krone-Martins}, {Lanzafame}, {Lebzelter}, {L{\"o}ffler},
  {Manteiga}, {Marrese}, {Mart{\'\i}n-Fleitas}, {Moitinho}, {Mora}, {Muinonen},
  {Osinde}, {Pancino}, {Pauwels}, {Petit}, {Recio-Blanco}, {Richards},
  {Rimoldini}, {Robin}, {Sarro}, {Siopis}, {Smith}, {Sozzetti}, {S{\"u}veges},
  {Torra}, {van Reeven}, {Abbas}, {Abreu Aramburu}, {Accart}, {Aerts},
  {Altavilla}, {{\'A}lvarez}, {Alvarez}, {Alves}, {Anderson}, {Andrei},
  {Anglada Varela}, {Antiche}, {Antoja}, {Arcay}, {Astraatmadja}, {Bach},
  {Baker}, {Balaguer-N{\'u}{\~n}ez}, {Balm}, {Barache}, {Barata}, {Barbato},
  {Barblan}, {Barklem}, {Barrado}, {Barros}, {Barstow}, {Bartholom{\'e}
  Mu{\~n}oz}, {Bassilana}, {Becciani}, {Bellazzini}, {Berihuete}, {Bertone},
  {Bianchi}, {Bienaym{\'e}}, {Blanco-Cuaresma}, {Boch}, {Boeche}, {Bombrun},
  {Borrachero}, {Bossini}, {Bouquillon}, {Bourda}, {Bragaglia}, {Bramante},
  {Breddels}, {Bressan}, {Brouillet}, {Br{\"u}semeister}, {Brugaletta},
  {Bucciarelli}, {Burlacu}, {Busonero}, {Butkevich}, {Buzzi}, {Caffau},
  {Cancelliere}, {Cannizzaro}, {Cantat-Gaudin}, {Carballo}, {Carlucci},
  {Carrasco}, {Casamiquela}, {Castellani}, {Castro-Ginard}, {Charlot},
  {Chemin}, {Chiavassa}, {Cocozza}, {Costigan}, {Cowell}, {Crifo}, {Crosta},
  {Crowley}, {Cuypers}, {Dafonte}, {Damerdji}, {Dapergolas}, {David}, {David},
  {de Laverny}, {De Luise}, {De March}, {de Martino}, {de Souza}, {de Torres},
  {Debosscher}, {del Pozo}, {Delbo}, {Delgado}, {Delgado}, {Di Matteo},
  {Diakite}, {Diener}, {Distefano}, {Dolding}, {Drazinos}, {Dur{\'a}n},
  {Edvardsson}, {Enke}, {Eriksson}, {Esquej}, {Eynard Bontemps}, {Fabre},
  {Fabrizio}, {Faigler}, {Falc{\~a}o}, {Farr{\`a}s Casas}, {Federici},
  {Fedorets}, {Fernique}, {Figueras}, {Filippi}, {Findeisen}, {Fonti},
  {Fraile}, {Fraser}, {Fr{\'e}zouls}, {Gai}, {Galleti}, {Garabato},
  {Garc{\'\i}a-Sedano}, {Garofalo}, {Garralda}, {Gavel}, {Gavras}, {Gerssen},
  {Geyer}, {Giacobbe}, {Gilmore}, {Girona}, {Giuffrida}, {Glass}, {Gomes},
  {Granvik}, {Gueguen}, {Guerrier}, {Guiraud}, {Guti{\'e}rrez-S{\'a}nchez},
  {Haigron}, {Hatzidimitriou}, {Hauser}, {Haywood}, {Heiter}, {Helmi}, {Heu},
  {Hilger}, {Hobbs}, {Hofmann}, {Holland}, {Huckle}, {Hypki}, {Icardi},
  {Jan{\ss}en}, {Jevardat de Fombelle}, {Jonker}, {Juh{\'a}sz}, {Julbe},
  {Karampelas}, {Kewley}, {Klar}, {Kochoska}, {Kohley}, {Kolenberg},
  {Kontizas}, {Kontizas}, {Koposov}, {Kordopatis}, {Kostrzewa-Rutkowska},
  {Koubsky}, {Lambert}, {Lanza}, {Lasne}, {Lavigne}, {Le Fustec}, {Le
  Poncin-Lafitte}, {Lebreton}, {Leccia}, {Leclerc}, {Lecoeur-Taibi},
  {Lenhardt}, {Leroux}, {Liao}, {Licata}, {Lindstr{\o}m}, {Lister}, {Livanou},
  {Lobel}, {L{\'o}pez}, {Managau}, {Mann}, {Mantelet}, {Marchal}, {Marchant},
  {Marconi}, {Marinoni}, {Marschalk{\'o}}, {Marshall}, {Martino}, {Marton},
  {Mary}, {Massari}, {Matijevi{\v{c}}}, {Mazeh}, {McMillan}, {Messina},
  {Michalik}, {Millar}, {Molina}, {Molinaro}, {Moln{\'a}r}, {Montegriffo},
  {Mor}, {Morbidelli}, {Morel}, {Morris}, {Mulone}, {Muraveva}, {Musella},
  {Nelemans}, {Nicastro}, {Noval}, {O'Mullane}, {Ord{\'e}novic},
  {Ord{\'o}{\~n}ez-Blanco}, {Osborne}, {Pagani}, {Pagano}, {Pailler},
  {Palacin}, {Palaversa}, {Panahi}, {Pawlak}, {Piersimoni}, {Pineau}, {Plachy},
  {Plum}, {Poggio}, {Poujoulet}, {Pr{\v{s}}a}, {Pulone}, {Racero}, {Ragaini},
  {Rambaux}, {Ramos-Lerate}, {Regibo}, {Reyl{\'e}}, {Riclet}, {Ripepi}, {Riva},
  {Rivard}, {Rixon}, {Roegiers}, {Roelens}, {Romero-G{\'o}mez}, {Rowell},
  {Royer}, {Ruiz-Dern}, {Sadowski}, {Sagrist{\`a} Sell{\'e}s}, {Sahlmann},
  {Salgado}, {Salguero}, {Sanna}, {Santana-Ros}, {Sarasso}, {Savietto},
  {Schultheis}, {Sciacca}, {Segol}, {Segovia}, {S{\'e}gransan}, {Shih},
  {Siltala}, {Silva}, {Smart}, {Smith}, {Solano}, {Solitro}, {Sordo}, {Soria
  Nieto}, {Souchay}, {Spagna}, {Spoto}, {Stampa}, {Steele},
  {Steidelm{\"u}ller}, {Stephenson}, {Stoev}, {Suess}, {Surdej}, {Szabados},
  {Szegedi-Elek}, {Tapiador}, {Taris}, {Tauran}, {Taylor}, {Teixeira},
  {Terrett}, {Teyssand ier}, {Thuillot}, {Titarenko}, {Torra Clotet}, {Turon},
  {Ulla}, {Utrilla}, {Uzzi}, {Vaillant}, {Valentini}, {Valette}, {van Elteren},
  {Van Hemelryck}, {van Leeuwen}, {Vaschetto}, {Vecchiato}, {Veljanoski},
  {Viala}, {Vicente}, {Vogt}, {von Essen}, {Voss}, {Votruba}, {Voutsinas},
  {Walmsley}, {Weiler}, {Wertz}, {Wevers}, {Wyrzykowski}, {Yoldas},
  {{\v{Z}}erjal}, {Ziaeepour}, {Zorec}, {Zschocke}, {Zucker}, {Zurbach}, \&
  {Zwitter}}]{gaiadr2}
{Gaia Collaboration}, {Brown}, A.~G.~A., {Vallenari}, A., {et~al.} 2018, \aap,
  616, A1, \dodoi{10.1051/0004-6361/201833051}

\bibitem[{{Gramajo} {et~al.}(2014){Gramajo}, {Rod{\'o}n}, \&
  {G{\'o}mez}}]{gramajo2014}
{Gramajo}, L.~V., {Rod{\'o}n}, J.~A., \& {G{\'o}mez}, M. 2014, \aj, 147, 140,
  \dodoi{10.1088/0004-6256/147/6/140}

\bibitem[{{Green} {et~al.}(2006){Green}, {Hartmann}, {Calvet}, {Watson},
  {Ibrahimov}, {Furlan}, {Sargent}, \& {Forrest}}]{green2006}
{Green}, J.~D., {Hartmann}, L., {Calvet}, N., {et~al.} 2006, \apj, 648, 1099,
  \dodoi{10.1086/505932}

\bibitem[{{Green} {et~al.}(2016){Green}, {Kraus}, {Rizzuto}, {Ireland},
  {Dupuy}, {Mann}, \& {Kuruwita}}]{green2016}
{Green}, J.~D., {Kraus}, A.~L., {Rizzuto}, A.~C., {et~al.} 2016, \apj, 830, 29,
  \dodoi{10.3847/0004-637X/830/1/29}

\bibitem[{{Green} {et~al.}(2013){Green}, {Evans}, {K{\'o}sp{\'a}l}, {Herczeg},
  {Quanz}, {Henning}, {van Kempen}, {Lee}, {Dunham}, {Meeus}, {Bouwman},
  {Chen}, {G{\"u}del}, {Skinner}, {Liebhart}, \& {Merello}}]{green2013}
{Green}, J.~D., {Evans}, Neal~J., I., {K{\'o}sp{\'a}l}, {\'A}., {et~al.} 2013,
  \apj, 772, 117, \dodoi{10.1088/0004-637X/772/2/117}

\bibitem[{{G\"uver} \& {{\"O}zel}(2009)}]{guver2009}
{G\"uver}, T., \& {{\"O}zel}, F. 2009, \mnras, 400, 2050,
  \dodoi{10.1111/j.1365-2966.2009.15598.x}

\bibitem[{{Hanuschik}(2003)}]{hanuschik03}
{Hanuschik}, R.~W. 2003, \aap, 407, 1157, \dodoi{10.1051/0004-6361:20030885}

\bibitem[{{Hartigan} {et~al.}(1995){Hartigan}, {Edwards}, \&
  {Ghandour}}]{hartigan1995}
{Hartigan}, P., {Edwards}, S., \& {Ghandour}, L. 1995, \apj, 452, 736,
  \dodoi{10.1086/176344}

\bibitem[{{Hartman} \& {Bakos}(2016)}]{hartman2016}
{Hartman}, J.~D., \& {Bakos}, G.~A. 2016, Astronomy and Computing, 17, 1,
  \dodoi{10.1016/j.ascom.2016.05.006}

\bibitem[{{Hartmann}(2009)}]{hartmann2009}
{Hartmann}, L. 2009, {Accretion Processes in Star Formation: Second Edition}
  (Cambridge University Press)

\bibitem[{{Hartmann} \& {Kenyon}(1987{\natexlab{a}})}]{kenyon-and-hartmann1987}
{Hartmann}, L., \& {Kenyon}, S.~J. 1987{\natexlab{a}}, \apj, 322, 393,
  \dodoi{10.1086/165737}

\bibitem[{{Hartmann} \& {Kenyon}(1987{\natexlab{b}})}]{hartmann1987}
---. 1987{\natexlab{b}}, \apj, 322, 393, \dodoi{10.1086/165737}

\bibitem[{{Hartmann} \& {Kenyon}(1996)}]{kenyon&hartmann1996}
---. 1996, \araa, 34, 207, \dodoi{10.1146/annurev.astro.34.1.207}

\bibitem[{{Henden} {et~al.}(2016){Henden}, {Templeton}, {Terrell}, {Smith},
  {Levine}, \& {Welch}}]{henden2016}
{Henden}, A.~A., {Templeton}, M., {Terrell}, D., {et~al.} 2016, VizieR Online
  Data Catalog, II/336

\bibitem[{{Herbig}(1966)}]{herbig1966}
{Herbig}, G.~H. 1966, Vistas in Astronomy, 8, 109,
  \dodoi{10.1016/0083-6656(66)90025-0}

\bibitem[{{Herbig}(1977)}]{herbig1977}
---. 1977, \apj, 217, 693, \dodoi{10.1086/155615}

\bibitem[{{Herbig}(2009)}]{herbig2009}
---. 2009, \aj, 138, 448, \dodoi{10.1088/0004-6256/138/2/448}

\bibitem[{{Herbig} {et~al.}(2003){Herbig}, {Petrov}, \&
  {Duemmler}}]{herbig2003}
{Herbig}, G.~H., {Petrov}, P.~P., \& {Duemmler}, R. 2003, \apj, 595, 384,
  \dodoi{10.1086/377194}

\bibitem[{{Herter} {et~al.}(2013){Herter}, {Vacca}, {Adams}, {Keller},
  {Schoenwald}, {Hirsch}, {Wang}, {De Buizer}, {Helton}, \&
  {Llorens}}]{herter13}
{Herter}, T.~L., {Vacca}, W.~D., {Adams}, J.~D., {et~al.} 2013, \pasp, 125,
  1393, \dodoi{10.1086/674144}

\bibitem[{{Ibragimov}(1997)}]{ibragimov1997}
{Ibragimov}, M.~A. 1997, Astronomy Letters, 23, 103

\bibitem[{{Jordi} {et~al.}(2006){Jordi}, {Grebel}, \& {Ammon}}]{jordi2006}
{Jordi}, K., {Grebel}, E.~K., \& {Ammon}, K. 2006, \aap, 460, 339,
  \dodoi{10.1051/0004-6361:20066082}

\bibitem[{{Kadam} {et~al.}(2020){Kadam}, {Vorobyov}, {Reg{\'a}ly},
  {K{\'o}sp{\'a}l}, \& {{\'A}brah{\'a}m}}]{kadam2020}
{Kadam}, K., {Vorobyov}, E., {Reg{\'a}ly}, Z., {K{\'o}sp{\'a}l}, {\'A}., \&
  {{\'A}brah{\'a}m}, P. 2020, \apj, 895, 41, \dodoi{10.3847/1538-4357/ab8bd8}

\bibitem[{{Kausch} {et~al.}(2015){Kausch}, {Noll}, {Smette}, {Kimeswenger},
  {Barden}, {Szyszka}, {Jones}, {Sana}, {Horst}, \& {Kerber}}]{kausch2015}
{Kausch}, W., {Noll}, S., {Smette}, A., {et~al.} 2015, \aap, 576, A78,
  \dodoi{10.1051/0004-6361/201423909}

\bibitem[{{Kenyon} {et~al.}(1988){Kenyon}, {Hartmann}, \&
  {Hewett}}]{kenyon-and-hartmann1988}
{Kenyon}, S.~J., {Hartmann}, L., \& {Hewett}, R. 1988, \apj, 325, 231,
  \dodoi{10.1086/165999}

\bibitem[{{Kenyon} \& {Hartmann}(1991{\natexlab{a}})}]{kenyon&hartmann1991}
{Kenyon}, S.~J., \& {Hartmann}, L.~W. 1991{\natexlab{a}}, \apj, 383, 664,
  \dodoi{10.1086/170823}

\bibitem[{{Kenyon} \& {Hartmann}(1991{\natexlab{b}})}]{kenyon1991}
---. 1991{\natexlab{b}}, \apj, 383, 664, \dodoi{10.1086/170823}

\bibitem[{{Kenyon} \& {Hartmann}(1991{\natexlab{c}})}]{KH91}
---. 1991{\natexlab{c}}, \apj, 383, 664, \dodoi{10.1086/170823}

\bibitem[{{Kim} {et~al.}(2002){Kim}, {Jang}, {Han}, {Jang}, {Sung}, {Chun},
  {Hyung}, {Yoon}, \& {Vogt}}]{kim2002}
{Kim}, K.-M., {Jang}, B.-H., {Han}, I., {et~al.} 2002, Journal of Korean
  Astronomical Society, 35, 221, \dodoi{10.5303/JKAS.2002.35.4.221}

\bibitem[{{Kochanek} {et~al.}(2017){Kochanek}, {Shappee}, {Stanek}, {Holoien},
  {Thompson}, {Prieto}, {Dong}, {Shields}, {Will}, {Britt}, {Perzanowski}, \&
  {Pojma{\'n}ski}}]{asas-sn-2}
{Kochanek}, C.~S., {Shappee}, B.~J., {Stanek}, K.~Z., {et~al.} 2017, \pasp,
  129, 104502, \dodoi{10.1088/1538-3873/aa80d9}

\bibitem[{{Kolotilov}(1990)}]{kolotilov1990}
{Kolotilov}, E.~A. 1990, Soviet Astronomy Letters, 16, 12

\bibitem[{{Kolotilov} \& {Kenyon}(1997)}]{kolotilov&kenyon1997}
{Kolotilov}, E.~A., \& {Kenyon}, S.~J. 1997, Information Bulletin on Variable
  Stars, 4494, 1

\bibitem[{{Kopatskaya} {et~al.}(2002){Kopatskaya}, {Grinin}, {Shakhovskoi}, \&
  {Shulov}}]{kopatskaya2002}
{Kopatskaya}, E.~N., {Grinin}, V.~P., {Shakhovskoi}, D.~N., \& {Shulov}, O.~S.
  2002, Astrophysics, 45, 143, \dodoi{10.1023/A:1016052529802}

\bibitem[{{Kopatskaya} {et~al.}(2013){Kopatskaya}, {Kolotilov}, \&
  {Arkharov}}]{kopatskaya2013}
{Kopatskaya}, E.~N., {Kolotilov}, E.~A., \& {Arkharov}, A.~A. 2013, \mnras,
  434, 38, \dodoi{10.1093/mnras/stt963}

\bibitem[{{K{\'o}sp{\'a}l} {et~al.}(2017){K{\'o}sp{\'a}l}, {{\'A}brah{\'a}m},
  {Westhues}, \& {Haas}}]{kospal2017a}
{K{\'o}sp{\'a}l}, {\'A}., {{\'A}brah{\'a}m}, P., {Westhues}, C., \& {Haas}, M.
  2017, \aap, 597, L10, \dodoi{10.1051/0004-6361/201629447}

\bibitem[{K{\'{o}}sp{\'{a}}l {et~al.}(2020)K{\'{o}}sp{\'{a}}l, Szab{\'{o}},
  {\'{A}}brah{\'{a}}m, Kraus, Takami, Lucas, Pe{\~{n}}a, \&
  Udalski}]{kospal2020}
K{\'{o}}sp{\'{a}}l, {\'{A}}., Szab{\'{o}}, Z.~M., {\'{A}}brah{\'{a}}m, P.,
  {et~al.} 2020, The Astrophysical Journal, 889, 148,
  \dodoi{10.3847/1538-4357/ab6174}

\bibitem[{{K{\'o}sp{\'a}l} {et~al.}(2011){K{\'o}sp{\'a}l}, {{\'A}brah{\'a}m},
  {Acosta-Pulido}, {Ar{\'e}valo Morales}, {Carnerero}, {Elek}, {Kelemen},
  {Kun}, {P{\'a}l}, {Szak{\'a}ts}, \& {Vida}}]{kospal2011}
{K{\'o}sp{\'a}l}, {\'A}., {{\'A}brah{\'a}m}, P., {Acosta-Pulido}, J.~A.,
  {et~al.} 2011, \aap, 527, A133, \dodoi{10.1051/0004-6361/201016160}

\bibitem[{{K{\'o}sp{\'a}l} {et~al.}(2016){K{\'o}sp{\'a}l}, {{\'A}brah{\'a}m},
  {Acosta-Pulido}, {Dunham}, {Garc{\'\i}a-{\'A}lvarez}, {Hogerheijde}, {Kun},
  {Mo{\'o}r}, {Farkas}, {Hajdu}, {Hodos{\'a}n}, {Kov{\'a}cs}, {Kriskovics},
  {Marton}, {Moln{\'a}r}, {P{\'a}l}, {S{\'a}rneczky}, {S{\'o}dor},
  {Szak{\'a}ts}, {Szalai}, {Szegedi-Elek}, {Szing}, {T{\'o}th}, {Vida}, \&
  {Vink{\'o}}}]{kospal2016}
---. 2016, \aap, 596, A52, \dodoi{10.1051/0004-6361/201528061}

\bibitem[{{Kuhn} {et~al.}(2020){Kuhn}, {Hillenbrand}, {Carpenter}, \& {Avelar
  Menendez}}]{kuhn2020}
{Kuhn}, M.~A., {Hillenbrand}, L.~A., {Carpenter}, J.~M., \& {Avelar Menendez},
  A.~R. 2020, arXiv e-prints, arXiv:2006.08622.
\newblock \doarXiv{2006.08622}

\bibitem[{{Kun} {et~al.}(2019){Kun}, {{\'A}brah{\'a}m}, {Acosta Pulido},
  {Mo{\'o}r}, \& {Prusti}}]{kun2019}
{Kun}, M., {{\'A}brah{\'a}m}, P., {Acosta Pulido}, J.~A., {Mo{\'o}r}, A., \&
  {Prusti}, T. 2019, \mnras, 483, 4424, \dodoi{10.1093/mnras/sty3425}

\bibitem[{{Landolt}(1975)}]{landolt1975}
{Landolt}, A.~U. 1975, \pasp, 87, 379, \dodoi{10.1086/129778}

\bibitem[{{Landolt}(1977)}]{landolt1977}
---. 1977, \pasp, 89, 704, \dodoi{10.1086/130213}

\bibitem[{{Laugalys} {et~al.}(2006){Laugalys}, {Strai{\v{z}}ys}, {Vrba},
  {Boyle}, {Philip}, \& {Kazlauskas}}]{laugalys2006}
{Laugalys}, V., {Strai{\v{z}}ys}, V., {Vrba}, F.~J., {et~al.} 2006, Baltic
  Astronomy, 15, 483

\bibitem[{{Lee} {et~al.}(2015){Lee}, {Park}, {Green}, {Cochran}, {Kang}, {Lee},
  \& {Sung}}]{lee2015}
{Lee}, J.-E., {Park}, S., {Green}, J.~D., {et~al.} 2015, \apj, 807, 84,
  \dodoi{10.1088/0004-637X/807/1/84}

\bibitem[{{Lin} \& {Papaloizou}(1985)}]{lin1985}
{Lin}, D.~N.~C., \& {Papaloizou}, J. 1985, in Protostars and Planets II, ed.
  D.~C. {Black} \& M.~S. {Matthews}, 981--1072

\bibitem[{{Liu} {et~al.}(2018){Liu}, {Dunham}, {Pascucci}, {Bourke}, {Hirano},
  {Longmore}, {Andrews}, {Carrasco-Gonz{\'a}lez}, {Forbrich},
  {Galv{\'a}n-Madrid}, {Girart}, {Green}, {Ju{\'a}rez}, {K{\'o}sp{\'a}l},
  {Manara}, {Palau}, {Takami}, {Testi}, \& {Vorobyov}}]{liu2018}
{Liu}, H.~B., {Dunham}, M.~M., {Pascucci}, I., {et~al.} 2018, \aap, 612, A54,
  \dodoi{10.1051/0004-6361/201731951}

\bibitem[{{Lodato} \& {Clarke}(2004)}]{lodato2004}
{Lodato}, G., \& {Clarke}, C.~J. 2004, \mnras, 353, 841–852,
  \dodoi{10.1111/j.1365-2966.2004.08112.x}

\bibitem[{{Loomis} {et~al.}(2017){Loomis}, {Öberg}, {Andrews}, \&
  {MacGregor}}]{Loomis2017}
{Loomis}, R.~A., {Öberg}, K.~I., {Andrews}, S.~M., \& {MacGregor}, M.~A. 2017,
  \apj, 840, 8, \dodoi{10.3847/1538-4357/aa6c63}

\bibitem[{{Magakian} {et~al.}(2013){Magakian}, {Nikogossian}, {Movsessian},
  {Moiseev}, {Aspin}, {Davis}, {Pyo}, {Khanzadyan}, {Froebrich}, {Smith},
  {Moriarty-Schieven}, \& {Beck}}]{magakian2013}
{Magakian}, T.~Y., {Nikogossian}, E.~H., {Movsessian}, T., {et~al.} 2013,
  \mnras, 432, 2685, \dodoi{10.1093/mnras/stt626}

\bibitem[{{McGinnis} {et~al.}(2015){McGinnis}, {Alencar}, {Guimarães},
  {Sousa}, {Stauffer}, {Bouvier}, {Rebull}, {Fonseca}, {Venuti}, {Hillenbrand},
  {Cody}, {Teixeira}, {Aigrain}, {Favata}, {Fűrész}, {Vrba}, {Flaccomio},
  {Turner}, {Gameiro}, {Dougados}, {Herbst}, {Morales-Calderón}, \&
  {Micela}}]{McGinnis2015}
{McGinnis}, P.~T., {Alencar}, S. H.~P., {Guimarães}, M.~M., {et~al.} 2015,
  \aap, 577, \dodoi{10.1051/0004-6361/201425475}

\bibitem[{{Mendoza}(1971)}]{mendoza1971}
{Mendoza}, E.~E. 1971, Boletin de los Observatorios Tonantzintla y Tacubaya, 6,
  135

\bibitem[{{Miller} {et~al.}(2011){Miller}, {Hillenbrand}, {Covey}, {Poznanski},
  {Silverman}, {Kleiser}, {Rojas-Ayala}, {Muirhead}, {Cenko}, {Bloom},
  {Kasliwal}, {Filippenko}, {Law}, {Ofek}, {Dekany}, {Rahmer}, {Hale}, {Smith},
  {Quimby}, {Nugent}, {Jacobsen}, {Zolkower}, {Velur}, {Walters}, {Henning},
  {Bui}, {McKenna}, {Kulkarni}, {Klein}, {Kandrashoff}, \&
  {Morton}}]{miller2011}
{Miller}, A.~A., {Hillenbrand}, L.~A., {Covey}, K.~R., {et~al.} 2011, \apj,
  730, 80, \dodoi{10.1088/0004-637X/730/2/80}

\bibitem[{{Milliner} {et~al.}(2019){Milliner}, {Matthews}, {Long}, \&
  {Hartmann}}]{milliner2019}
{Milliner}, K., {Matthews}, J.~H., {Long}, K.~S., \& {Hartmann}, L. 2019,
  \mnras, 483, 1663–1673, \dodoi{10.1093/mnras/sty3197}

\bibitem[{{Mould} {et~al.}(1978){Mould}, {Hall}, {Ridgway}, {Hintzen}, \&
  {Aaronson}}]{mould1978}
{Mould}, J.~R., {Hall}, D.~N.~B., {Ridgway}, S.~T., {Hintzen}, P., \&
  {Aaronson}, M. 1978, \apjl, 222, L123, \dodoi{10.1086/182706}

\bibitem[{{Paczynski}(1976)}]{paczynski1976}
{Paczynski}, B. 1976, in IAU Symposium, Vol.~73, Structure and Evolution of
  Close Binary Systems, ed. P.~{Eggleton}, S.~{Mitton}, \& J.~{Whelan}, 75

\bibitem[{{P{\'a}l}(2012)}]{pal2012}
{P{\'a}l}, A. 2012, \mnras, 421, 1825, \dodoi{10.1111/j.1365-2966.2011.19813.x}

\bibitem[{{P{\'a}l} {et~al.}(2020){P{\'a}l}, {Szak{\'a}ts}, {Kiss}, {B{\'o}di},
  {Bogn{\'a}r}, {Kalup}, {Kiss}, {Marton}, {Moln{\'a}r}, {Plachy},
  {S{\'a}rneczky}, {Szab{\'o}}, \& {Szab{\'o}}}]{2020ApJS..247...26P}
{P{\'a}l}, A., {Szak{\'a}ts}, R., {Kiss}, C., {et~al.} 2020, \apjs, 247, 26,
  \dodoi{10.3847/1538-4365/ab64f0}

\bibitem[{{Park} {et~al.}(2020){Park}, {Lee}, {Pyo}, {Jaffe}, {Mace}, {Sung},
  {Lee}, {Kang}, {Oh}, {Yoon}, {Yoon}, \& {Green}}]{park2020}
{Park}, S., {Lee}, J.-E., {Pyo}, T.-S., {et~al.} 2020, \apj, 900, 36,
  \dodoi{10.3847/1538-4357/aba532}

\bibitem[{{Powell} {et~al.}(2012){Powell}, {Irwin}, {Bouvier}, \&
  {Clarke}}]{powell2012}
{Powell}, S.~L., {Irwin}, M., {Bouvier}, J., \& {Clarke}, C.~J. 2012, \mnras,
  426, 3315, \dodoi{10.1111/j.1365-2966.2012.21898.x}

\bibitem[{{Press}(1978)}]{press1978}
{Press}, W.~H. 1978, Comments on Modern Physics, Part C - Comments on
  Astrophysics, 7, 103

\bibitem[{{Rebull} {et~al.}(2011){Rebull}, {Guieu}, {Stauffer}, {Hillenbrand },
  {Noriega-Crespo}, {Stapelfeldt}, {Carey}, {Carpenter}, {Cole}, {Padgett},
  {Strom}, \& {Wolff}}]{rebull2011}
{Rebull}, L.~M., {Guieu}, S., {Stauffer}, J.~R., {et~al.} 2011, \apjs, 193, 25,
  \dodoi{10.1088/0067-0049/193/2/25}

\bibitem[{{Reggiani} {et~al.}(2014){Reggiani}, {Quanz}, {Meyer}, {Pueyo},
  {Absiland}, \& et~al.}]{Reggiani2014}
{Reggiani}, M., {Quanz}, S.~P., {Meyer}, M.~R., {et~al.} 2014, \apjl, 792, 5,
  \dodoi{10.1088/2041-8205/792/1/L23}

\bibitem[{{Reipurth} \& {Aspin}(2010)}]{reipurth2010}
{Reipurth}, B., \& {Aspin}, C. 2010, in Evolution of Cosmic Objects through
  their Physical Activity, ed. H.~A. {Harutyunian}, A.~M. {Mickaelian}, \&
  Y.~{Terzian}, 19--38

\bibitem[{{Ricker} {et~al.}(2015){Ricker}, {Winn}, {Vanderspek}, {Latham},
  {Bakos}, {Bean}, {Berta-Thompson}, {Brown}, {Buchhave}, {Butler}, {Butler},
  {Chaplin}, {Charbonneau}, {Christensen-Dalsgaard}, {Clampin}, {Deming},
  {Doty}, {De Lee}, {Dressing}, {Dunham}, {Endl}, {Fressin}, {Ge}, {Henning},
  {Holman}, {Howard}, {Ida}, {Jenkins}, {Jernigan}, {Johnson}, {Kaltenegger},
  {Kawai}, {Kjeldsen}, {Laughlin}, {Levine}, {Lin}, {Lissauer}, {MacQueen},
  {Marcy}, {McCullough}, {Morton}, {Narita}, {Paegert}, {Palle}, {Pepe},
  {Pepper}, {Quirrenbach}, {Rinehart}, {Sasselov}, {Sato}, {Seager},
  {Sozzetti}, {Stassun}, {Sullivan}, {Szentgyorgyi}, {Torres}, {Udry}, \&
  {Villasenor}}]{ricker2015}
{Ricker}, G.~R., {Winn}, J.~N., {Vanderspek}, R., {et~al.} 2015, Journal of
  Astronomical Telescopes, Instruments, and Systems, 1, 014003,
  \dodoi{10.1117/1.JATIS.1.1.014003}

\bibitem[{{Rieke} {et~al.}(1972){Rieke}, {Lee}, \& {Coyne}}]{rieke1972}
{Rieke}, G., {Lee}, T., \& {Coyne}, G. 1972, \pasp, 84, 37,
  \dodoi{10.1086/129242}

\bibitem[{{Rucinski} {et~al.}(2008){Rucinski}, {Matthews}, {Kuschnig}, G.,
  {Rowe}, {Guenther}, {Moffat}, {Sasselov}, {Walker}, \& {Weiss}}]{ruc08}
{Rucinski}, S.~M., {Matthews}, J.~M., {Kuschnig}, R., {et~al.} 2008, \mnras,
  391, 11, \dodoi{doi:10.1111/j.1365-2966.2008.14014.x}

\bibitem[{{Schwarzenberg-Czerny}(1997)}]{Czerny1997}
{Schwarzenberg-Czerny}, A. 1997, \apj, 489, 941–945

\bibitem[{{Shappee} {et~al.}(2014){Shappee}, {Prieto}, {Grupe}, {Kochanek},
  {Stanek}, {De Rosa}, {Mathur}, {Zu}, {Peterson}, {Pogge}, {Komossa}, {Im},
  {Jencson}, {Holoien}, {Basu}, {Beacom}, {Szczygie{\l}}, {Brimacombe},
  {Adams}, {Campillay}, {Choi}, {Contreras}, {Dietrich}, {Dubberley},
  {Elphick}, {Foale}, {Giustini}, {Gonzalez}, {Hawkins}, {Howell}, {Hsiao},
  {Koss}, {Leighly}, {Morrell}, {Mudd}, {Mullins}, {Nugent}, {Parrent},
  {Phillips}, {Pojmanski}, {Rosing}, {Ross}, {Sand}, {Terndrup}, {Valenti},
  {Walker}, \& {Yoon}}]{asas-sn-1}
{Shappee}, B.~J., {Prieto}, J.~L., {Grupe}, D., {et~al.} 2014, \apj, 788, 48,
  \dodoi{10.1088/0004-637X/788/1/48}

\bibitem[{{Sicilia-Aguilar} {et~al.}(2020){Sicilia-Aguilar}, {Bouvier},
  {Dougados}, {Grankin}, \& {Donati}}]{Sicilia-Aguilar2020}
{Sicilia-Aguilar}, A., {Bouvier}, J., {Dougados}, C., {Grankin}, K., \&
  {Donati}, J.~F. 2020, \aap, 643, A29, \dodoi{10.1051/0004-6361/202038489}

\bibitem[{{Simon}(1975)}]{simon1975}
{Simon}, T. 1975, \pasp, 87, 317, \dodoi{10.1086/129761}

\bibitem[{{Simon} \& {Joyce}(1988)}]{simon1988}
{Simon}, T., \& {Joyce}, R.~R. 1988, \pasp, 100, 1549, \dodoi{10.1086/132364}

\bibitem[{{Simon} {et~al.}(1982){Simon}, {Wolstencroft}, {Dyck}, \&
  {Joyce}}]{simon1982}
{Simon}, T., {Wolstencroft}, R.~D., {Dyck}, H.~M., \& {Joyce}, R.~R. 1982,
  Information Bulletin on Variable Stars, 2155, 1

\bibitem[{{Siwak} {et~al.}(2020){Siwak}, {Og{\l}oza}, \&
  {Krzesi{\'n}ski}}]{siwak2020}
{Siwak}, M., {Og{\l}oza}, W., \& {Krzesi{\'n}ski}, J. 2020, \aap, 644,
  \dodoi{10.1051/0004-6361/202037607}

\bibitem[{{Siwak} {et~al.}(2018){Siwak}, {Winiarski}, {Og{\l}oza},
  {Dro{\.z}d{\.z}}, {Zo{\l}a}, {Moffat}, {Stachowski}, {Rucinski}, {Cameron},
  {Matthews}, {Weiss}, {Kuschnig}, {Rowe}, {Guenther}, \&
  {Sasselov}}]{siwak2018}
{Siwak}, M., {Winiarski}, M., {Og{\l}oza}, W., {et~al.} 2018, \aap, 618,
  \dodoi{10.1051/0004-6361/201833401}

\bibitem[{{Skinner} {et~al.}(2009){Skinner}, {Sokal}, {G{\"u}del}, \&
  {Briggs}}]{skinner2009}
{Skinner}, S.~L., {Sokal}, K.~R., {G{\"u}del}, M., \& {Briggs}, K.~R. 2009,
  \apj, 696, 766, \dodoi{10.1088/0004-637X/696/1/766}

\bibitem[{{Smette} {et~al.}(2015){Smette}, {Sana}, {Noll}, {Horst}, {Kausch},
  {Kimeswenger}, {Barden}, {Szyszka}, {Jones}, {Gallenne}, {Vinther},
  {Ballester}, \& {Taylor}}]{smette2015}
{Smette}, A., {Sana}, H., {Noll}, S., {et~al.} 2015, \aap, 576, A77,
  \dodoi{10.1051/0004-6361/201423932}

\bibitem[{{Stellingwerf}(1978)}]{stellingwerf1978}
{Stellingwerf}, R.~F. 1978, \apj, 224, 953, \dodoi{10.1086/156444}

\bibitem[{{Szegedi-Elek} {et~al.}(2020){Szegedi-Elek}, {{\'A}brah{\'a}m},
  {Wyrzykowski}, {Kun}, {K{\'o}sp{\'a}l}, {Chen}, {Marton}, {Mo{\'o}r}, {Kiss},
  {P{\'a}l}, {Szabados}, {Varga}, {Varga-Vereb{\'e}lyi}, {Andreas}, {Bachelet},
  {Bischoff}, {B{\'o}di}, {Breedt}, {Burgaz}, {Butterley}, {Carrasco},
  {{\v{C}}epas}, {Damljanovic}, {Gezer}, {Godunova}, {Gromadzki}, {Gurgul},
  {Hardy}, {Hildebrandt}, {Hoffmann}, {Hundertmark}, {Ihanec}, {Janulis},
  {Kalup}, {Kaczmarek}, {K{\"o}nyves-T{\'o}th}, {Krezinger}, {Kruszy{\'n}ska},
  {Littlefair}, {Maskoli{\={u}}nas}, {M{\'e}sz{\'a}ros}, {Miko{\l}ajczyk},
  {Mugrauer}, {Netzel}, {Ordasi}, {Pak{\v{s}}tien{\.{e}}}, {Rybicki},
  {S{\'a}rneczky}, {Seli}, {Simon}, {{\v{S}}i{\v{s}}kauskait{\.{e}}},
  {S{\'o}dor}, {Sokolovsky}, {Stenglein}, {Street}, {Szak{\'a}ts}, {Tomasella},
  {Tsapras}, {Vida}, {Zdanavi{\v{c}}ius}, {Zieli{\'n}ski}, {Zieli{\'n}ski}, \&
  {Zi{\'o}{\l}kowska}}]{szegedi-elek2020}
{Szegedi-Elek}, E., {{\'A}brah{\'a}m}, P., {Wyrzykowski}, {\L}., {et~al.} 2020,
  \apj, 899, 130, \dodoi{10.3847/1538-4357/aba129}

\bibitem[{{Szul{\'a}gyi} \& {Mordasini}(2017)}]{Szulagyi2017}
{Szul{\'a}gyi}, J., \& {Mordasini}, C. 2017, \mnras, 465, L64–L68,
  \dodoi{10.1093/mnrasl/slw212}

\bibitem[{{Takagi} {et~al.}(2018){Takagi}, {Honda}, {Arai}, {Morihana},
  {Takahashi}, {Oasa}, \& {Itoh}}]{takagi18}
{Takagi}, Y., {Honda}, S., {Arai}, A., {et~al.} 2018, \aj, 155, 101,
  \dodoi{10.3847/1538-3881/aaa545}

\bibitem[{{Tody}(1986)}]{tody1986}
{Tody}, D. 1986, in Society of Photo-Optical Instrumentation Engineers (SPIE)
  Conference Series, Vol. 627, Instrumentation in astronomy VI, ed. D.~L.
  {Crawford}, 733, \dodoi{10.1117/12.968154}

\bibitem[{{Tody}(1993)}]{tody1993}
{Tody}, D. 1993, in Astronomical Society of the Pacific Conference Series,
  Vol.~52, Astronomical Data Analysis Software and Systems II, ed. R.~J.
  {Hanisch}, R.~J.~V. {Brissenden}, \& J.~{Barnes}, 173

\bibitem[{{Turner} {et~al.}(1997){Turner}, {Bodenheimer}, \&
  {Bell}}]{turner1997}
{Turner}, N.~J.~J., {Bodenheimer}, P., \& {Bell}, K.~R. 1997, \apj, 480, 754,
  \dodoi{10.1086/303983}

\bibitem[{{Vorobyov} \& {Elbakyan}(2018)}]{vorobyov2018}
{Vorobyov}, E.~I., \& {Elbakyan}, V.~G. 2018, \aap, 618,
  \dodoi{10.1051/0004-6361/201833226}

\bibitem[{{Vorobyov} {et~al.}(2021){Vorobyov}, {Elbakyan}, {Liu}, \&
  {Takami}}]{vorobyov2021}
{Vorobyov}, E.~I., {Elbakyan}, V.~G., {Liu}, H.~B., \& {Takami}, M. 2021, \aap,
  647, \dodoi{10.1051/0004-6361/202039391}

\bibitem[{{Wachmann}(1954)}]{wachmann1954}
{Wachmann}, A. 1954, \zap, 35, 74

\bibitem[{{Welin}(1971{\natexlab{a}})}]{welin1971a}
{Welin}, G. 1971{\natexlab{a}}, \aap, 12, 312

\bibitem[{{Welin}(1971{\natexlab{b}})}]{welin1971b}
---. 1971{\natexlab{b}}, Information Bulletin on Variable Stars, 581, 1

\bibitem[{{Welin}(1975)}]{welin1975}
---. 1975, Information Bulletin on Variable Stars, 1057, 1

\bibitem[{{Welin}(1976)}]{welin1976}
---. 1976, Information Bulletin on Variable Stars, 1195, 1

\bibitem[{{Wendker}(1983)}]{wendker1983}
{Wendker}, H.~J. 1983, in Wissenschaftsberichte aus der Universit\&auml;t
  Hamburg: 150 Jahre Hamburger Sternwarte, 49--54

\bibitem[{{Wright} {et~al.}(2010){Wright}, {Eisenhardt}, {Mainzer}, {Ressler},
  {Cutri}, {Jarrett}, {Kirkpatrick}, {Padgett}, {McMillan}, {Skrutskie},
  {Stanford}, {Cohen}, {Walker}, {Mather}, {Leisawitz}, {Gautier}, {McLean},
  {Benford}, {Lonsdale}, {Blain}, {Mendez}, {Irace}, {Duval}, {Liu}, {Royer},
  {Heinrichsen}, {Howard}, {Shannon}, {Kendall}, {Walsh}, {Larsen}, {Cardon},
  {Schick}, {Schwalm}, {Abid}, {Fabinsky}, {Naes}, \& {Tsai}}]{wright2010}
{Wright}, E.~L., {Eisenhardt}, P. R.~M., {Mainzer}, A.~K., {et~al.} 2010, \aj,
  140, 1868, \dodoi{10.1088/0004-6256/140/6/1868}

\bibitem[{{Young} {et~al.}(2012){Young}, {Becklin}, {Marcum}, {Roellig}, {De
  Buizer}, {Herter}, {G{\"u}sten}, {Dunham}, {Temi}, {Andersson}, {Backman},
  {Burgdorf}, {Caroff}, {Casey}, {Davidson}, {Erickson}, {Gehrz}, {Harper},
  {Harvey}, {Helton}, {Horner}, {Howard}, {Klein}, {Krabbe}, {McLean}, {Meyer},
  {Miles}, {Morris}, {Reach}, {Rho}, {Richter}, {Roeser}, {Sandell}, {Sankrit},
  {Savage}, {Smith}, {Shuping}, {Vacca}, {Vaillancourt}, {Wolf}, \&
  {Zinnecker}}]{young12}
{Young}, E.~T., {Becklin}, E.~E., {Marcum}, P.~M., {et~al.} 2012, \apjl, 749,
  L17, \dodoi{10.1088/2041-8205/749/2/L17}

\bibitem[{{Zhu} {et~al.}(2008){Zhu}, {Hartmann}, {Calvet}, {Hernand ez},
  {Tannirkulam}, \& {D'Alessio}}]{zhu2008}
{Zhu}, Z., {Hartmann}, L., {Calvet}, N., {et~al.} 2008, \apj, 684, 1281,
  \dodoi{10.1086/590241}

\end{thebibliography}
\bibliographystyle{aasjournal}

\end{document}